\documentclass[preprint,12pt, 3p]{elsarticle}

\journal{Journal of Computational Physics}
\usepackage{graphics}
\usepackage{setspace}
\usepackage{verbatim}
 \usepackage{graphicx}
 \usepackage{subfig}
 \usepackage{float}
\usepackage{booktabs}
\usepackage{amssymb}
\usepackage{bm}
\usepackage{amsfonts}
\usepackage{amsmath}
\newcommand*\colvec[3][]{    \begin{pmatrix}\ifx\relax#1\relax\else#1\\\fi#2\\#3\end{pmatrix}}

\providecommand{\tabularnewline}{\\}

\usepackage{color}
\usepackage{amssymb}

\def\spacce#1{\hskip #1pt}
\def\drawline#1#2{\raise 2.5pt\vbox{\hrule width #1pt height #2pt}}

\def\solid{\drawline{22}{1.0}\nobreak\ }

\def\tdash{\hbox{\drawline{4}{0.8}\spacce{2}}}
\def\dashed{\tdash \tdash \tdash \tdash \nobreak\ }

\def\tdot{\hbox{\drawline{1}{1.0}\spacce{2}}}
\def\dotted{\tdot \tdot \tdot \tdot \tdot \tdot \tdot \tdot \nobreak\ }

\def\dashdot{\tdash \tdot \tdash \tdot \tdash \nobreak\ }

\def\ssquare{${\vcenter{\hrule height 0.9pt
       \hbox{\vrule width 0.9pt height 10pt \kern 10pt
       \vrule width 0.9pt}
       \hrule height 0.9pt}}$\nobreak\ }
       
\def\ssquareb{$\Box$\nobreak\ }       
       
\def\blackssquare{$\scriptstyle\blacksquare$\nobreak\ }
       
\def\circle{$\circ$\nobreak\ }

\def\blackcircle{$\bullet$\nobreak\ }

\def\bigcircle{$\bigcirc$\nobreak\ }

\def\losange{$\Diamond$\nobreak\ }

\def\blacklosange{$\blacklozenge$\nobreak\ }

\def\trianup{\raise 1.25pt\hbox{$\triangle$}\nobreak\ }

\def\blacktrianup{\raise 1.25pt\hbox{$\blacktriangle$}\nobreak\ }

\def\bigtriandown{\raise 1.25pt\hbox{$\bigtriangledown$}\nobreak\ } 

\def\plus{$+$ \nobreak}

\def\asterix{$\ast$ \nobreak}

\def\solidopencircle{\solid \nobreak\spacce{-19.5}\raise
 -0.5pt\hbox{\circle} \nobreak\ }
 
\def\solidblackcircle{\solid \nobreak\spacce{-21}\raise
 -0pt\hbox{\blackcircle} \nobreak\ }

\def\solidopensquare{\solid \nobreak\spacce{-19}\raise
 -1.0pt\hbox{\ssquare} \nobreak\ }
 
\def\solidblacksquare{\solid \nobreak\spacce{-19}\raise
 -1.0pt\hbox{\blackssquare} \nobreak\ }
 
\def\solidopenlosange{\solid \nobreak\spacce{-19}\raise
 -1.0pt\hbox{\losange} \nobreak\ }
 
\def\solidblacklosange{\solid \nobreak\spacce{-21}\raise
 -2.0pt\hbox{\blacklosange} \nobreak\ }

\def\solidopentriangleup{\solid \nobreak\spacce{-19}\raise
 -1.pt\hbox{\trianup} \nobreak\ }
 
\def\solidblacktriangleup{\solid \nobreak\spacce{-21.5}\raise
 -2.0pt\hbox{\blacktrianup} \nobreak\ }

\def\solidopencross{\solid \nobreak\spacce{-22}\raise
 -1.0pt\hbox{\cross} \nobreak\ }
 
\def\solidopenplus{\solid \nobreak\spacce{-21}\raise
 -1.5pt\hbox{\plus} \nobreak\ }
 
\def\solidopenasterix{\solid \nobreak\spacce{-21}\raise
 -1.5pt\hbox{\asterix} \nobreak\ }

\def\dashedopencircle{\dashed \nobreak\spacce{-22}\raise
 -1.0pt\hbox{\circle} \nobreak\ }
 
\def\dashedblackcircle{\dashed \nobreak\spacce{-23}\raise
 -1.4pt\hbox{\blackcircle} \nobreak\ }
 
\def\dashedopensquare{\dashed \nobreak\spacce{-22}\raise
 -1.0pt\hbox{\ssquare} \nobreak\ }
 
\def\dashedblacksquare{\dashed \nobreak\spacce{-23}\raise
 -1.0pt\hbox{\blackssquare} \nobreak\ }
 
\def\dashedopenlosange{\dashed \nobreak\spacce{-22}\raise
 -1.0pt\hbox{\losange} \nobreak\ }
 
\def\dashedblacklosange{\dashed \nobreak\spacce{-23}\raise
 -2.0pt\hbox{\blacklosange} \nobreak\ }
 
\def\dashedopentriangleup{\dashed \nobreak\spacce{-22}\raise
 -1.0pt\hbox{\trianup} \nobreak\ }
 
\def\dashedblacktriangleup{\dashed \nobreak\spacce{-24}\raise
 -2.0pt\hbox{\blacktrianup} \nobreak\ }
 
\def\dashedopencross{\dashed \nobreak\spacce{-24}\raise
 -1.0pt\hbox{\cross} \nobreak\ }
 
\def\dashedopenplus{\dashed \nobreak\spacce{-24}\raise
 -1.0pt\hbox{\plus} \nobreak\ }

\def\dashedopenasterix{\dashed \nobreak\spacce{-22}\raise
 -1.5pt\hbox{\asterix} \nobreak\ }

\def\dottedopencircle{\dotted \nobreak\spacce{-22}\raise
 -1.0pt\hbox{\circle} \nobreak\ }
 
\def\dottedblackcircle{\dotted \nobreak\spacce{-23}\raise
 -1.4pt\hbox{\blackcircle} \nobreak\ }

\def\dottedopensquare{\dotted \nobreak\spacce{-22}\raise
 -1.0pt\hbox{\ssquare} \nobreak\ }
 
\def\dotted\mathbfuare{\dotted \nobreak\spacce{-23}\raise
 -1.0pt\hbox{\blackssquare} \nobreak\ }
 
\def\dottedopenlosange{\dotted \nobreak\spacce{-22}\raise
 -1.0pt\hbox{\losange} \nobreak\ }
 
\def\dottedblacklosange{\dotted \nobreak\spacce{-24}\raise
 -2.0pt\hbox{\blacklosange} \nobreak\ }
 
\def\dottedopentriangleup{\dotted \nobreak\spacce{-22}\raise
 -1.0pt\hbox{\trianup} \nobreak\ }
 
\def\dottedblacktriangleup{\dotted \nobreak\spacce{-25}\raise
 -2.0pt\hbox{\blacktrianup} \nobreak\ }
 
\def\dottedopencross{\dotted \nobreak\spacce{-23.5}\raise
 -1.0pt\hbox{\cross} \nobreak\ }
 
\def\dottedopenplus{\dotted \nobreak\spacce{-23.5}\raise
 -1.0pt\hbox{\plus} \nobreak\ }

\def\dottedopenasterix{\dotted \nobreak\spacce{-23.5}\raise
 -1.0pt\hbox{\asterix} \nobreak\ }

\def\whitehisto{${\vcenter{\color{black} \hrule height 1.0pt
       \hbox{\vrule width 1.0pt height 4pt \kern 8pt
       \vrule width 1.0pt}
       \hrule height 1.0pt}}$\nobreak\ }
       
\def\histosymb#1{${\vcenter{\color{#1} \hrule height 0.0pt
       \hbox{\vrule width 11.0pt height 6pt \kern 0pt 
       \vrule width 0.0pt}
       \hrule height 0.0pt}}$\nobreak\ }

\def\symbol#1#2#3#4#5#6{\color{#5}#1 \nobreak\spacce{-#3}\raise
 -#4pt\hbox{\color{#6}#2}\color{black} \nobreak\ }

\journal{Journal Computational Physics}

\begin{document}
\begin{frontmatter}

\title{Large-scale simulation of steady and time-dependent active suspensions with the force-coupling method}

\author[label1,label2]{Blaise Delmotte \corref{cor1}}

\ead{blaise.delmotte@imft.fr}
\address[label1]{University of Toulouse – INPT-UPS: Institut de M\'ecanique des Fluides, Toulouse, France.}
\address[label2]{IMFT - CNRS, UMR 5502 1, All\'ee du Professeur Camille Soula, 31 400 Toulouse, France.}

\author[label3]{Eric E. Keaveny \corref{cor1}}
\ead{e.keaveny@imperial.ac.uk}
\address[label3]{Department of Mathematics, Imperial College London, South Kensington Campus, London, SW7 2AZ, UK}

\cortext[cor1]{Corresponding authors: Blaise Delmotte, Eric E. Keaveny}

\author[label1,label2]{Franck Plourabou\'e}
\ead{franck.plouraboue@imft.fr}

\author[label1,label2]{Eric Climent}
\ead{eric.climent@imft.fr}

\begin{abstract}
We present a new development of the force-coupling method (FCM) to address the accurate simulation of a large number of interacting micro-swimmers.  Our approach is based on the squirmer model, which we adapt to the FCM framework, resulting in a method that is suitable for simulating semi-dilute squirmer suspensions.  Other effects, such as steric interactions, are considered with our model.  We test our method by comparing the velocity field around a single squirmer and the pairwise interactions between two squirmers with exact solutions to the Stokes equations and results given by other numerical methods.  We also illustrate our method's ability to describe spheroidal swimmer shapes and biologically-relevant time-dependent swimming gaits.  We detail the numerical algorithm used to compute  the  hydrodynamic coupling between a  large collection ($10^4-10 ^5$) of micro-swimmers.  Using this methodology,  we investigate the emergence of polar order in a suspension of squirmers and show that for large domains, both the steady-state polar order parameter and the growth rate of instability are independent of system size.  These results demonstrate the effectiveness of our approach to achieve near continuum-level results, allowing for better comparison with experimental measurements while complementing and informing continuum models.  
\end{abstract}


\end{frontmatter}

\section{Introduction}

Suspensions of active, self-propelled particles arise in both biological systems, such as populations of micro-organisms \cite{dombrowski_04, Zhang2010,Sokolov_2012,ryan_2013} and synthetic, colloidal systems \cite{Thutupalli2011}.  These suspensions can exhibit the formation of coherent structures and complex flow patterns which may lead to enhanced mixing of chemicals in the surrounding fluid, the alteration of suspension rheology, or, in the biological case, increased nutrient uptake by a population of micro-organisms.  In addition to promising applications such as algae biofuels \cite{Croze2013, Bees2014}, characterizing the collective dynamics found in these suspensions is of fundamental importance to understanding zooplankton dynamics \cite{Guasto2012,Kiorboe2014} and  mammal fertility \cite{Riedel2005,Creppy2013}.

The mathematical modeling of active suspensions entails describing how individual swimmers move and interact in response to the flow fields that they generate \cite{Lauga2009,Ishikawa2009,Saintillan2013}.  It is particularly important for these models to be able to handle a large collection of swimmers in order to obtain suspension properties at the lab/\emph{in situ} scale.  The modeling of the collective behavior of active matter has been a vibrant area of research during the last decade \cite{Ramaswamy_10,Koch2011,Saintillan2013, Marchetti2014}, to cite only a few recent reviews.  Generally speaking, the modelling approaches can be sorted into two categories: continuum theories and particle-based simulations.
Most of the continuum models are generally valid for dilute suspensions where the hydrodynamic disturbances are given by a mean-field description of far-field hydrodynamic interactions \cite{Saintillan2008,Baskaran2009,Koch2011}.  Recent advances towards more concentrated suspensions include steric interactions \cite{Ezhilan2013}, but the inclusion of high-order singularities due to particle size remains outstanding.  Despite this, these models are very attractive as they naturally provide a description of the dynamics at the population level and the resulting equations can be analysed using a wide range of analytical and numerical techniques.  

Particle-based simulations resolve the dynamics of each individual swimmer and from their positions and orientations, construct a picture of the dynamics of the suspension as a whole.  As discussed in \cite{Koch2011}, particle-based models provide opportunities to (i) test continuum theories, (ii) analyse finite-size effects resulting from a discrete number of swimmers,  (iii) explore more complex interactions between swimmers and/or boundaries, and in some cases, (iv) reveal the effects of short-range hydrodynamic interactions and/or steric repulsion.
Various models have been proposed in this context, each using different approximations to address the difficult problems of resolving the hydrodynamic interactions and incorporating the geometry of the swimmers.  
Some of the first such models used point force distributions to create dumbbell-shaped swimmers \cite{hernandez-Ortiz_2005,hernandez-Ortiz_2007,underhill_2005}, slender-body theory to model a slip velocity along the surfaces of rod-like swimmers \cite{Saintillan_Shelley_2007,saintillan_Shelley_2012}, or the squirmer model \cite{Lighthill1952,Blake1971} to examine the interactions between spherical swimmers \cite{Ishikawa2006}.  These initial studies provided important fundamental results connecting the properties of the individual swimmers to the emergence of collective dynamics.  Based on their success, these models have been more recently incorporated into a number of numerical approaches for suspension and fluid-structure interaction simulations including Stokesian dynamics \cite{Ishikawa_2007,Ishikawa_2008,Mehandia2008}, the immersed boundary method \cite{Lushi2013, Lambert2013}, Lattice Boltzmann methods \cite{Alarcon2013,Pagonabarraga2013}, and hybrid finite element/penalization schemes \cite{Decoene2011}.  This has allowed for both increased swimmer numbers as well as the incorporation of other effects such as steric interactions, external boundaries, and aligning torques.

In this paper, we introduce an extension of the force coupling method (FCM) \cite{Maxey2001,Lomholt2003}, an approach for the large-scale simulation of passive particles, to capture the many-body interactions between active particles.  FCM relies on a regularized, rather than a singular, multipole expansion to account for the hydrodynamic interactions between the particles.  It includes a higher-order correction due to particle rigidity by enforcing the constraint of zero-averaged strain rate in the vicinity of each particle.  Since the force distributions have been regularized, the total particle force, including that associated with the constraints, can be projected onto a grid over which the fluid flow can be found numerically.  This allows the hydrodynamic interactions for all particles to be resolved simultaneously.  This mesh can be structured and simple such that an efficient parallel Stokes solver can be used to find the hydrodynamic interactions.

We extend FCM to active particles by introducing the regularized singularities in the FCM multipole expansion that have a direct correspondence to the surface velocity modes of the squirmer model \cite{Blake1971}.  With these terms included, we then rely on the usual FCM framework to resolve the hydrodynamic interactions in a very efficient manner.  We show that by using the full capacity afforded by FCM, we are able to accurately simulate active particle suspensions in the semi-dilute limit with $O(10^4 - 10^5)$ swimmers.  Using this method, we examine the influence of domain size on the steady-state polar order observed for squirmer suspensions.  At the same time, we show that our method is quite versatile, being able to handle time-dependent swimming gaits, ellipsoidal swimmer shapes, and steric interactions, each at a minimal additional computational cost.  We explore in detail how to incorporate biologically-relevant, time-dependent swimming gaits by tuning our model to the recent measurements of the oscillatory flow around \emph{Chlamydomonas Rheinardtii} \cite{Guasto2010}.  These experiments revealed that considering time-averaged flows for such micro-organisms may oversimplify the hydrodynamic interactions between neighbors.  Time-dependency is also closely associated with the way zooplankton feed, mix the surrounding fluid, and interact with each other \cite{Croze2013,Kiorboe2014}.  As stated in \cite{Saintillan2010}, modelling micro-swimmers with a time-dependent swimming gait might be more realistic and should be included in mathematical models and computer simulations. We show that time-dependence can indeed affect the overall organization of the suspension.  

We organize our paper as follows: In section \ref{Squirmer_FCM}, we review FCM and present the theoretical background for its adaptation to active particles. 
Section \ref{num_meth} details the numerical method, its algorithmic implementation and how the computational work scales with the particle number. 
In Section \ref{valid}, we validate the method and test its accuracy by comparing flow fields, trajectories, and pairwise interactions with previous results available in the literature.
The effectiveness of our approach is demonstrated in Section \ref{simul_res} where we present results from large-scale simulations of active particle suspensions. Finally, extensions of FCM to more complex scenarios are introduced in Section \ref{sec:Spheroid_Time_dep}.  We simulate suspensions of spheroidal swimmers and demonstrate the new implementation of time-dependent swimming gaits.  Here, we also present preliminary results showing the effect of time-dependence on suspension properties.

\section{Squirmers using FCM}
\label{Squirmer_FCM}

The force-coupling method (FCM) developed by Maxey and collaborators \cite{Maxey2001,Lomholt2003} is an effective approach for the large-scale simulation of particulate suspensions, especially for moderately concentrated suspensions at low Reynolds number.  
In this context, it has been used to address a variety of problems in microfluidics \cite{Climent2004}, biofluid dynamics \cite{Pivkin2006}, and micron-scale locomotion \cite{Keaveny2008I,Keaveny2008II,Majmudar2012}.  
FCM has also been extended to incorporate finite Reynolds number effects \cite{Xu2002}, thermal fluctuations \cite{Keaveny2014}, near contact lubrication hydrodynamics \cite{Dance2003,Yeo2010}, and ellipsoidal particle shapes \cite{Liu2009}.   
With these additional features, FCM has been used to address questions in fundamental fluid dynamics in regimes where inertial effects are important and/or there is a high volume fraction of particles \cite{Yeo2010,Yeo2010jfm}.  At the the same time, FCM has been used to address problems of technological importance, such as micro-bubble drag reduction \cite{Xu2002} and 
the dynamics of colloidal particles \cite{Keaveny2014}.  
In this section, we expand on \cite{KeavenyPhD} and develop the theoretical underpinnings of FCM's further extension to active particle suspensions using the squirmer model proposed by Lighthill \cite{Lighthill1952}, advanced by Blake \cite{Blake1971}, and employed by Ishikawa {\it et al.} \cite{Ishikawa2006}.  
To begin this presentation, we give an overview of FCM, establishing also the notation that will be used throughout the paper.

\subsection{FCM for passive particles}
\label{FCM_passive}
Consider a suspension of $N_p$ rigid spherical particles, each having radius $a$.  
Each particle $n$, $(n = 1,\dots, N_p)$, is centered at $\mathbf{Y}_n$ and subject to force $\mathbf{F}_n$ and torque $\bm{\tau}_n$.  
To determine their motion through the surrounding fluid, we first represent each particle by a low order, finite-force multipole expansion in the Stokes equations
\begin{eqnarray}
	\bm{\nabla} p - \eta \nabla^2\mathbf{u}&=&\sum_n \mathbf{F}_n \Delta_n(\mathbf{x})  + \frac{1}{2}\bm{\tau}_n \times \bm{\nabla} \Theta_n(\mathbf{x}) + \mathbf{S}_{n} \cdot \bm{\nabla}\Theta_n(\mathbf{x})\nonumber\\
	\bm{\nabla} \cdot \mathbf{u} &=& 0.
	\label{eq:FCM1}
\end{eqnarray}
In Eq. (\ref{eq:FCM1}), $\mathbf{S}_{n}$ are the particle stresslets determined through a constraint on the local rate-of-strain as described below.  
Also in Eq. (\ref{eq:FCM1}) are the two Gaussian envelopes,
\begin{eqnarray}
\Delta_n(\mathbf{x})&=&(2\pi\sigma_{\Delta}^2)^{-3/2}\textrm{e}^{-|\mathbf{x} - \mathbf{Y}_n|^2/2\sigma_{\Delta}^2} \nonumber\\
\Theta_n(\mathbf{x})&=&(2\pi\sigma_{\Theta}^2)^{-3/2}\textrm{e}^{-|\mathbf{x} - \mathbf{Y}_n|^2/2\sigma_{\Theta}^2},
\label{eq:FCM2}
\end{eqnarray} 
used to project the particle forces onto the fluid.  

After solving Eq. (\ref{eq:FCM1}), the velocity, $\mathbf{V}_n$, angular velocity, $\bm{\Omega}_n$, and local rate-of-strain, $\mathbf{E}_n$, of each particle $n$ are found by volume averaging of the resulting fluid flow,
\begin{eqnarray}
\mathbf{V}_n&=&\int\mathbf{u}\Delta_n(\mathbf{x})d^3\mathbf{x} \label{eq:FCM3a}\\
\bm{\Omega}_n&=&\frac{1}{2}\int\left[\bm{\nabla}\times\mathbf{u}\right] \Theta_n(\mathbf{x})d^3\mathbf{x} \label{eq:FCM3b}, \\
\mathbf{E}_n&=&\frac{1}{2}\int \left[\bm{\nabla}\mathbf{u} + (\bm{\nabla}\mathbf{u})^T\right]\Theta_n(\mathbf{x})d^3\mathbf{x},
\label{eq:FCM3c}
\end{eqnarray}
where the integration is performed over the entire domain.  
In order for Eqs. (\ref{eq:FCM3a}) --  (\ref{eq:FCM3c}) to recover the correct mobility relations for a single, isolated sphere, namely that $\mathbf{V} = \mathbf{F}/(6\pi a \eta)$ and $\bm{\Omega} = \bm{\tau}/(8\pi a^3 \eta)$, the envelope length scales need to be $\sigma_{\Delta} = a/\sqrt{\pi}$ and $\sigma_{\Theta} = a/\left(6\sqrt{\pi}\right)^{1/3}$.  
As the particles are rigid, the stresslets are found by enforcing the constraint that $\mathbf{E}_n = \mathbf{0}$ for each particle $n$ \cite{Lomholt2003}.

\subsection{Squirmer model}

In addition to undergoing rigid body motion in the absence of applied forces or torques, active and self-propelled particles are also characterized by the flows they generate.  
To model such particles, we will need to incorporate these flows into FCM.  
We accomplish this by adapting the axisymmetric squirmer model \cite{Lighthill1952,Blake1971,Ishikawa2006} to the FCM framework, though we note that a more general squirmer model \cite{Pak2014} with non-axysimmetric surface motion could also be considered.\\

\begin{figure}[H]
\centering
\includegraphics[height=5cm]{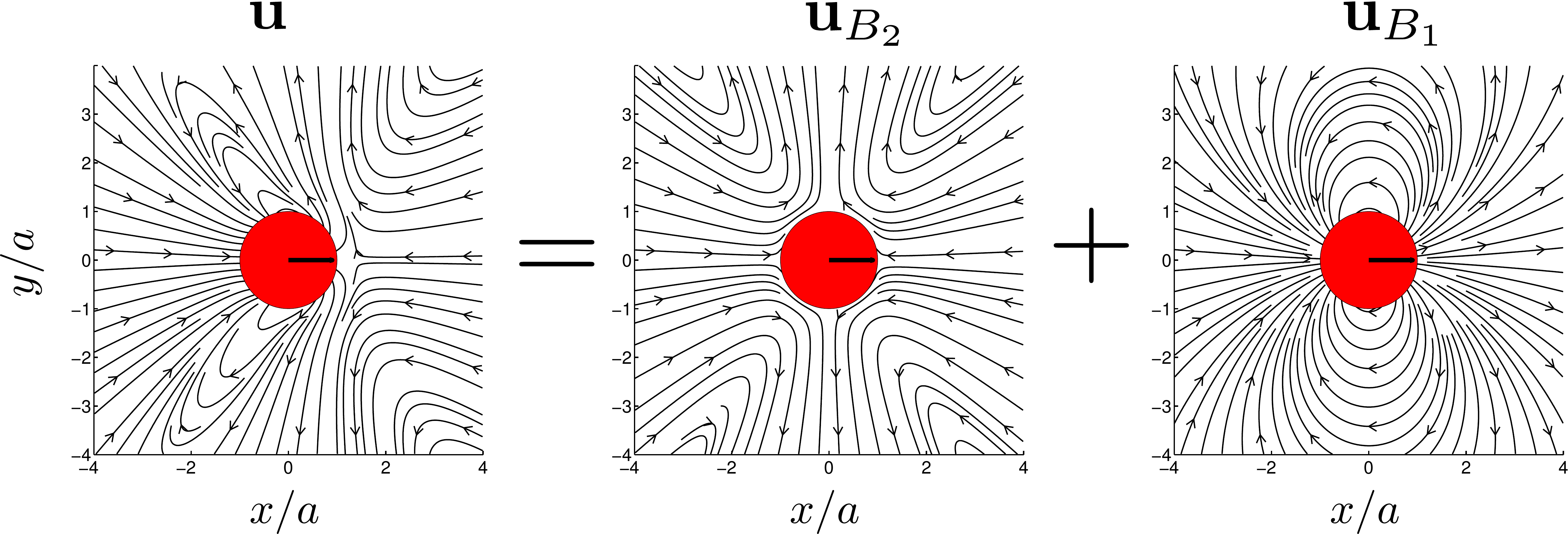}
\caption{Decomposition of squirmer velocity field for $\beta = 1$.}
\label{fig:sqflow}
\end{figure}

The squirmer model consists of a spherically shaped, self-propelled particle that utilizes axisymmetric surface distortions to move through fluid with speed $U$ in the direction $\mathbf{p}$.  If the amplitude of the distortions is small compared to the radius, $a$, of the squirmer, their effect can be represented by the surface velocity, $\mathbf{v}(r = a) = v_r\hat{\mathbf{r}} + v_\theta \hat{\bm{\theta}}$ where
\begin{eqnarray}
v_{r} &=& U\cos\theta + \sum_{n = 0}^{\infty}A_{n}(t)P_{n}(\cos \theta),\\
v_{\theta} &=& -U\sin\theta - \sum_{n = 1}^{\infty}B_{n}(t)V_{n}(\cos \theta).
\end{eqnarray}
Here, $P_{n}(x)$ are the Legendre polynomials,
\begin{equation}
V_{n}(\cos \theta) = \frac{2}{n(n+1)} \sin \theta P'_{n}(\cos \theta),
\end{equation}
the angle $\theta$ is measured with respect to the swimming direction $\mathbf{p}$, and $P'_{n}(x) = dP_n/dx$.  In order for the squirmer to be force-free, we have
\begin{equation}
U = \frac{1}{3}(2B_1 - A_1).
\label{eq:UB1A1}
\end{equation}
Following Ishikawa \cite{Ishikawa2006}, we consider a reduced squirmer model where $A_{n} = 0$ for all $n$ and $B_{n} = 0$ for all $n > 2$.  
We therefore only have the first two terms of the series.  
For this case, the resulting flow field in the frame moving with the swimmer is given by
\begin{eqnarray}
\mathbf{u}_{r}(r,\theta) &=& \frac{2}{3}B_1\frac{a^3}{r^3}P_1(\cos \theta) + \left(\frac{a^4}{r^4} - \frac{a^2}{r^2}\right)B_{2}P_{2}(\cos\theta) \\
\mathbf{u}_{\theta}(r, \theta) &=& \frac{1}{3}B_1\frac{a^3}{r^3}V_{1}(\cos \theta) + \frac{a^4}{r^4} B_{2}V_{2}(\cos\theta)
\end{eqnarray}
where we have used $U = 2B_1/3$ from Eq. (\ref{eq:UB1A1}).  
In terms of $\mathbf{p}$, $\mathbf{x}$, and $r$, this becomes
\begin{eqnarray}
\mathbf{u}(\mathbf{x})&=& -\frac{B_1}{3}\frac{a^3}{r^3}\left(\mathbf{I} - 3\frac{\mathbf{xx}^T}{r^2}\right)\mathbf{p} + \left(\frac{a^4}{r^4} - \frac{a^2}{r^2}\right)B_{2}P_{2}\left(\frac{\mathbf{p}\cdot\mathbf{x}}{r}\right)\frac{\mathbf{x}}{r} \nonumber\\
&&- 3\frac{a^4}{r^4} B_{2}\left(\frac{\mathbf{p}\cdot\mathbf{x}}{r}\right)\left(\mathbf{I} - \frac{\mathbf{xx}^T}{r^2}\right)\mathbf{p}. \label{eq:squ}\\
&=& \mathbf{u}_{B_1} + \mathbf{u}_{B_2} \nonumber
\end{eqnarray}
Fig. \ref{fig:sqflow} shows an example of a flow field given by Eq. (\ref{eq:squ}), as well as the flows $\mathbf{u}_{B_1}$ and $\mathbf{u}_{B_2}$ related to the $B_1$ and $B_2$ contributions.
While we have already seen that $B_1$ is related to the swimming speed, it can be shown \cite{Ishikawa2006} that $B_2$ is directly related to the stresslet 
\begin{eqnarray}
\mathbf{G} &=& \frac{4}{3}\pi\eta a^2\left(3\mathbf{p}\mathbf{p} - \mathbf{I}\right)B_2 \label{eq:sqstresslet} 
\end{eqnarray}
generated by the surface distortions.  This term sets the leading-order flow field that decays like $r^{-2}$.  
We can introduce the parameter $\beta = B_2/B_1$ which describes the relative stresslet strength.  
In addition, if $\beta > 0$, the squirmer behaves like a `puller,' bringing fluid in along $\mathbf{p}$ and expelling it laterally, whereas if $\beta < 0$, the squirmer is a `pusher', expelling fluid along $\mathbf{p}$ and bringing it in laterally.

To adapt this model to the FCM framework, we first recognize that the flow given by Eq. (\ref{eq:squ}) can be represented by the following singularity system in the Stokes equations 
\begin{eqnarray}
\bm{\nabla} p - \eta \nabla^2 \mathbf{u} & = &\mathbf{G}\cdot\bm{\nabla}\left(\delta(\mathbf{x})+\frac{a^2}{6}\nabla^2\delta(\mathbf{x})\right) + \mathbf{H} \nabla^2\delta(\mathbf{x}) \label{eq:Stokessquirm}\\
\bm{\nabla}\cdot \mathbf{u} &=&0
\end{eqnarray}
where the degenerate quadrupole is related to $B_1$ through
\begin{eqnarray}
\mathbf{H} &=& -\frac{4}{3}\pi\eta a^3 B_1 \mathbf{p}.
\label{eq:sqdegquad}
\end{eqnarray}
and the stresslet $\mathbf{G}$ is given by Eq. (\ref{eq:sqstresslet}).  
We can draw a parallel between these singularities and the regularized singularities used with FCM.
  The stresslet term in Eq. (\ref{eq:Stokessquirm}) is the gradient of the singularity system for a single sphere subject to an applied force.  Accordingly, the corresponding regularized singularity in FCM is $\bm{\nabla} \Delta(\mathbf{x})$, where $\Delta(\mathbf{x})$ is given by Eq. (\ref{eq:FCM2}).  
It is important to note that even though we replace two singular force distributions with the one regularized FCM distribution, the particular choice of $\Delta(\mathbf{x})$ will yield flows that are asymptotic to both singular flow fields \cite{Maxey2001}.  
For the degenerate quadrupole, however, there is not a corresponding natural choice for the regularized distribution.  Following \cite{Keaveny2008III}, we choose a Gaussian envelope with a length-scale small enough to yield an accurate representation of the singular flow, but not so small as to 
significantly increase the resolution needed in a numerical simulation (see Section \ref{sec:Comparison_Blake}).  
We therefore employ the FCM envelope for the force dipole and replace the singular distribution by $\nabla^2\Theta(\mathbf{x})$.  
Thus, for a single squirmer, the Stokes equations with the FCM squirmer force distribution are 
\begin{eqnarray}
\bm{\nabla} p - \eta \nabla^2 \mathbf{u} & = &\mathbf{G} \cdot \bm{\nabla} \Delta(\mathbf{x}) + \mathbf{H}\nabla^2\Theta(\mathbf{x}) \nonumber \\
\bm{\nabla}\cdot \mathbf{u}& = &0. \label{eq:FCMStokesSquirm}
\end{eqnarray}
As we show Section \ref{valid}, the flow satisfying Eq. (\ref{eq:FCMStokesSquirm}) closely matches that obtained using the original singularity distribution, Eq. (\ref{eq:Stokessquirm}).

\subsection{Squirmer interactions and motion}

\label{sq_interactions}
Using FCM, the task of computing the interactions between squirmers is relatively straightforward.  
We now consider $N_p$ independent squirmers where each squirmer has swimming dipole $\mathbf{G}_n$ and degenerate quadrupole $\mathbf{H}_n$.  
The squirmers may also be subject to external forces $\mathbf{F}_{n}$ and torques $\bm{\tau}_n$.  
Using the linearity of the Stokes equations, we combine Eqs. (\ref{eq:FCM1}) and (\ref{eq:FCMStokesSquirm}) to obtain 
\begin{eqnarray}
\bm{\nabla} p - \eta \nabla^2 \mathbf{u} & =& \sum_n \mathbf{F}_n \Delta_n(\mathbf{x})  + \frac{1}{2}\bm{\tau}_n \times \bm{\nabla} \Theta_n(\mathbf{x}) + \mathbf{S}_{n} \cdot \bm{\nabla}\Theta_n(\mathbf{x}) \nonumber\\
&&+ \mathbf{G}_{n} \cdot \bm{\nabla} \Delta_n(\mathbf{x}) + \mathbf{H}_n\nabla^2\Theta_n(\mathbf{x}) \label{eq:FCMStokesSquirmFull1}\\
\bm{\nabla}\cdot \mathbf{u}& = &0 \label{eq:FCMStokesSquirmFull2}
\end{eqnarray}
for the flow field generated by the suspension.  

After finding the flow field, we determine the motion of the squirmers using Eqs. (\ref{eq:FCM3a}) -- (\ref{eq:FCM3c}) with two modifications.  
First, we need to add the swimming velocity, $U\mathbf{p}_n$, to Eq. (\ref{eq:FCM3a}).  
Second, we must subtract the artificial, self-induced velocity and the local rate-of-strain due to the squirming modes.  
The self-induced velocity is given by 
\begin{equation}\label{eq:potdipint}
\mathbf{W}_{n} = \int \mathbf{A}\cdot \mathbf{H}_{n} \Delta(\mathbf{x}) d^3\mathbf{x}
\end{equation}
where the tensor $\mathbf{A}$, in index notation, is \cite{Maxey2001} 
\begin{eqnarray}\label{eq:pij}
A_{ij}(\mathbf{x})&=&\frac{1}{4\pi\eta r^3}\left[\delta_{ij}-\frac{3x_{i}x_j}{r^2}\right]\mathrm{erf}\left(\frac{r}{\sigma_{\Theta}\sqrt{2}}\right)\\
& &-\frac{1}{\eta}\left[\left(\delta_{ij}-\frac{x_{i}x_j}{r^2}\right)+\left(\delta_{ij}-\frac{3x_{i}x_j}{r^2}\right)\left(\frac{\sigma_{\Theta}}{r}\right)^2\right]\Theta(\mathbf{x}).
\end{eqnarray}
The self-induced rate-of-strain is given by
\begin{equation}\label{eq:forcedipint}
\mathbf{K}_n = \int \frac{1}{2}\left(\bm{\nabla}\mathbf{R}\cdot\mathbf{G}_{n} + (\bm{\nabla}\mathbf{R}\cdot\mathbf{G}_{n})^T\right)\Theta(\mathbf{x}) d^3\mathbf{x}
\end{equation}
where the expression for the third rank tensor $\mathbf{R}$ can be found in \cite{Lomholt2003}.  
Taking these self-induced effects into account, the motion of a squirmer $n$ is given by
\begin{eqnarray}
\mathbf{V}_n &=& U\mathbf{p}_n - \mathbf{W}_n + \int \mathbf{u} \Delta_n(\mathbf{x}) d^3\mathbf{x} \label{eq:part_vel_self_ind}\\
\bm{\Omega}_n &=& \frac{1}{2}\int\left[\bm{\nabla}\times\mathbf{u}\right] \Theta_n(\mathbf{x})d^3\mathbf{x} \label{eq:part_rot}\\
\mathbf{E}_n&=&-\mathbf{K}_n + \frac{1}{2}\int \left[\bm{\nabla}\mathbf{u} + (\bm{\nabla}\mathbf{u})^T\right]\Theta_n(\mathbf{x})d^3\mathbf{x}. \label{eq:ROS_self_ind}
\end{eqnarray}
As before, the stresslets $\mathbf{S}_n$ due to squirmer rigidity are obtained from the usual constraint on the local rate-of-strain, namely $\mathbf{E}_{n} = \mathbf{0}$ for all $n$.  Squirmer positions $\mathbf{Y}_{n}$ and orientations $\mathbf{p}_{n}$ are then updated with the Lagrangian equations
\begin{equation}
\dfrac{d\mathbf{Y}_{n}}{dt}=\mathbf{V}_{n},
\label{eq:lag_pos}
\end{equation}
\begin{equation}
\dfrac{d\mathbf{p}_{n}}{dt}=\boldsymbol{\Omega}_{n}\times\mathbf{p}_{n}.
\label{eq:lag_orient}
\end{equation}  

In Section \ref{valid}, we show through a comparison with the boundary element simulations from \cite{Ishikawa2006} that our FCM squirmer model recovers the velocities, angular velocities, stresslets ($\mathbf{S}_n$) and trajectories for two interacting squirmers for a wide range of separations.

\subsubsection{Including additional features}
\label{add_features}
With FCM, additional effects can readily be incorporated into the squirmer model and in our subsequent simulations, we consider several of them to demonstrate the versatility of our approach.  
For example, forces due to steric interactions and external torques experienced by magnetotactic or gyrotactic organisms can be considered by including them in $\mathbf{F}_n$ and $\bm{\tau}_n$ in Eq. (\ref{eq:FCMStokesSquirmFull1}).  
Additionally, we can extend the FCM squirmer model to ellipsoidal shapes by using the ellipsoidal FCM Gaussian distributions.  Such a model can be carefully tuned by comparing with results from \cite{Kanevsky2010} and \cite{Leshansky2007}.  
The effects of particle aspect ratio on suspension properties can then be explored systematically while still accounting for particle size effects such as Jeffery orbits.  
Finally, we are not limited to constant values for $B_1$ and $B_2$.  
By allowing these parameters to be functions of time, the FCM squirmer model can be used to explore how the swimmers' strokes affect overall suspension dynamics.
These extensions are addressed in Section \ref{sec:Spheroid_Time_dep}.   

\section{Numerical Methods}

\label{num_meth}
\subsection{Fluid solver}

The smoothness of the Gaussian force distributions allows FCM to be used with a variety of numerical methods to discretize the Stokes equations. It has been implemented with spectral and spectral element methods  \cite{Pivkin2006,Liu2009, Yeo2010} and finite volume methods \cite{Loisel2013,Agbangla2014} in both simple and complex domain geometries. 

Here, we use a Fourier spectral method with Fast Fourier Transforms (FFTs) to solve the Stokes equations, Eqs. \eqref{eq:FCMStokesSquirmFull1} -- \eqref{eq:FCMStokesSquirmFull2}, for the fluid flow in a three-dimensional periodic domain. 
 We set the zeroth Fourier mode of the velocity to zero, $\hat{\mathbf{u}}(\mathbf{k}=\mathbf{0}) = \mathbf{0}$, to ensure no mean fluid flow through the unit cell.  As the fluid encompasses the entire domain, this is equivalent to adding a uniform pressure gradient to balance the net force the particles exert on the fluid in order to maintain the system in equilibrium \cite{Hasimoto1959,Maxey2001}. This also ensures we have a convergent evaluation of the stresslet interactions \cite{Brady1988}.  As the squirmers can move without exerting a force on the fluid, we do permit a net material flux of squirmers through the unit cell.  This corresponds directly to the far-field Stokesian Dynamics computations performed in \cite{Ishikawa_2008} for squirmer suspensions.
 The FFTs are parallelized with the MPI library P3DFFT.
This library uses 2D decomposition of the 3D domain, introducing a better scalabilty than FFT libraries that implement a 1D decomposition. 
This decomposition has shown good scalability up to $N_c = 32,768$ cores in Direct Numerical Simulation (DNS) of turbulence \cite{Pekurovsky2012}.

\subsection{Computational work}
\begin{figure}
\centering
\includegraphics[height=7cm]{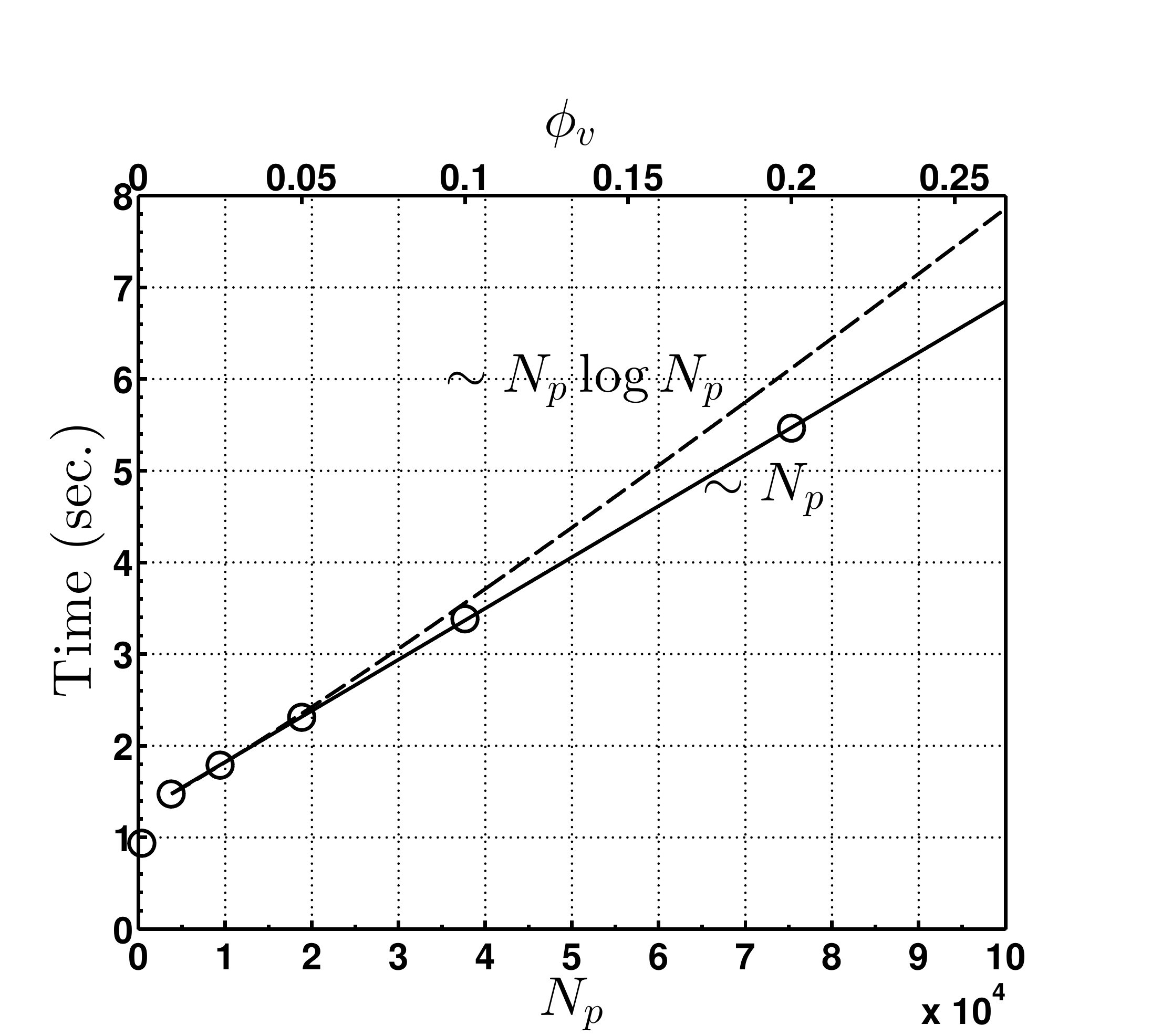}
\caption{Scaling of the FCM with the number of squirmers $N_p$. Computational time per time-step versus $N_p$, or equivalently, versus volumetric fraction $\phi_v$. $N_c = 256$ cores work in parallel for a cubic domain with $384^3$ grid points.}
\label{fig:FCM_scaling} 
\end{figure}

As explained in \cite{Yeo2010}, the number of floating point operations for FCM scales linearly with the number of particles, $N_p$. We find the same scaling for our implementation of FCM.  Fig. \ref{fig:FCM_scaling} shows the computational time per times-step for $N_p$ up to $80,000$ particles with $384^3 \sim 6\cdot10^7$ grid points and $N_c=256$ cores.

\subsection{Steric interactions}
\label{sec:steric}
Including steric repulsion is straightforward with FCM. These forces are introduced to both prevent  particles from  overlapping during the finite time-step and to account for contact forces. 
For spherical particles, we use the steric barrier described in \cite{Dance2004}.  For particles $n$ and $m$, let $\mathbf{r}_{nm} =  \mathbf{Y}_{m} - \mathbf{Y}_{n}$ and $r_{nm} = \|\mathbf{r}_{nm}\|$. The repulsive force experienced by $n$ due to steric interactions is
\begin{equation}
\mathbf{F}_{n}^{b}=
\begin{cases}
-\dfrac{F_{\mbox{ref}}}{2a}\left[\dfrac{R_{\mbox{ref}}^{2}-r_{nm}^{2}}{R_{\mbox{ref}}^{2}-4a^{2}}\right]^{2\gamma}\mathbf{r}_{nm}, & \mbox{for }r_{nm}<R_{\mbox{ref}},\\
0, \mbox{  otherwise.} & 
\end{cases}
\label{eq:barrier_dance}
\end{equation}
where $F_{\mbox{ref}}$ is the magnitude of the force, the cut-off distance $R_{\mbox{ref}}$ sets the distance over which the force acts, and the exponent $\gamma$ can be adjusted to control the stiffness of the force.  Unless specified, all the simulations are run with $F_{\mbox{ref}}/6\pi\eta a U = 4$, $R_{\mbox{ref}} = 2.2a$ and $\gamma=2$.  From Newton's third law, we obtain $\mathbf{F}_{m}^{b} = - \mathbf{F}_{n}^{b}$.  
 It is important to note that this force barrier is generic and it is not intended to model any particular physics.  While we could instead use an exponential DLVO-like potential \cite{Agbangla2014}, the force barrier can be evaluated more rapidly than an exponential potential and by adjusting its parameters we can incorporate repulsion of a similar strength and length scale as DLVO forces.  A thorough study on the effect of this force barrier on the dynamics of particulate suspensions is provided in \cite{Dance2004}.  Steric repulsion for spheroidal particles requires different modelling. In our study, steric forces and torques are introduced by using a soft repulsive potential with the surface-to-surface distance approximated by the Berne-Pechukas range parameter \cite{Allen2006}.

Using a direct pairwise calculation, the evaluation of steric interactions between all particle pairs at each time step would require $O(N_p^2/(2N_c))$ computations per core. This cost is much greater than the $O(N_p)$ cost of the hydrodynamic aspects of FCM.
Therefore, instead of a direct calculation, we use the linked-list algorithm described in \cite{Allen1987}.  This method divides the computational domain into smaller sub-domains into which the particles are sorted.  The edge-length of each sub-domain is slightly larger than $R_{\mbox{ref}}$. These sub-domains are distributed over the cores where the steric interactions are evaluated.  For a homogeneous suspension, this results in a per core cost for steric interactions that is $O\left(14N_p^2/(2N_c N_s)\right)$, where $N_s$ is the total number of sub-domains. Since for large system sizes, we have $N_s \approx L^3/R_{\mbox{ref}}^3 \gg 14$ and $N_c\gg 1$, the linked-list algorithm for steric interactions is much more efficient than the direct computation.

\subsection{Algorithm}
\label{algorithm}
We summarize the overall procedure to simulate large populations of microswimmers in Stokes flow with the FCM:

\begin{itemize}
 \item Initialize particle positions $\mathbf{Y}^{(0)}_n$ and orientations $\mathbf{p}^{(0)}_n$,
 \item Start time loop: \emph{for} $k=1, ..., N_{it}$
 \begin{enumerate}
	\item Compute Gaussians $\Delta^{(k)}_n(\mathbf{x})$ and $\Theta^{(k)}_n(\mathbf{x})$, Eq. \eqref{eq:FCM2},
	\item Update swimming multipoles $\mathbf{G}^{(k)}_n$, Eq. \eqref{eq:sqstresslet}, and $\mathbf{H}^{(k)}_n$, Eq. \eqref{eq:sqdegquad}, which both depend on $\mathbf{p}_n^{(k)}$, 
	\item Compute steric interactions, Eq. \eqref{eq:barrier_dance}, with the linked-list algorithm,
	\item Add additional forcing if any (gyrotactic torques, magnetic dipoles,...),
	\item Project the Gaussian distributions onto the grid (RHS of Eq. \eqref{eq:FCMStokesSquirmFull2}),
	\item Solve Stokes equations, Eqs. \eqref{eq:FCMStokesSquirmFull1} -- \eqref{eq:FCMStokesSquirmFull2}, to obtain the fluid velocity field $\mathbf{u}^{(k)}(\mathbf{x})$,
	\item Compute particle rate of strains $\mathbf{E}^{(k)}_n$, Eq. \eqref{eq:ROS_self_ind},
	\item \emph{If}  $\| \mathbf{E}^{(k)}_n \| > \varepsilon$, compute stresslets $\mathbf{S}^{(k)}_n$ following \cite{Yeo2010},
	\begin{enumerate}[a)]
		\item Project all the multipoles onto the grid (RHS of Eq. \eqref{eq:FCMStokesSquirmFull2}),
		\item Solve for Stokes equations, Eqs. \eqref{eq:FCMStokesSquirmFull1} -- \eqref{eq:FCMStokesSquirmFull2}, to obtain the fluid velocity field $\mathbf{u}^{(k)}(\mathbf{x})$,
	\end{enumerate}	
	\item Compute particle velocities $\mathbf{V}^{(k)}_n$, Eq. \eqref{eq:part_vel_self_ind}, and rotations $\bm{\Omega}^{(k)}_n$, Eq. \eqref{eq:part_rot},
	\item Integrate Eqs. \eqref{eq:lag_pos} and \eqref{eq:lag_orient} using the fourth-order Adams-Bashforth scheme to obtain $\mathbf{Y}^{(k+1)}_n$ and $\mathbf{p}^{(k+1)}_n$.
 \end{enumerate}
\end{itemize}


\section{Validation}

\label{valid}

\subsection{Comparison with Blake's solution}
\label{sec:Comparison_Blake}

\begin{figure}
\centering
\subfloat[]{ \label{fig:usq_Blake} \includegraphics[height=4.5cm]{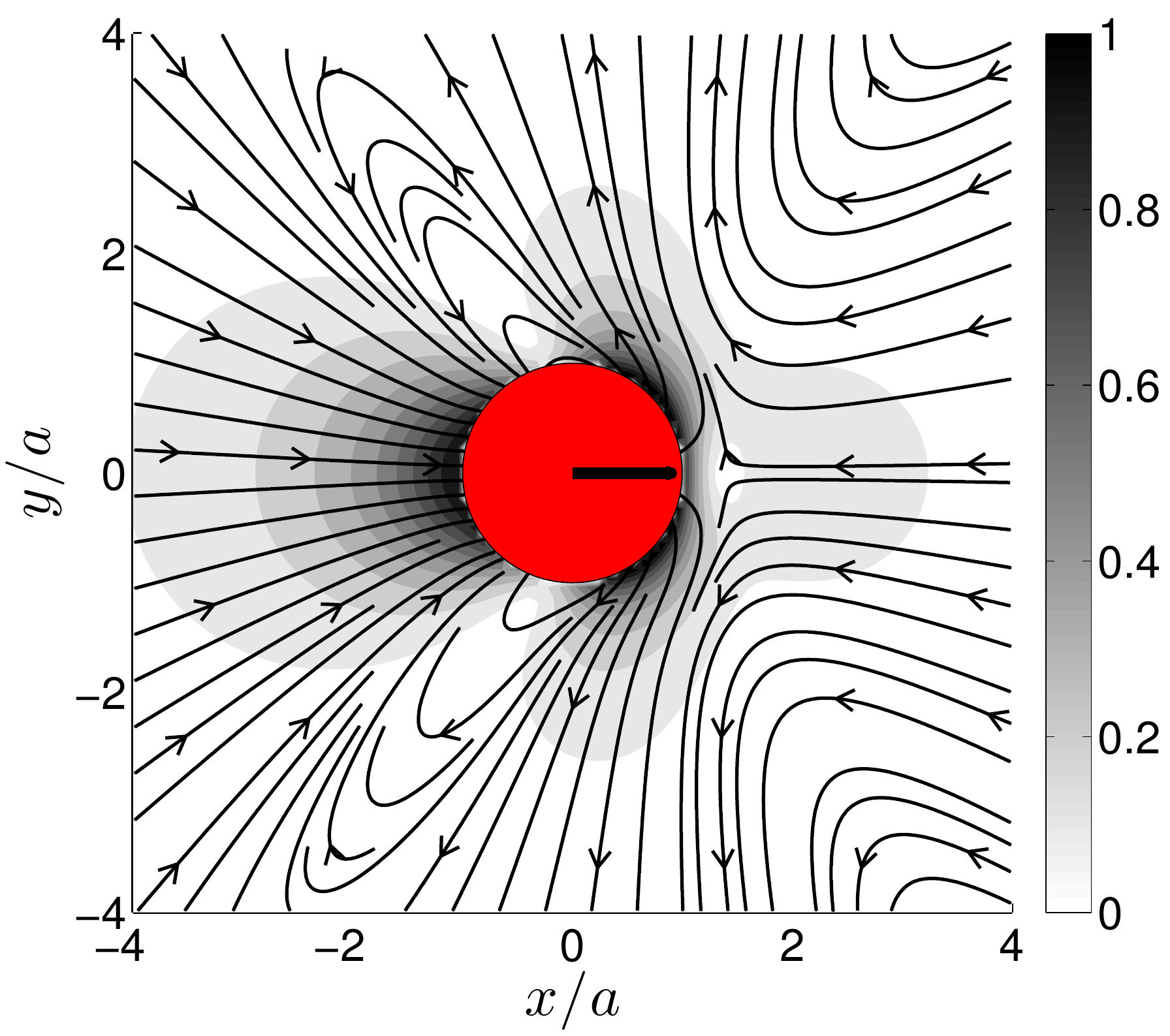}}
\subfloat[]{ \label{fig:usq_FCM} \includegraphics[height=4.5cm]{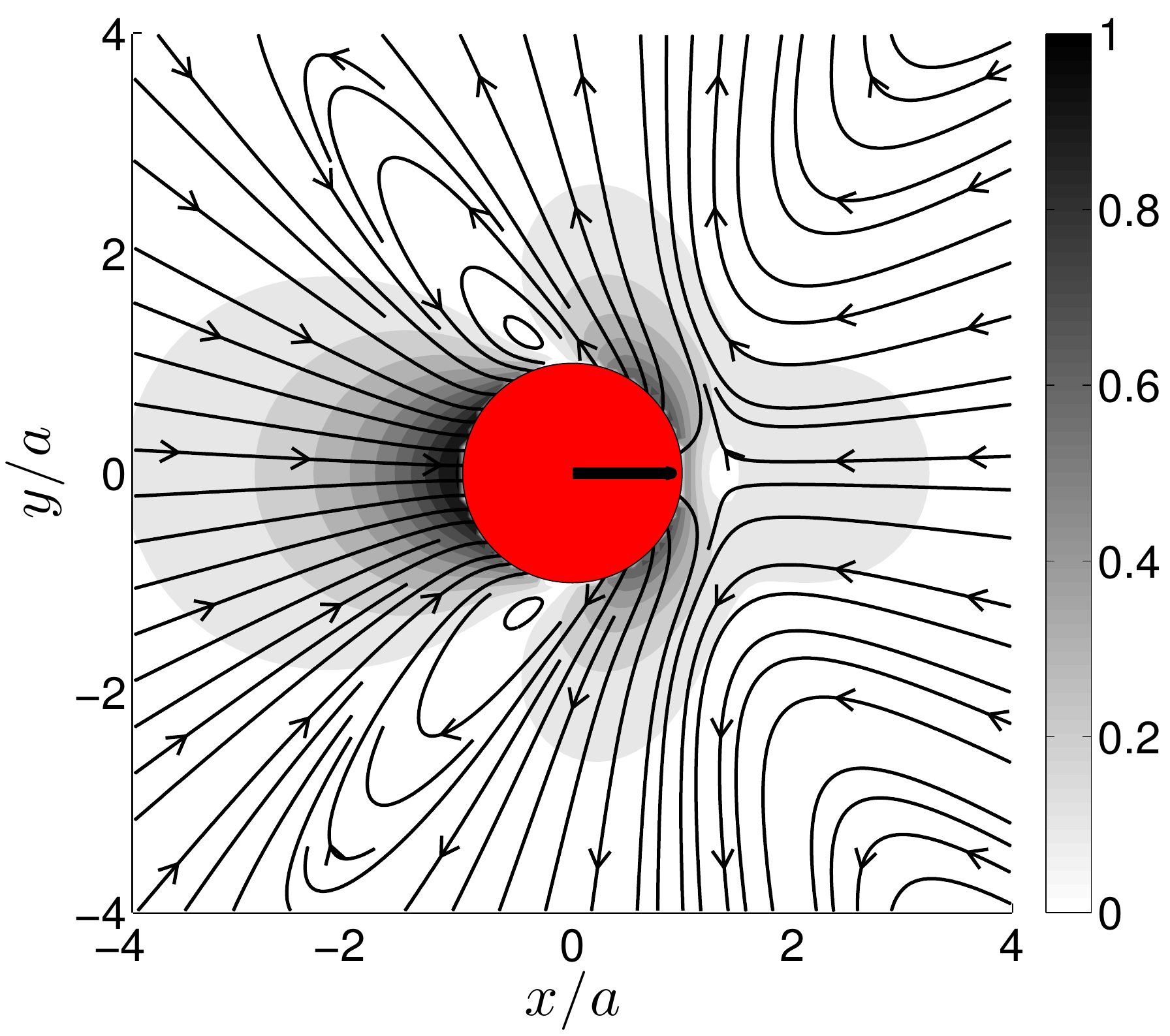}}
\subfloat[]{ \label{fig:usq_FCM_sig_dip_2} \includegraphics[height=4.5cm]{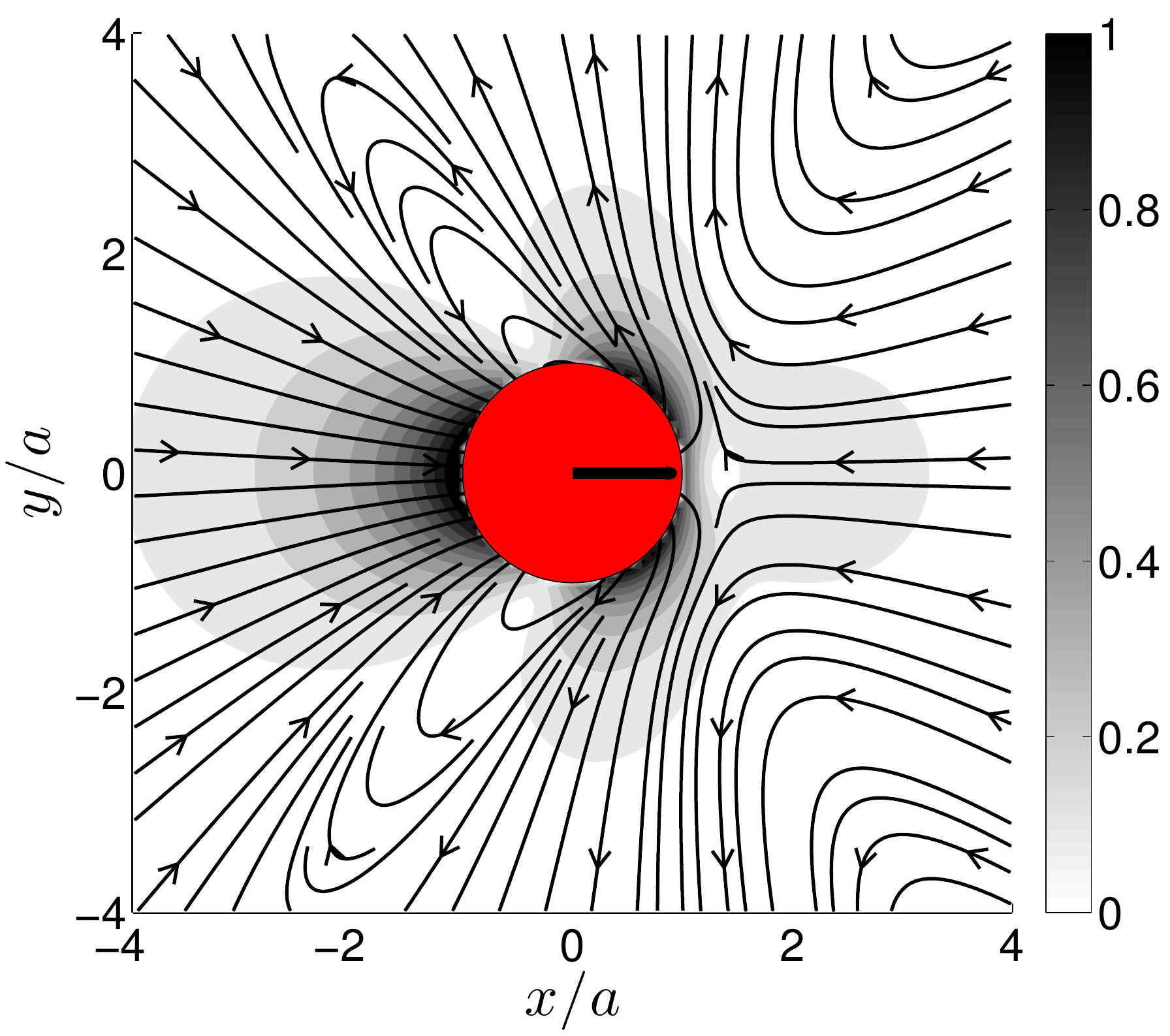}}
\caption{Velocity field $\mathbf{u}/U$ around a puller squirmer ($\beta = 1$) swimming to the right.
a) Blake's solution; 
b) FCM solution with $\sigma_{\Theta}$ for the degenerate quadrupole envelope. 
c) FCM solution with $\sigma_{\Theta}/2$ for the degenerate quadrupole envelope.
} 
\label{fig:Comparison_usq}
\end{figure}

As explained in Section \ref{sq_interactions}, in FCM the velocity field around an isolated squirmer $n$ in an infinite, quiescent fluid is given by
\begin{equation}
\label{eq:usq_FCM}
 \mathbf{u}^{FCM} = \mathbf{A}\cdot\mathbf{H}_n + \mathbf{R}:\mathbf{G}_n,
\end{equation}
where $\mathbf{A}$ is a second rank  tensor given by Eq. \eqref{eq:pij} and the third rank  tensor $\mathbf{R}$ can be found in \cite{Lomholt2003}. Fig. \ref{fig:Comparison_usq} shows, Blake's solution Eq. \eqref{eq:squ}, and the velocity field provided by Eq. \eqref{eq:usq_FCM} with two different values for the degenerate quadrupole Gaussian envelope size, $\sigma_{\Theta}$ and $\sigma_{\Theta}/2$, that appears in the tensor $\mathbf{A}$.  The streamlines are identical, except for a near-field recirculating region that appears when the width of the degenerate quadrupole envelope is $\sigma_{\Theta}$ (Fig. \ref{fig:usq_FCM}).  In this region, however, the magnitude of the velocity is small compared to the swimming speed.  This region should not significantly impact the squirmer-squirmer hydrodynamic interactions that we are aiming to resolve.

\begin{figure}
\centering
\subfloat[]{ \label{fig:Err_usq_sig_dip} \includegraphics[height=7cm]{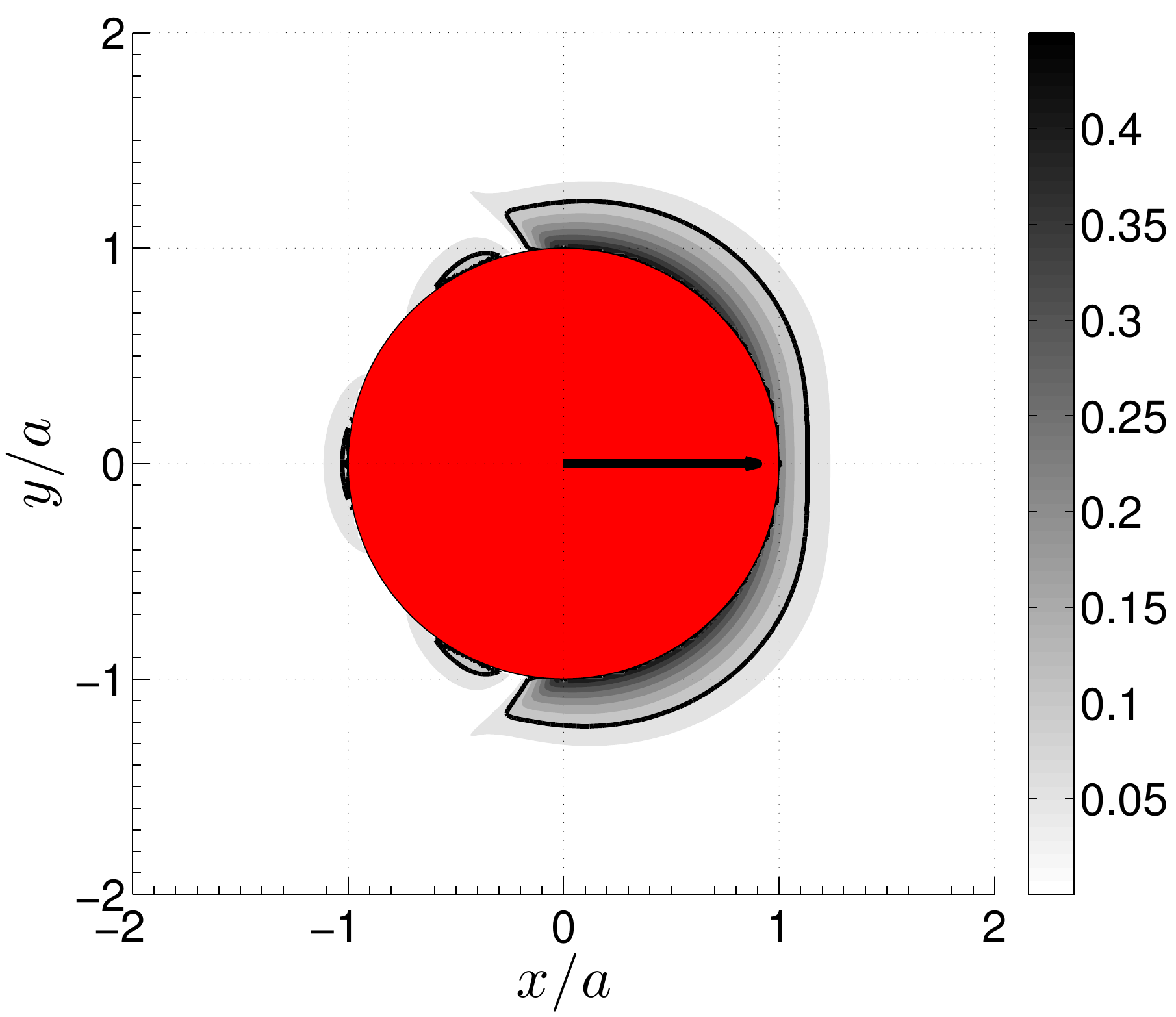}}
\subfloat[]{ \label{fig:u_x_comparison} \includegraphics[height=7cm]{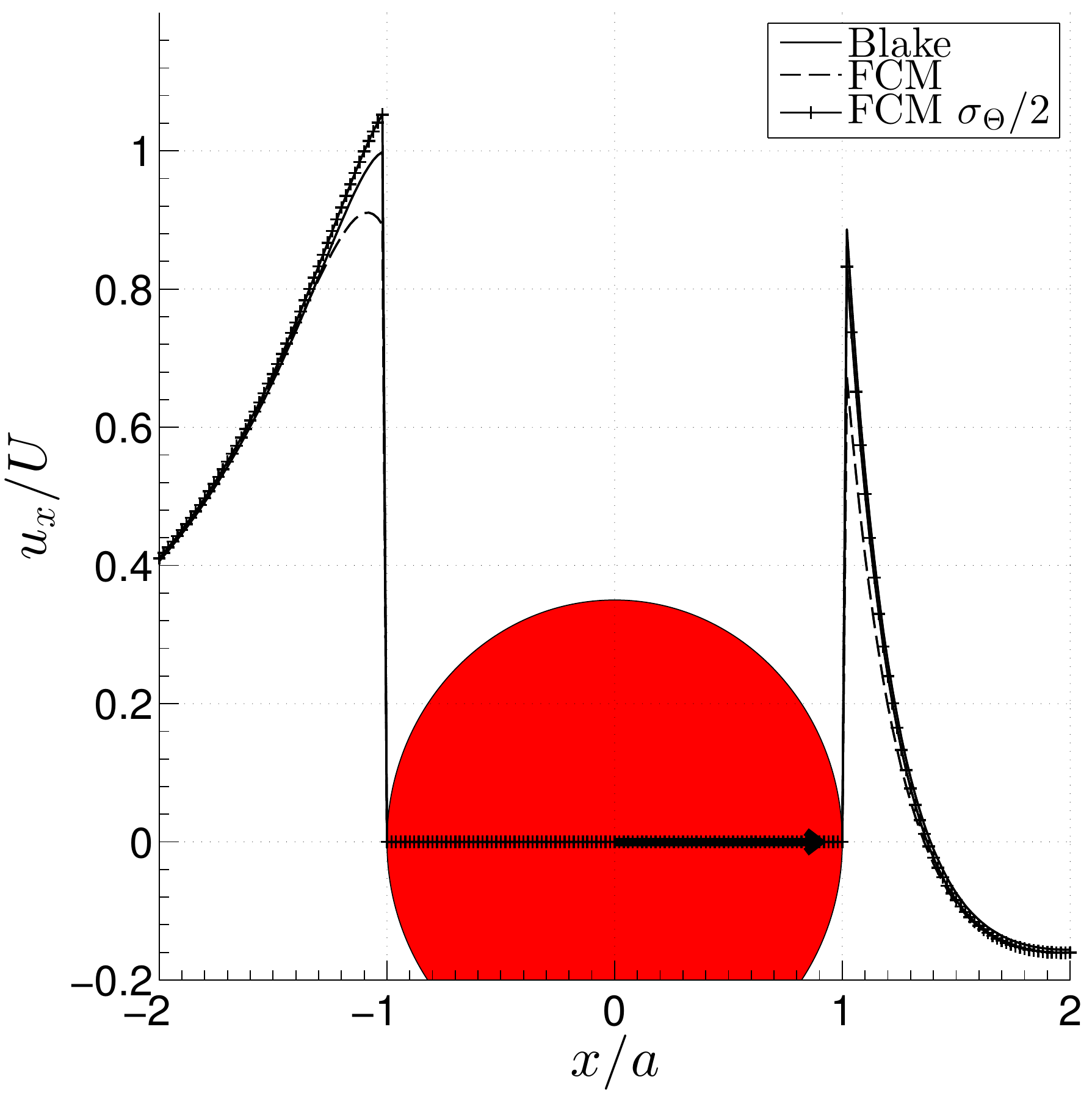}}
\caption{Comparison with Blake's solution, Eq. \eqref{eq:squ}.
a) Normalized difference,$\| \mathbf{u}^{\text{Blake}} - \mathbf{u}^{\text{FCM}}\|/U$, between FCM and Blake's solution for a puller squirmer ($\beta = 1$). The half-width for the degenerate quadrupole envelope is $\sigma_{\Theta}$ .
\symbol{\solid}{}{7}{0}{black}{black}: $10\%$ iso-value.
b) Velocity profile along the swimmer axis for Blake's solution and the FCM approximation for the two different degenerate quadrupole envelope sizes.
} 
\label{fig:Err_usq}
\end{figure}

A quantitative comparison of the velocity field is provided in Fig. \ref{fig:Err_usq}. The agreement with Blake's solution is very good for $r/a>1.25$ when using $\sigma_{\Theta}$ for the width of the degenerate quadrupole envelope.  As shown in Fig. \ref{fig:u_x_comparison}, the smaller envelope size ($\sigma_{\Theta}/2$) matches Blake's solution more closely for $r/a<1.2$, with clear improvement at the front and rear of the squirmer. For this envelope size, the velocity field induced by the degenerate quadrupole $\mathbf{u}^{FCM}_{B_1}$ matches exactly the analytical solution $\mathbf{u}_{B_1}$ in Eq. \eqref{eq:squ} (not shown here).  Below this width no quantitative improvement is observed as the remaining error comes from the dipolar contribution, $\mathbf{u}_{B_2}$.

While this quantitative comparison provides a nice way to choose the degenerate quadrupole envelope size, we must also keep in mind the computational cost associated with decreasing this length scale. Even though it would yield a flow field slightly more in register with Blake's solution, resolving the length scale $\sigma_{\Theta}/2$ in a 3D simulation would require a grid with 8 times as many points as that needed for $\sigma_{\Theta}$, the smallest length-scale already in FCM.  This would increase computation times by at least an order of magnitude.  In addition, the FCM volume averaging used to determine the squirmer translational and angular velocities, Eq. \eqref{eq:FCM3a} -- \eqref{eq:FCM3c}, will reduce the contribution of the localized velocity field discrepancies to the squirmer-squirmer interactions.  In our subsequent simulations, we therefore utilize $\sigma_{\Theta}$ for the degenerate quadrupole envelope size since reducing this length-scale would significantly increase the computational cost, but only provide a minimal improvement.  

\subsection{Pairwise interactions of squirmers}

In \cite{Ishikawa2006}, the authors performed a variety of simulations using the Boundary Element Method (BEM) to compute to high accuracy the pairwise interactions between a squirmer and an inert sphere, and between two squirmers.  Here, we consider the same scenarios as those authors and compare results from our FCM simulations with their BEM results.

\subsubsection{Interactions between a squirmer and an inert sphere}
\label{Sq_inert}
\begin{figure}
\centering
\includegraphics[height=6cm]{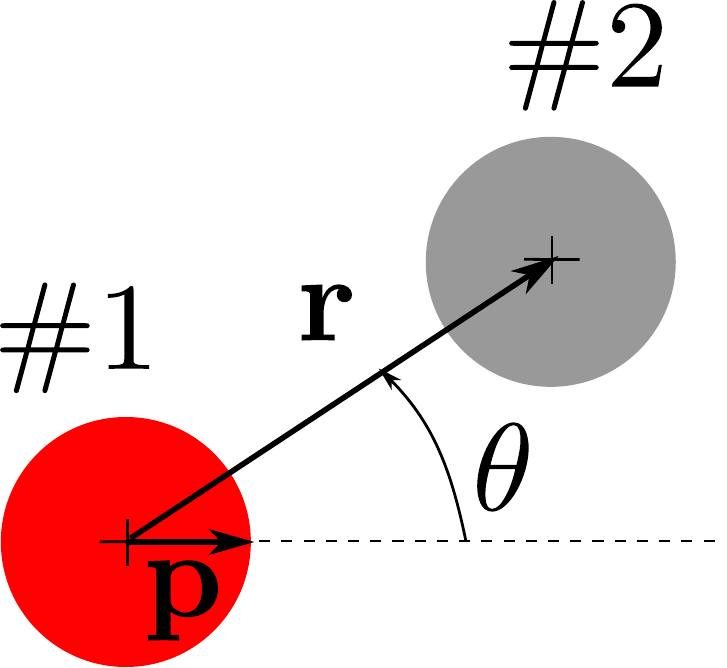}
\caption{Sketch showing the set-up of our computations of the interactions between squirmer ``1'' and inert sphere ``2''.} 
\label{fig:Sketch_Sq_inert}
\end{figure}

We first consider the interactions between an inert sphere (labelled ``$2$'') located at a point $\mathbf{r}$ from the center of a puller squirmer (labelled ``$1$'') with $\beta=5$.  The direction $\mathbf{r}/r$ forms the angle $\theta$ with the swimming direction $\mathbf{p}$ of the squirmer.  The problem setup is depicted in Fig. \ref{fig:Sketch_Sq_inert}.  

\begin{figure}
\centering
\subfloat[]{ \label{fig:Sq_inert_Ur} \includegraphics[height=6cm]{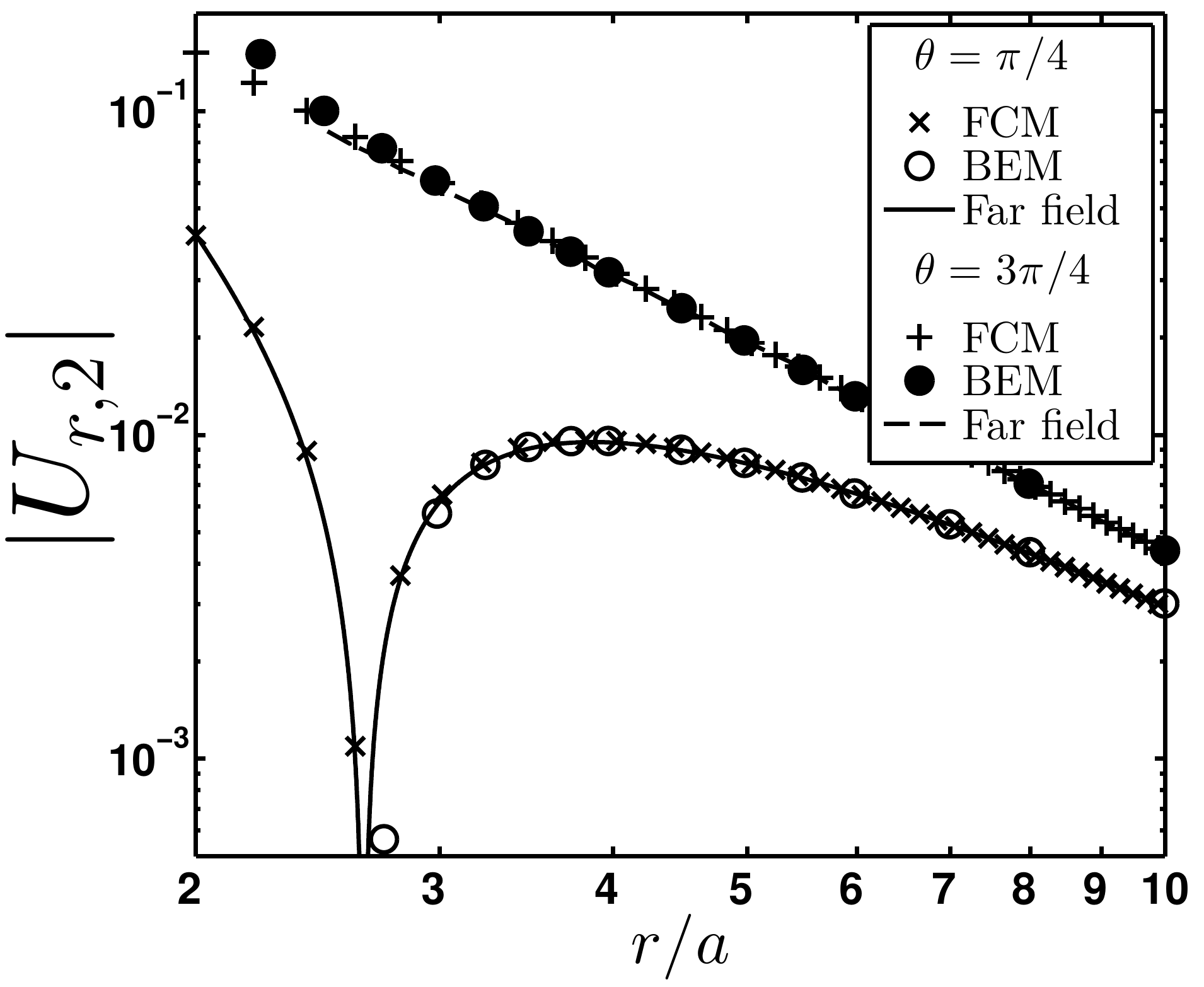}}
\subfloat[]{ \label{fig:Sq_inert_Omz} \includegraphics[height=6cm]{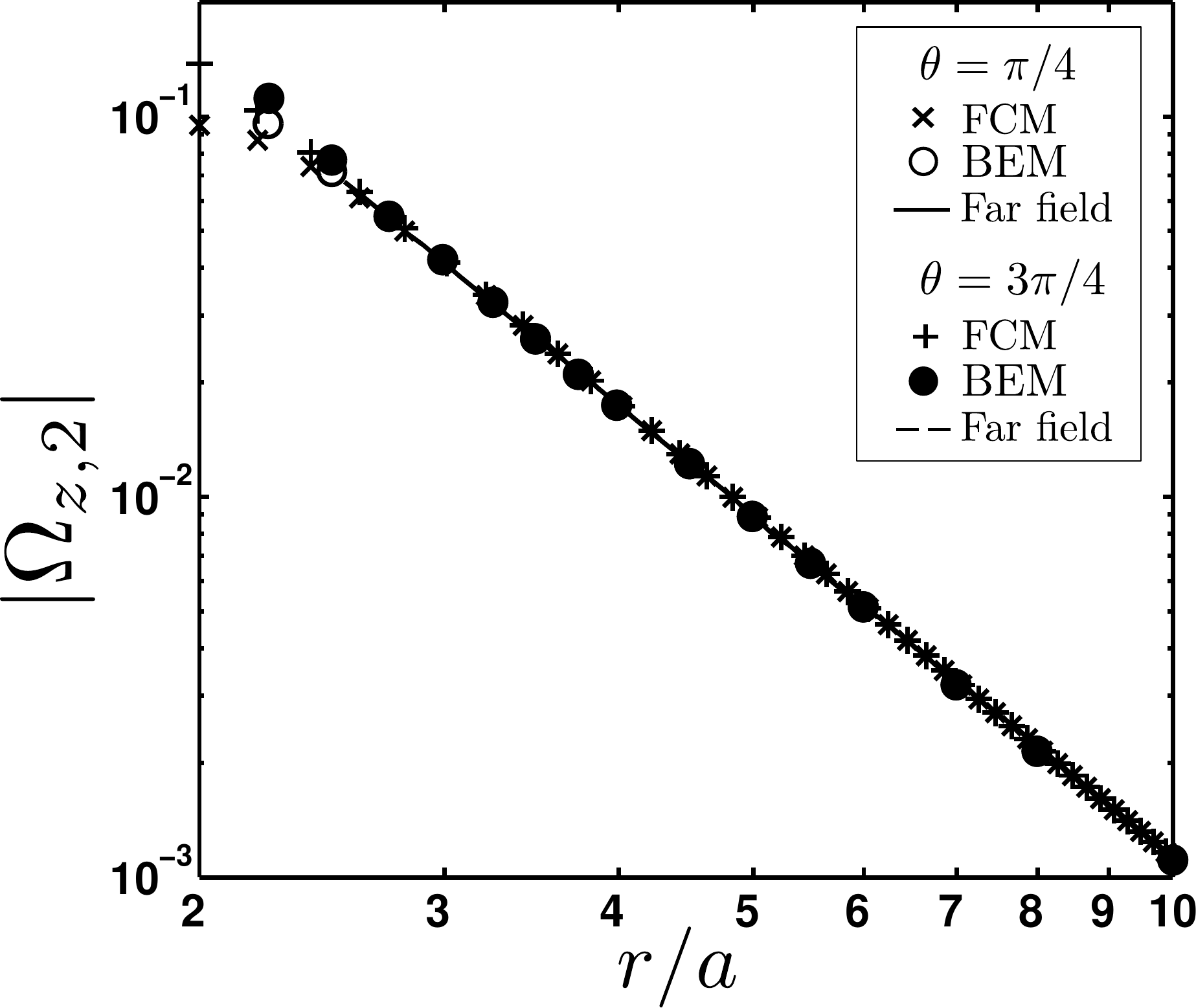}}
\\
\subfloat[]{ \label{fig:Sq_inert_Sxx} \includegraphics[height=6cm]{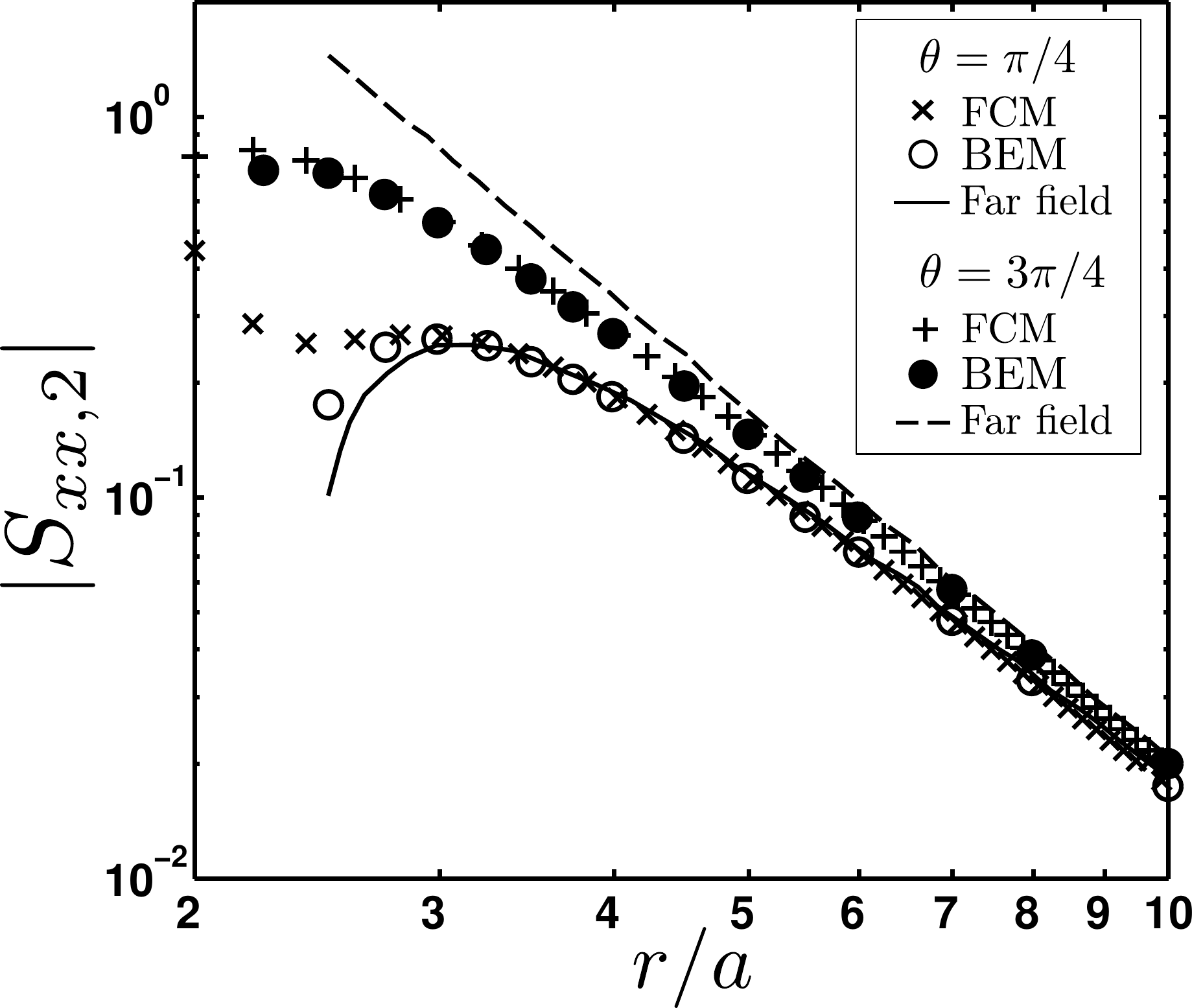}}
\subfloat[]{ \label{fig:Sq_inert_Sxy} \includegraphics[height=6cm]{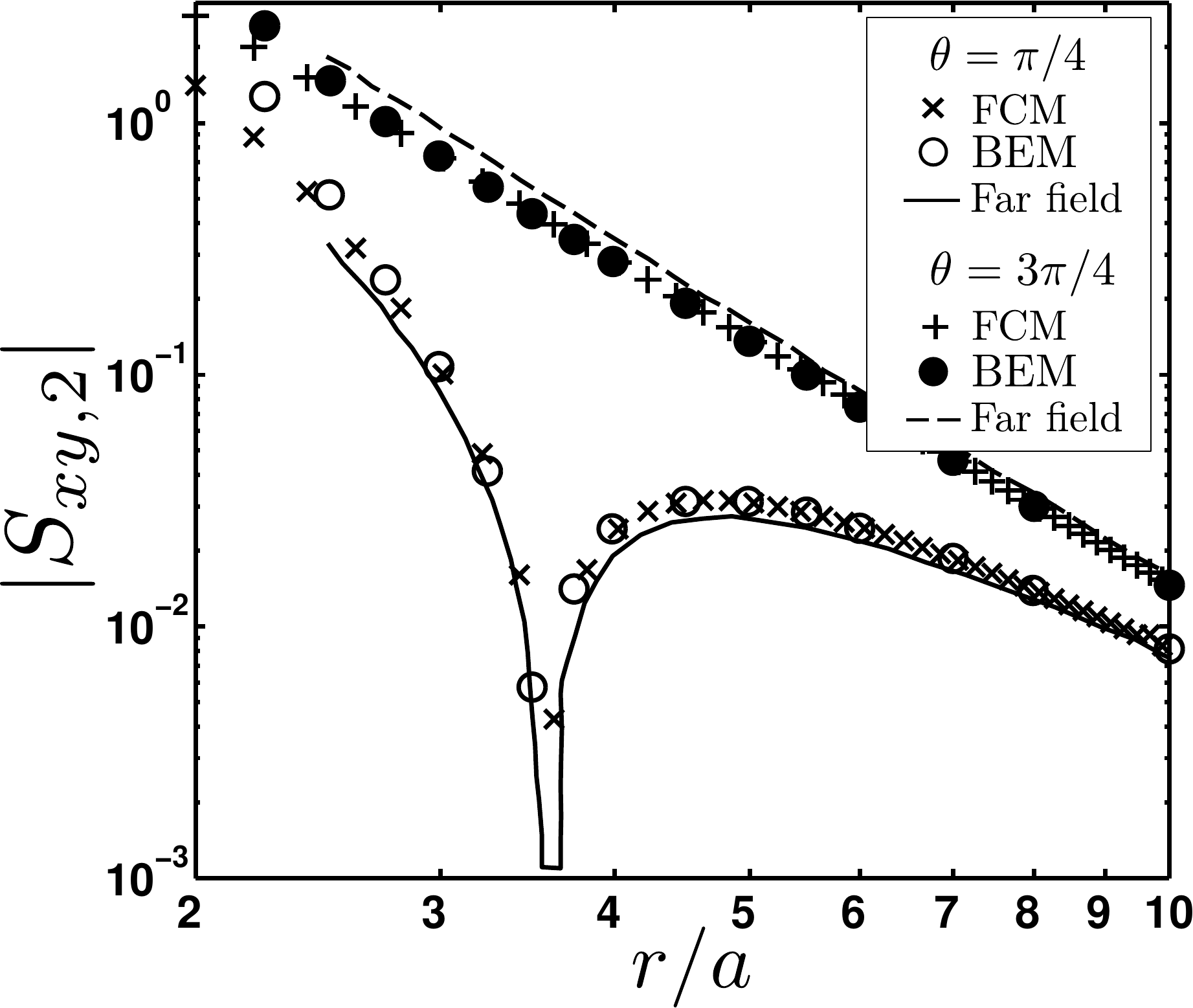}}
\caption{The
a) Radial velocity $|U_{r,2}|$, 
b) Angular velocity $|\Omega_{z,2}|$,
c) Stresslet component $|S_{xx,2}|$, and 
d) Stresslet component $|S_{xy,2}|$ for the inert sphere ``$2$" at a distance $r$ from a puller squirmer ($\beta=5$).
} 
\label{fig:Sq_inert}
\end{figure}

Fig. \ref{fig:Sq_inert} compares the velocity of the sphere obtained using FCM simulations with the BEM results and far-field analytical solutions from \cite{Ishikawa2006}.  We computed the far-field velocity in Fig. \ref{fig:Sq_inert_Ur} using the expressions provided in \cite{Ishikawa2006} since it was not plotted in Fig. 6a of \cite{Ishikawa2006} for $r/a<2.5$.  We suspect that the BEM results were also not plotted for this range.  As FCM also resolves the mutually induced particle stresslets, it provides a more accurate estimation than far-field approximation of \cite{Ishikawa2006}.  As a result, we see that the FCM results very closely match BEM, even in the range where $r/a<3$. Similar trends are observed for the angular velocity of the inert sphere (Fig. \ref{fig:Sq_inert_Omz}) and the stresslet components (Fig. \ref{fig:Sq_inert_Sxx}, \ref{fig:Sq_inert_Sxy}). These comparisons illustrate the accuracy of the results that can be obtained using FCM, which closely matches BEM but incurs a fraction of the computational cost.

\subsubsection{Trajectories of two interacting squirmers}
\label{Collision_squirmers}

\begin{figure}
\centering
\includegraphics[height=6cm]{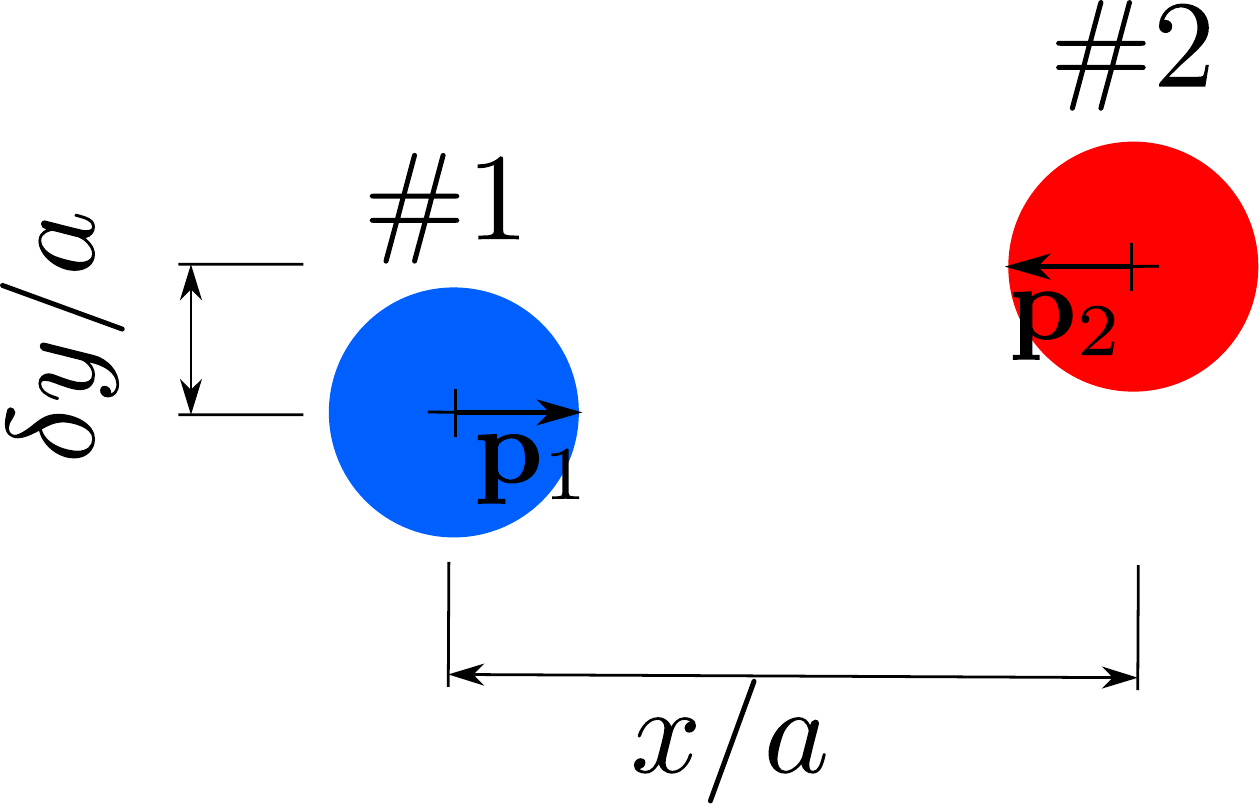}
\caption{Initial configuration of the squirmers in the trajectory simulations.} 
\label{fig:Sketch_Sq_coll}
\end{figure}

We compute the trajectories of two puller squirmers ($\beta=5$) and compare the results with the BEM simulations of \cite{Ishikawa2006}. One squirmer initially swims in the $x$-direction, $\mathbf{p}_1=\mathbf{e}_x$, and the other one in the opposite direction, $\mathbf{p}_2=-\mathbf{e}_x$.  They are placed with initial separation distance $\delta y = 1a,...,10a$ in the transverse direction and $\delta x = 10a$ in the $x$-direction.  The problem set-up is depicted in Fig. \ref{fig:Sketch_Sq_coll}.  Since the squirmers may collide, we also include steric interactions provided by the force barrier \eqref{eq:barrier_dance}. 
\\

\begin{figure}
\centering
\includegraphics[height=8cm]{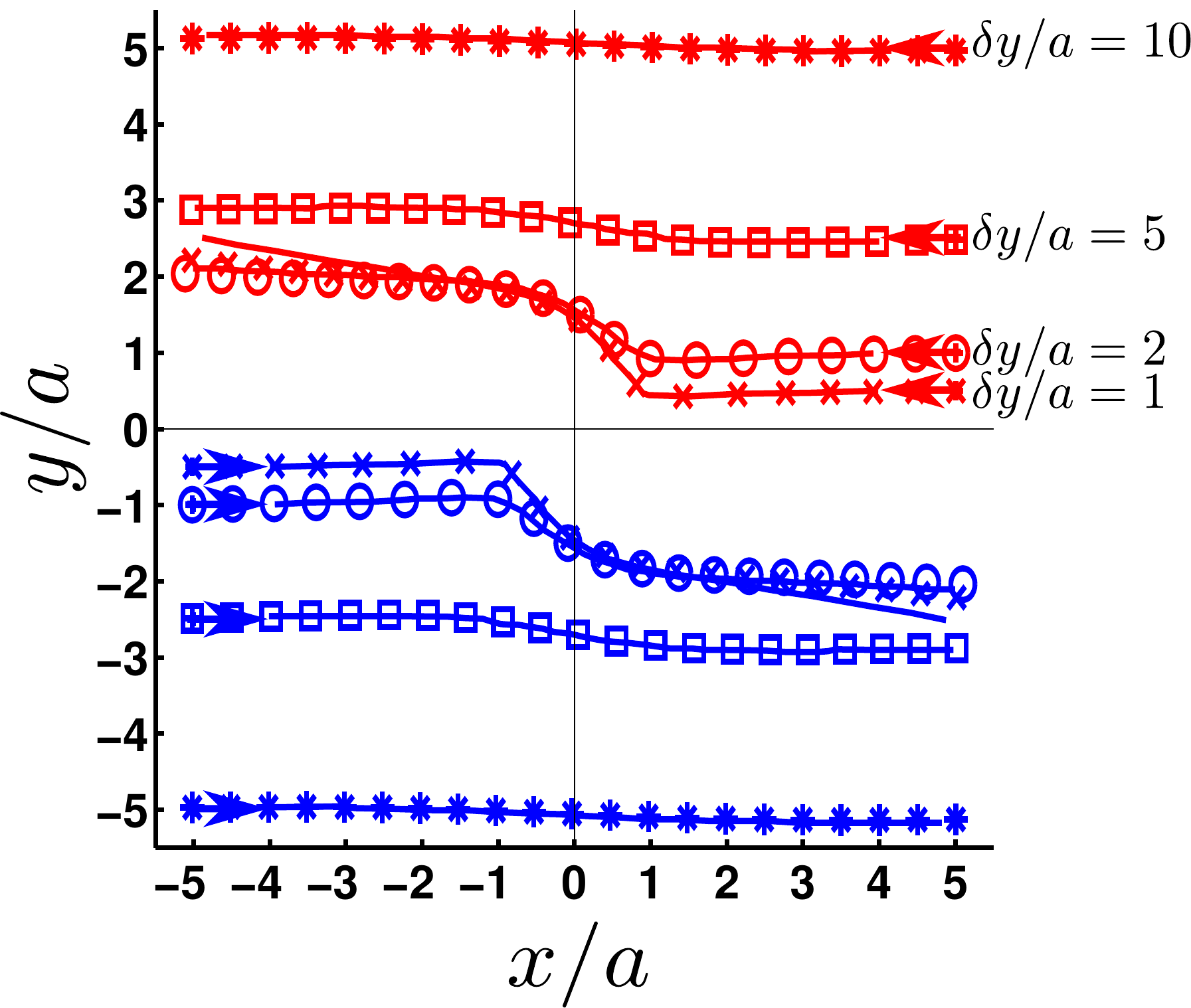} 
\caption{Trajectories of two squirmers swimming in opposite directions with transverse initial distance $\delta y =1a,...,10a$. 
Lines: data from \cite{Ishikawa2006}.
Symbols: FCM results.} 
 \label{fig:Traj_dz_1_10}
\end{figure}

\begin{figure}
\centering
\subfloat[]{ \label{fig:Traj_dz_1} \includegraphics[height=6cm]{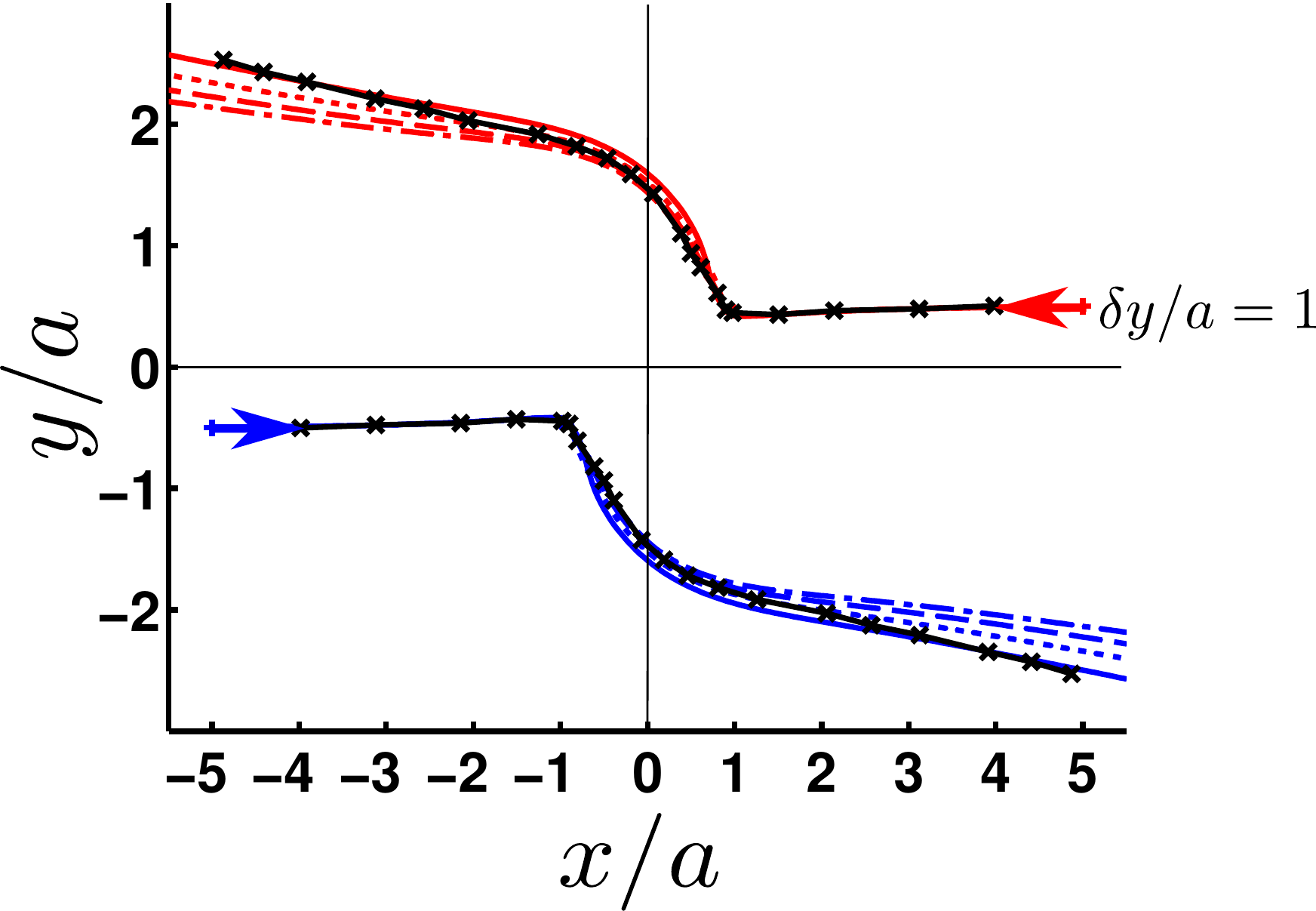} }
\subfloat[]{ \label{fig:Barriers_dz_1} \includegraphics[height=6cm]{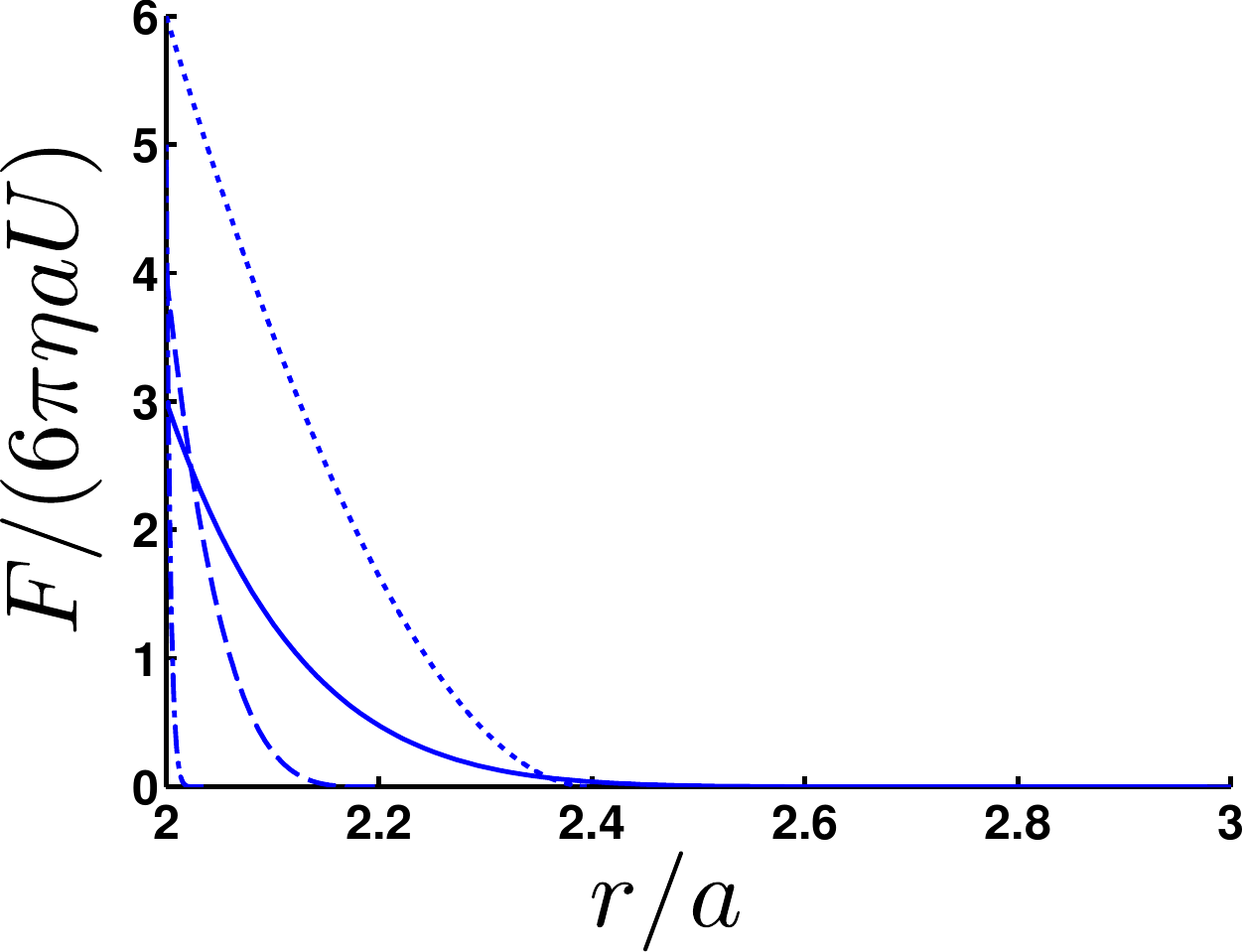} }
\caption{Influence of the collision barrier parameters on the squirmer trajectories for the case where initially $\delta y =1a$.
a) Trajectories. Crosses: data from \cite{Ishikawa2006}.
b) Force barrier profiles (see also Appendix B). The line styles correspond those showing the trajectories in a). \symbol{\dashed}{}{10}{0}{blue}{black}: barrier used in Fig. \ref{fig:Traj_dz_1_10}.} 
\label{fig:Barriers}
\end{figure}

As shown in Fig. \ref{fig:Traj_dz_1_10}, the trajectories match very well with the BEM results for $\delta y \ge 2a$.  When $\delta y =1a$, the collision barrier and near-field hydrodynamic interactions play an important role in determining the overall squirmer trajectories.  Fig. \ref{fig:Barriers} shows the effect of the steric repulsion parameters $F_{\mbox{ref}}$ and $R_{\mbox{ref}}$ on the squirmer trajectories.  The specific values of these parameters are provided in Appendix B.  We see that by varying the barrier parameters one can obtain trajectories that closely match the results of \cite{Ishikawa2006}.

\section{Simulation Results}

\label{simul_res}
\subsection{Large suspensions of swimming micro-organisms}

The squirmer model has been widely used to investigate both the behavior of single swimmers \cite{Doostmohammadi2012, Wang2012, Michelin2013, Zhu2013}, as well as their collective dynamics and interactions \cite{Ishikawa_2007,Ishikawa_2008,Evans2011,Alarcon2013,Lambert2013}. Simulations of suspensions revealed that the overall population dynamics depend strongly on the squirming parameter $\beta$.  In particular, when $|\beta|$ is small, the isotropic state for a periodic suspension has been shown to be unstable, and the suspension evolves to a polar steady-state with a non-zero value of the polar order parameter
\begin{equation}
 P(t)=\left|\dfrac{1}{N_{p}}\sum_{n=1}^{N_{p}}\mathbf{p}_{n}(t)\right|.
 \label{eq:pop}
\end{equation}
This instability has been studied numerically by \cite{Ishikawa_2008} and \cite{Evans2011} using Stokesian dynamics with $N_p=64$ swimmers for volume fractions $\phi_v=0.01 \rightarrow 0.5$.  Using the Lattice-Boltzmann method, \cite{Alarcon2013} observed the same behaviour for $N_p=2000$  and $\phi_v=0.1$.  

\subsubsection{Polar order parameter}

Using our FCM model, we study this instability and the resulting polar order of a squirmer suspension.  In particular, we examine how the domain size affects both the growth rate of the instability and the final steady-state.  The influence of domain size has not been addressed previously for squirmer suspensions though it has been observed in simulations of rod-like swimmers \cite{saintillan_Shelley_2012}.  We perform simulations of semi-dilute suspensions ($\phi_v = 0.1$) of puller squirmers ($\beta = 1$) in triply-periodic square domains with edge lengths ranging from $L/a = 14$ to $L/a=116$.  As the volume fraction $\phi_v$ is fixed, varying $L/a$ increases of the number of swimmers from $N_p = 64$ (as in \cite{Ishikawa_2008, Evans2011}) to $N_p = 37,659$.  We initialize a homogeneous, isotropic suspension by distributing the swimmer positions uniformly in the domain and the swimming directions uniformly over the unit sphere.  Depending on domain size, the simulations are run to final time $t_f = 1000 - 1500 a/U$ with a time step $\Delta t = 0.005a/U$.  Thus, each simulation requires between $2\times 10^5-3\times 10^5$ time-steps.

\begin{figure}
\centering
\includegraphics[width=16cm]{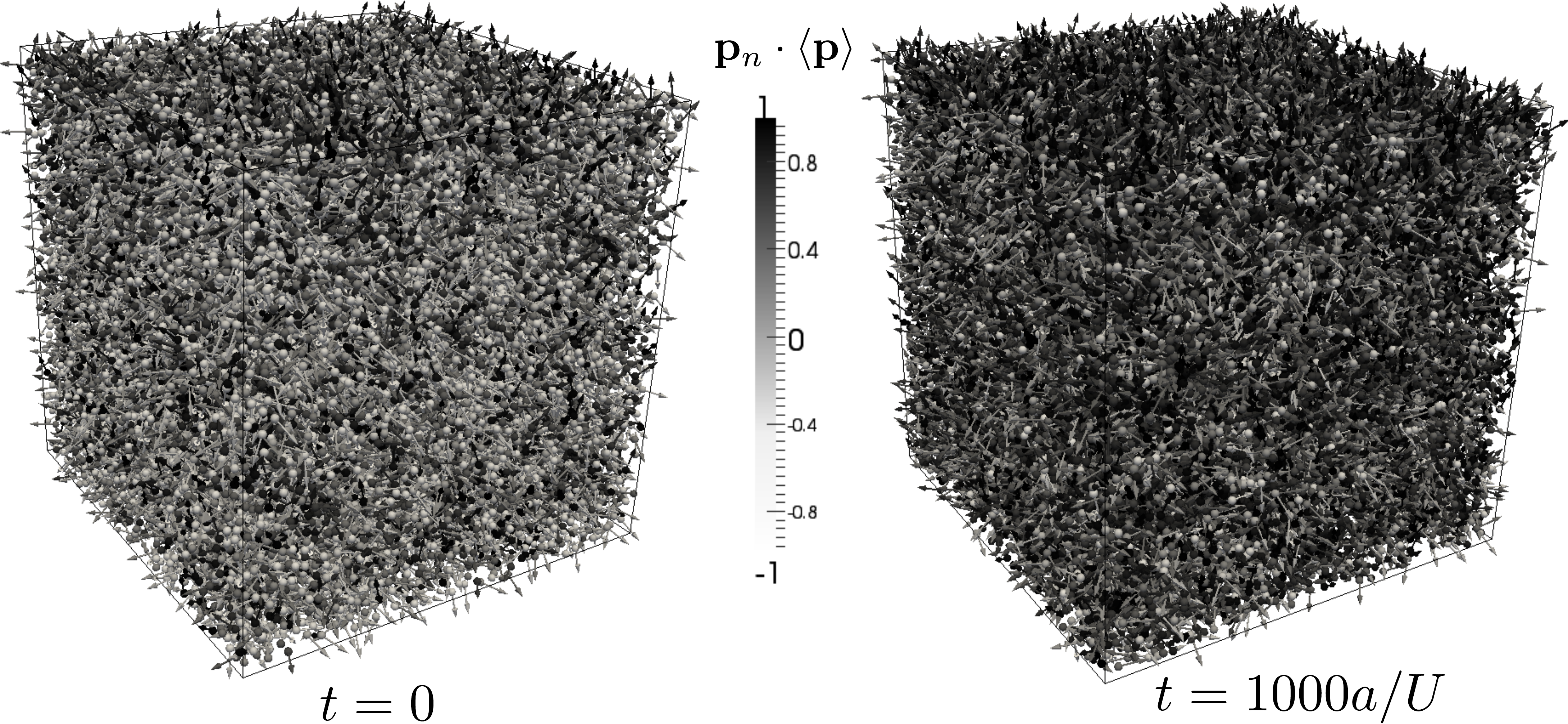}
\caption{\label{fig:snapshot_PO} Snapshots of the orientational state in a semi-dilute suspension ($\phi_v=0.1$) containing $N_p = 37,659$ swimmers with $\beta=1$. $\langle \mathbf{p} \rangle$ is the mean steady-state  orientation vector on the unit sphere defined in Eq. \eqref{eq:mean_dir}.  $\mathbf{p}_n \cdot \langle \mathbf{p} \rangle$ quantifies the degree of alignment with the mean direction $\langle \mathbf{p} \rangle$ of swimmer $n$.
} 
\end{figure}

Figure \ref{fig:snapshot_PO} illustrates the polar ordered state for a simulations with $L/a = 116,\,N_p = 37,659$. 
We quantify the degree of alignment of each swimmer $n$, $\mathbf{p}_n \cdot \langle \mathbf{p} \rangle$, with the mean steady-state orientation $\langle \mathbf{p}\rangle$ given by
\begin{equation}
\langle \mathbf{p} \rangle = \lim_{t\rightarrow\infty}\dfrac{1}{N_{p}}\sum_{n=1}^{N_{p}}\dfrac{\mathbf{p}_{n}(t)}{P(t)}.
\label{eq:mean_dir}
\end{equation}
where $\langle \mathbf{p} \rangle$ lies on the unit sphere $\mathbb{S}^2$.
At $t=0$, there is no clear mean orientation, while at $t = 1000a/U$, a significant proportion of particles are aligned with the mean direction $\langle \mathbf{p} \rangle$.

\begin{figure}
\centering
\subfloat[]{ \label{fig:P_t} \includegraphics[height=6.2cm]{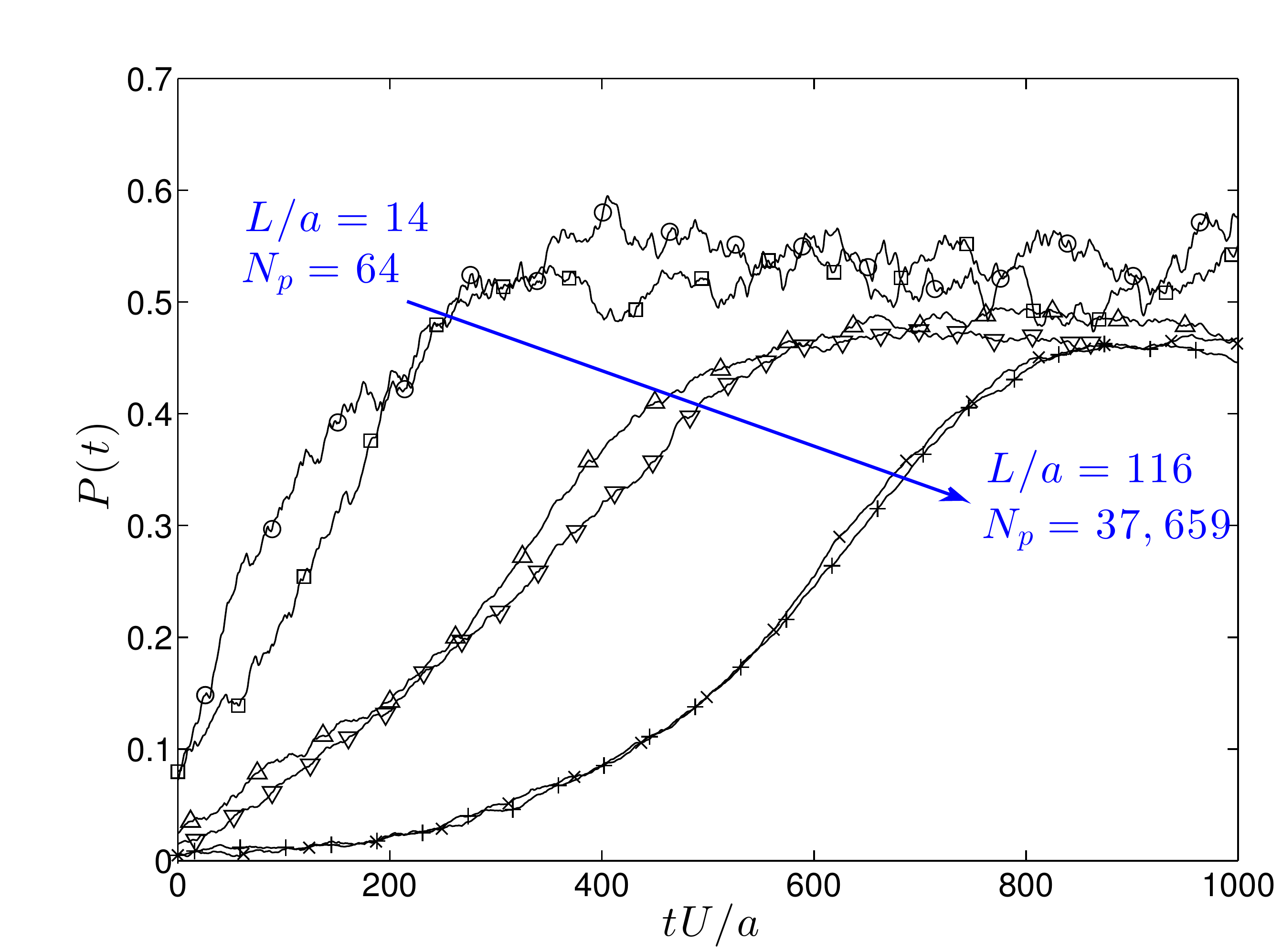}}
\subfloat[]{ \label{fig:P_inf} \includegraphics[height=5.8cm]{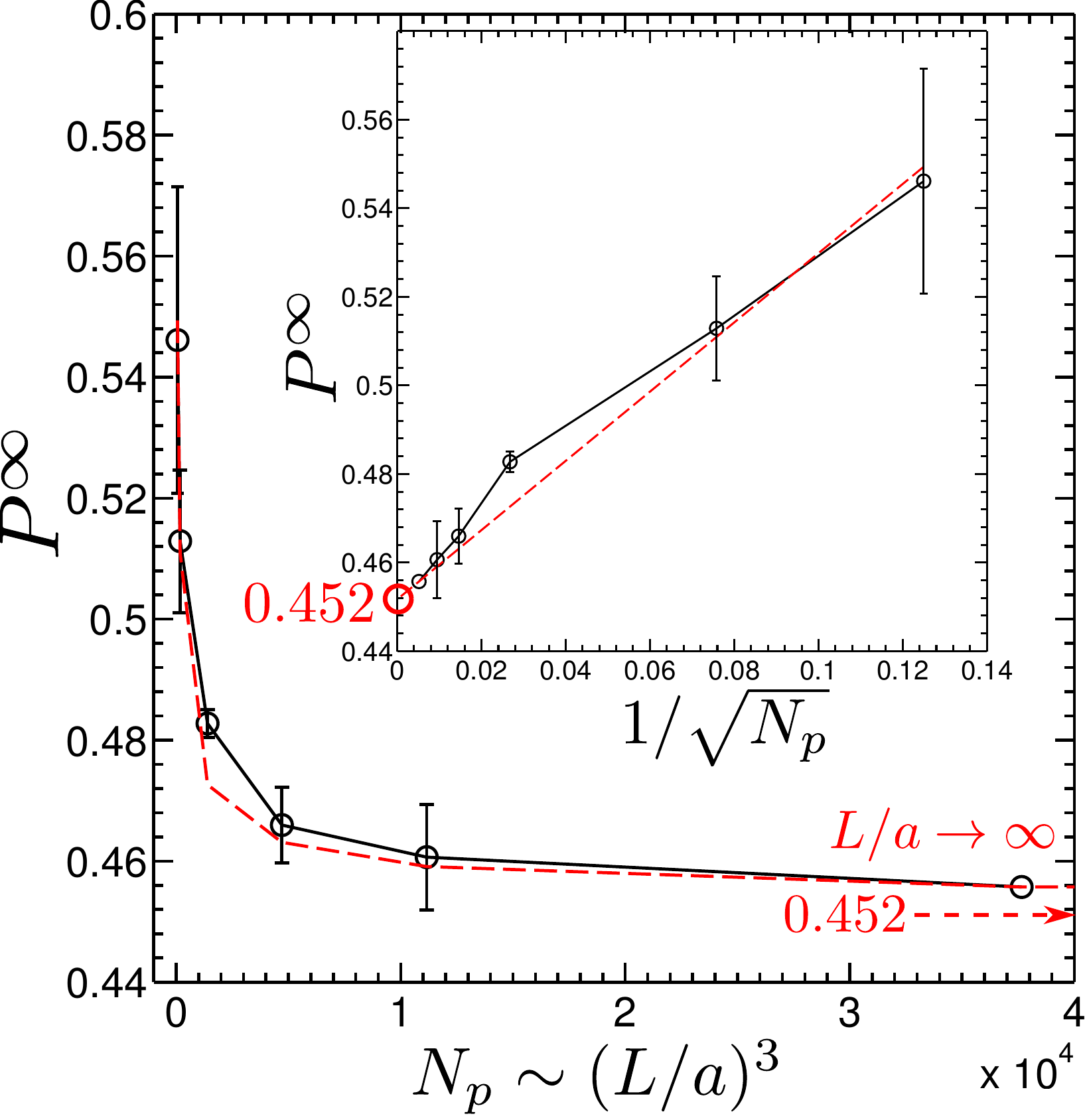}}

\caption{Polar order $P(t)$ in a semi-dilute suspension ($\phi_v=0.1$) of squirmer pullers ($\beta=1$). 
a) Time evolution of polar order depending on the number of swimmers.  
\symbol{\solid}{\bigcircle}{22}{0}{black}{black} : $L/a = 14, N_p = 64$; 
\symbol{\solid}{\ssquareb}{22}{1}{black}{black} : $L/a = 19, N_p = 174$; 
\symbol{\solid}{\trianup}{22}{1}{black}{black} : $L/a = 38, N_p = 1,395$; 
\symbol{\solid}{\bigtriandown}{22}{0}{black}{black} : $L/a = 58, N_p = 4,707$; 
\symbol{\solid}{\asterix}{22}{-0.5}{black}{black} : $L/a = 77, N_p = 11,158$; 
\symbol{\solid}{\plus}{22}{-0.2}{black}{black} : $L/a = 116, N_p = 37,659$. 
b) Steady state value $P^{\infty}$ depending on  $N_p \sim (L/a)^3$. 
\symbol{\dashed}{}{10}{0}{red}{black}: fit linear with $1/\sqrt{N_p}$.
(Inset): Dependence of  $P^{\infty}$ with $1/\sqrt{N_p}$.
} 
\label{fig:Polar_order}
\end{figure}

Fig. \ref{fig:P_t} shows $P(t)$, Eq. \eqref{eq:pop}, for the simulations with different domain sizes.  For each domain size, we see that the suspension evolves from the initial isotropic state to one that has polar order.  We observe, however, that the final value, $P^{\infty}$, of the polar order parameter depends on the domain size.  We find that it decreases as $L/a$ (and $N_p$) increases.  The data also shows that as $L/a \rightarrow \infty$, $P^{\infty}$ decays like $(L/a)^{-3/2}$ (or $N_p^{-1/2}$) and reaches an asymptotic value of $P^{\infty} \rightarrow 0.452$.  These results would indicate that polar order should also arise in an unbounded suspension.    We also observe that in the polar ordered state, the average swimming speed is 4\% less than its value in isotropic state which itself is nearly the free swimming value. Indeed, we remind that in our simulations, a zero net flux of the suspension is imposed, therefore, when the swimmers are aligned, they are slowed down by the backflow. 

\begin{figure}
\centering
\includegraphics[height=8cm]{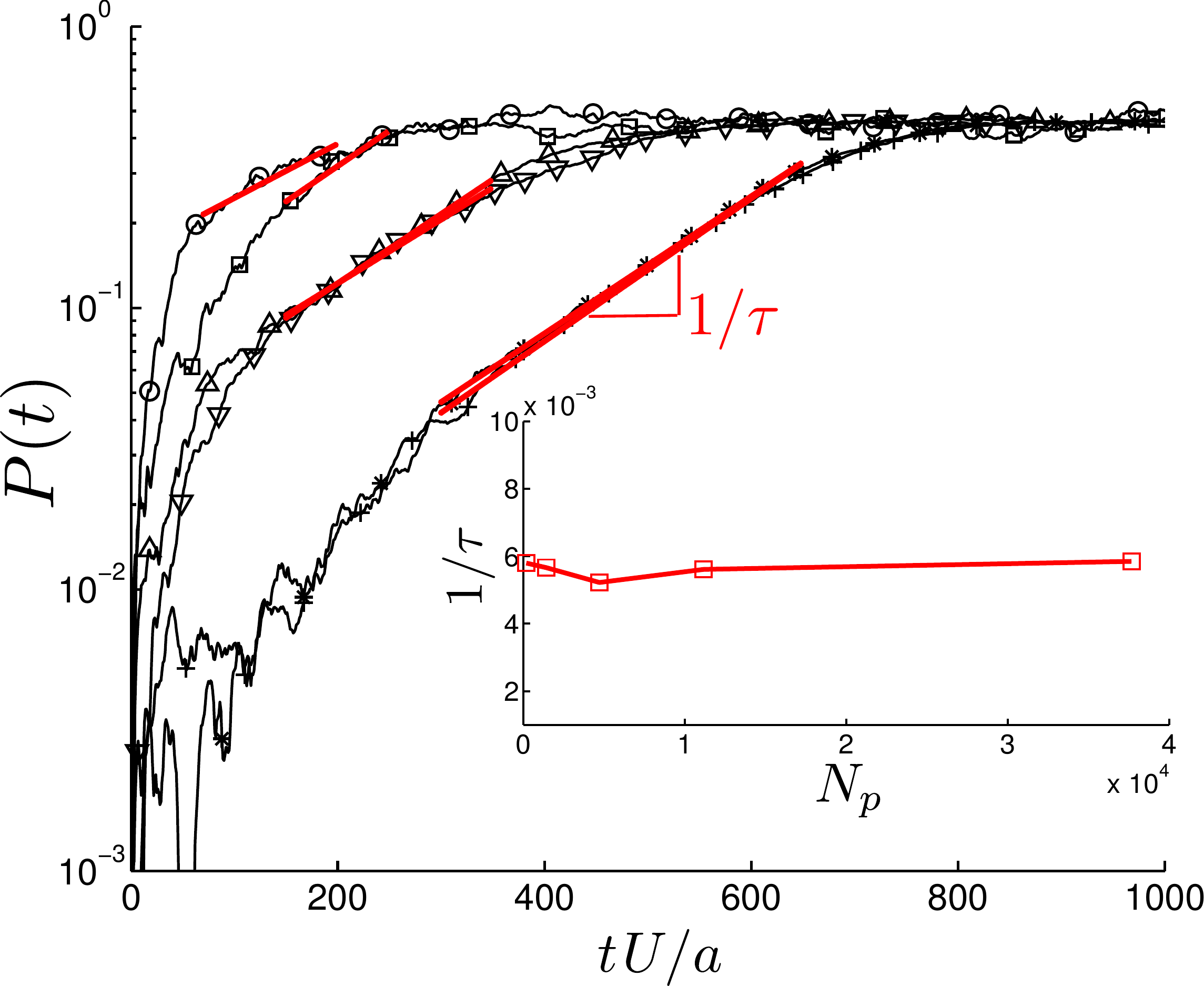}
\caption{Characterization of the polar instability. 
(Main figure): Time evolution of polar order in semilogarithmic scale suggests an exponential growth of the instability.
(Inset): Growth rate of the instability for different $N_p$ (and $L/a$). 
} 
\label{fig:growth_instability}
\end{figure}

From our simulations, we may also analyse how $L/a$ affects the time evolution of the instability.  In Fig. \ref{fig:P_t}, we clearly see that the time it takes to reach the final polar state increases with domain size.  We analyze this data in more detail in Fig. \ref{fig:growth_instability}, now plotting it in semilogarithmic scale.  For each case, we find that after an initial transient state, the instability grows exponentially and with a growth rate that is independent of the system size (see inset figure).  It would certainly be interesting to investigate if this result could be reproduced by a linear stability analysis of continuum model for a squirmer suspension.

\subsubsection{Orientational distribution}
\begin{figure}
\centering
\subfloat[]{ \label{fig:PDf_th_phi_90} \includegraphics[height=6cm]{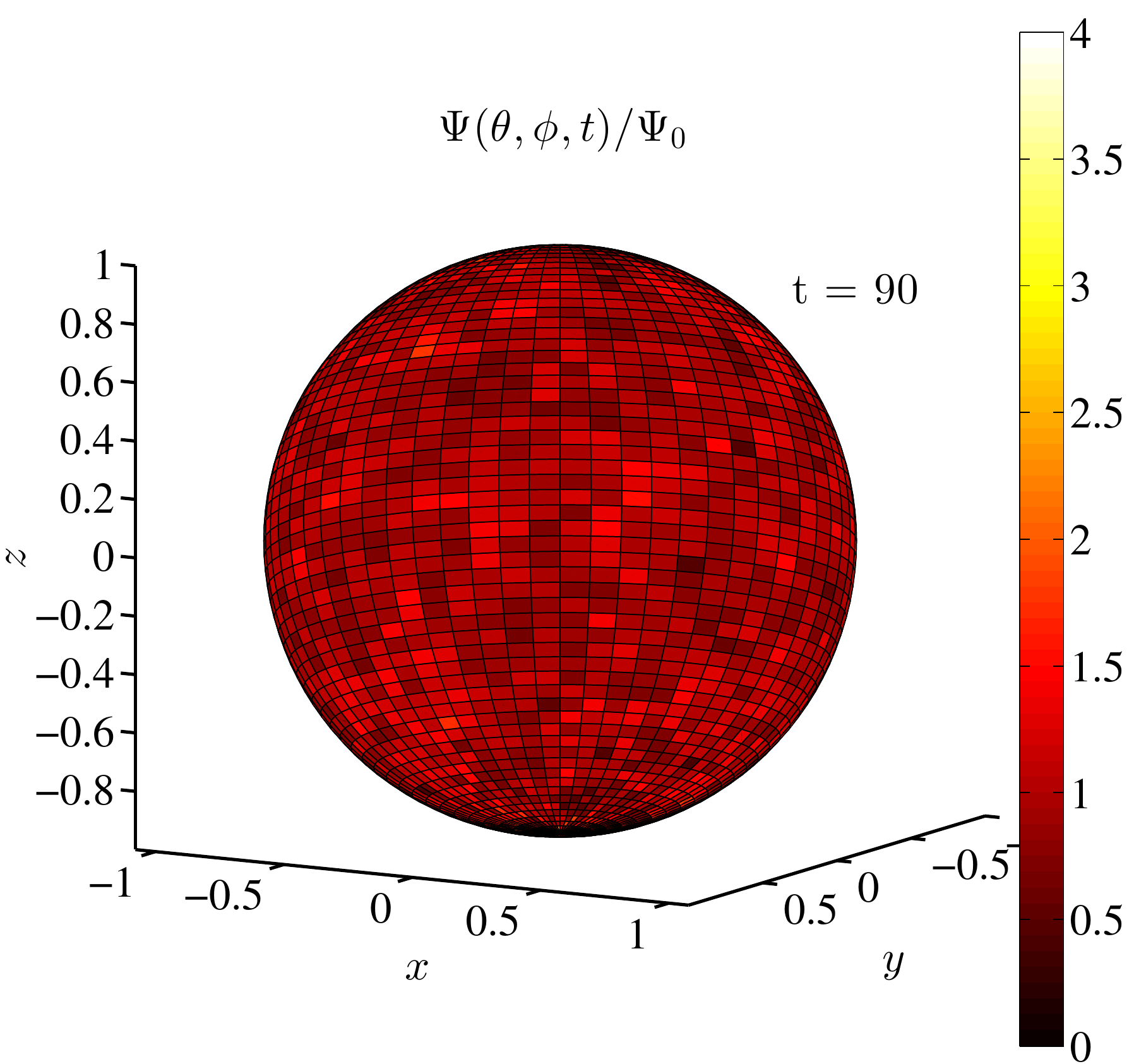}}
\subfloat[]{ \label{fig:PDf_th_phi_1110} \includegraphics[height=6cm]{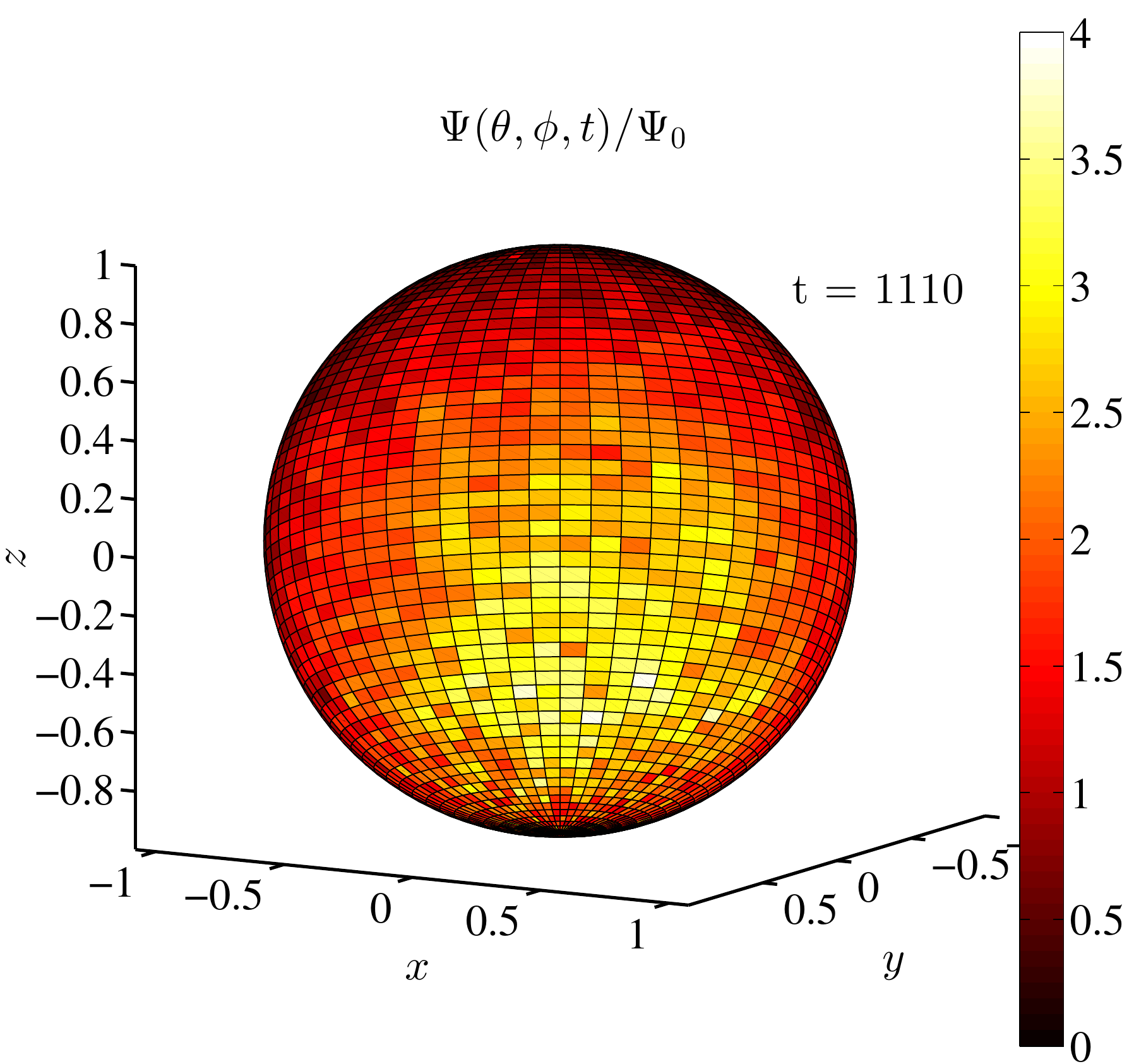}}
\caption{Orientational distribution on the unit sphere $(\phi,\theta)$. 
a) Distribution before the transition to polar order:  $tU/a=90$.
b) Distribution once the polar ordered state is reached:   $tU/a=1110$.
} 
\label{fig:Orientation_distrib_time}
\end{figure}

\begin{figure}
\centering
\subfloat[]{ \label{fig:PDf_th_phi_pmean_steady} \includegraphics[height=6cm]{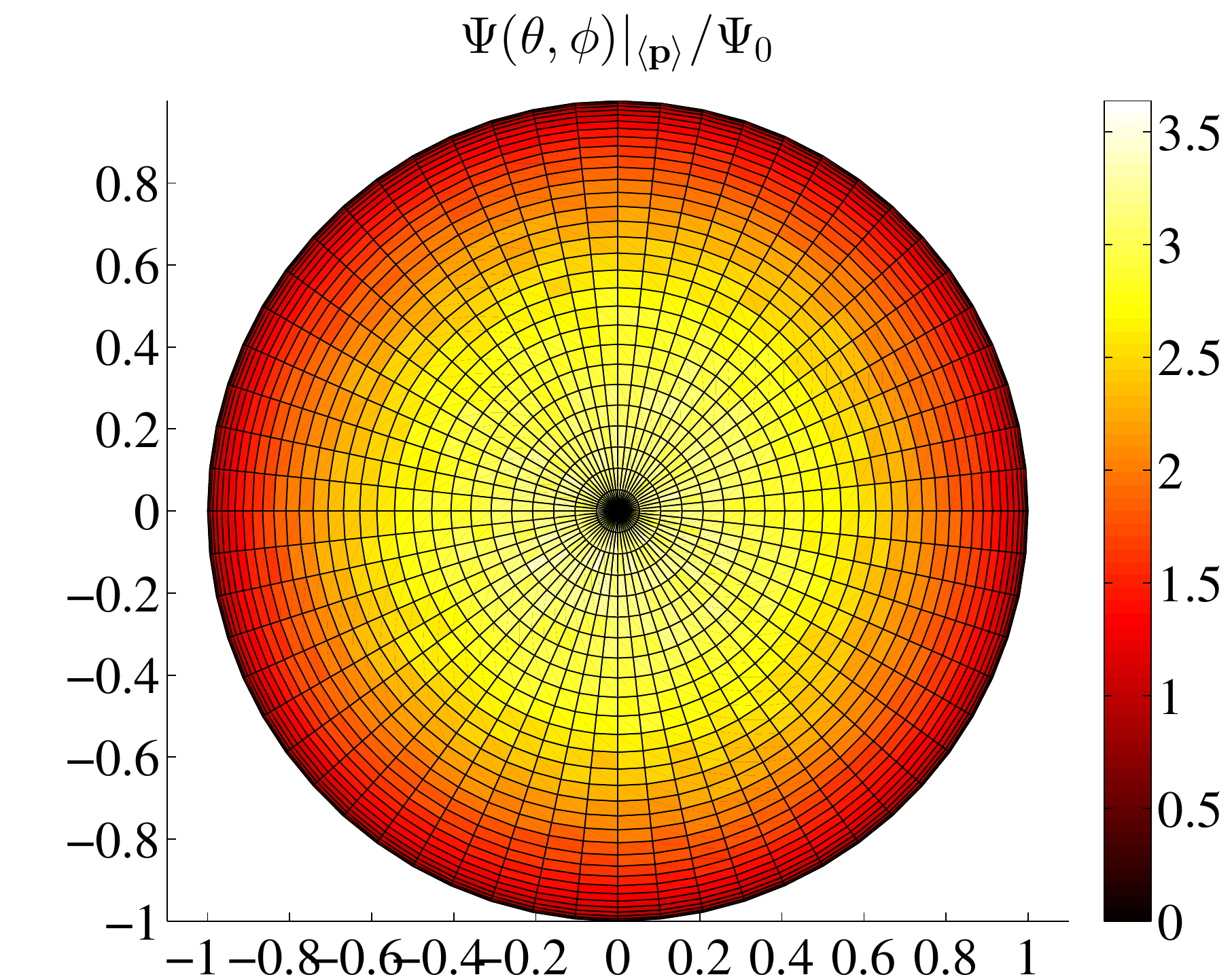}}
\hspace{1cm}
\subfloat[]{ \label{fig:PDf_1D_th_phi_pmean_steady} \includegraphics[height=6cm]{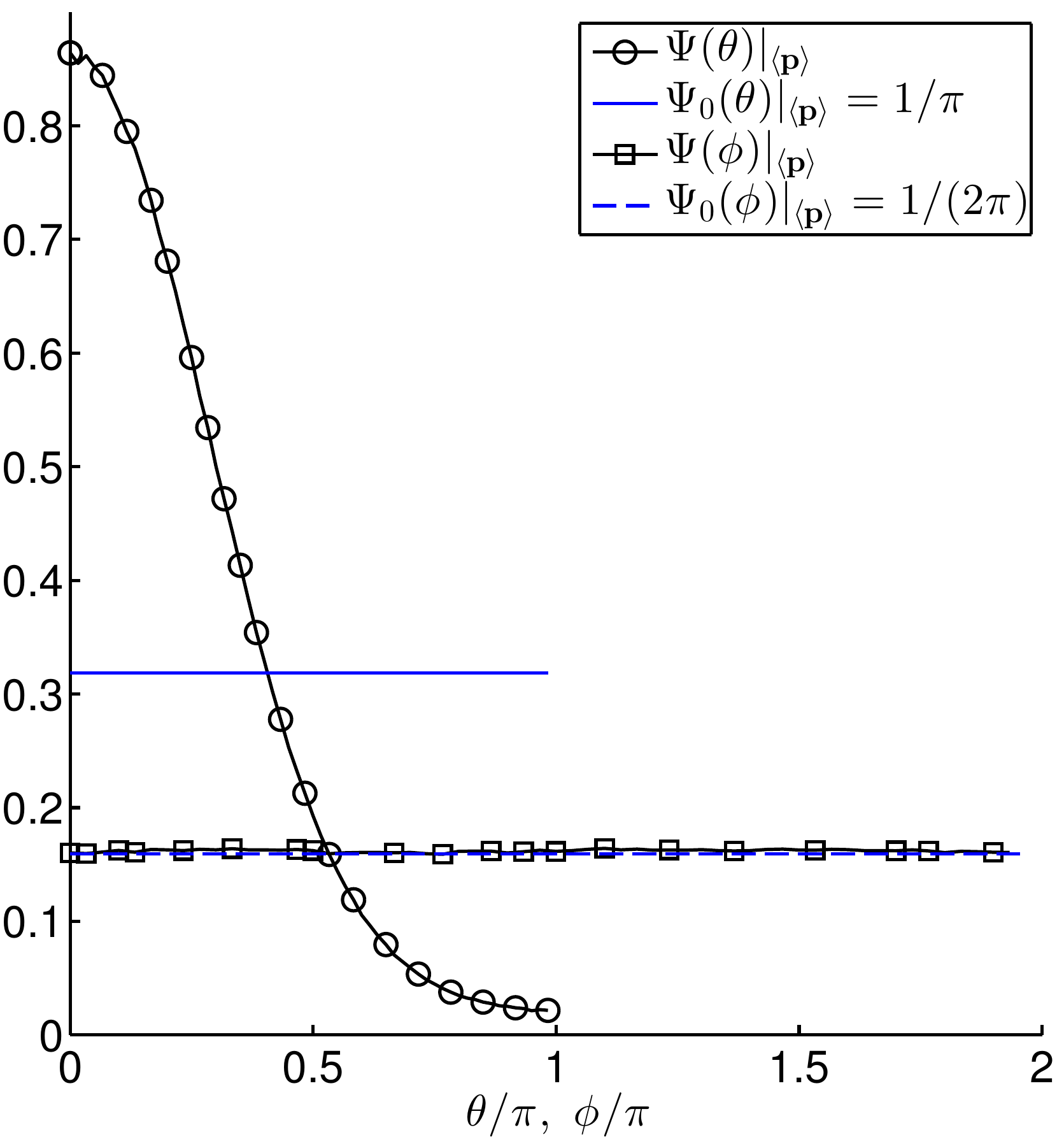}}
\caption{Time-averaged steady-state orientational distribution in the frame of the mean orientation vector $\left.\Psi(\theta,\phi)\right|_{\left\langle \mathbf{p}\right\rangle}$. 
a) Distribution over the unit sphere.
b) \symbol{\solid}{\bigcircle}{22}{-1}{black}{black} : distribution of elevation angle $\theta$ averaged over azimuthal angle $\phi$; 
\symbol{\solid}{}{10}{0}{blue}{black} : uniform distribution $\Psi_0(\theta) = 1/\pi$; 
\symbol{\solid}{\ssquareb}{22}{1}{black}{black} : distribution of azimutal angle $\phi$ averaged over elevation angle $\theta$; 
\symbol{\dashed}{}{10}{0}{blue}{black} : uniform distribution $\Psi_0(\phi) = 1/(2\pi)$.
} 
\label{fig:Orientation_distrib_steady}
\end{figure}

We can examine the polar order in more detail by computing the orientational distribution $\Psi(\theta,\phi,t)$ defined over the unit sphere \cite{Saintillan2010b}. Here, $\theta= \cos^{-1}(p_z)$ corresponds to the elevation angle while $\phi = \tan^{-1}(p_y/p_x)$ gives the azimuthal angle.  Fig. \ref{fig:Orientation_distrib_time} shows $\Psi(\theta,\phi,t)$ normalized by the isotropic distribution $\Psi_0 = 1/(4\pi)$ at times before and after the transition to polar order for the case where $N_p = 11,158$ and $L=78a$.  As expected, before the transition, the distribution is nearly uniform over the surface of the sphere.  After the polar state is reached, we see that $\Psi(\theta, \phi)$ is narrowly distributed around the mean direction $\langle \mathbf{p} \rangle$.  We note that the steady state mean direction depends both on the random initial seeding of swimmers and the domain geometry.  We find that system first aligns along an arbitrary direction due to the random initial seeding of swimmers, but as time goes on, the mean director tends to align with the normal to one of the periodic boundaries.  As a consequence, the steady state mean direction is biased by the domain shape that breaks radial symmetry.  We would like to stress that, however, as demonstrated here and in [30], the polar ordering itself is a consequence of squirmer hydrodynamic and steric interactions rather than the domain shape. Fig. \ref{fig:Orientation_distrib_steady} shows the steady-state averaged distribution $\left.\Psi(\theta,\phi)\right|_{\left\langle \mathbf{p}\right\rangle }$ in the frame where the mean direction is given by $\theta = 0$ and $\phi = 0$.  We see that the resulting distribution is axisymmetric in that it does not depend on $\phi$. This is not surprising as the flow field induced by a squirmer is axisymmetric. Orientational distributions could also be obtained from continuum models, as was done for swimmers in external flow fields \cite{Saintillan2010b}.  Our results could be compared with continuum models of squirmer suspensions, though, to the best of our knowledge, the continuum models that are currently available in the literature predict a stable isotropic state for spherical swimmer suspensions.  

\subsubsection{Spatial distribution}
\begin{figure}
\centering
\includegraphics[height=8cm]{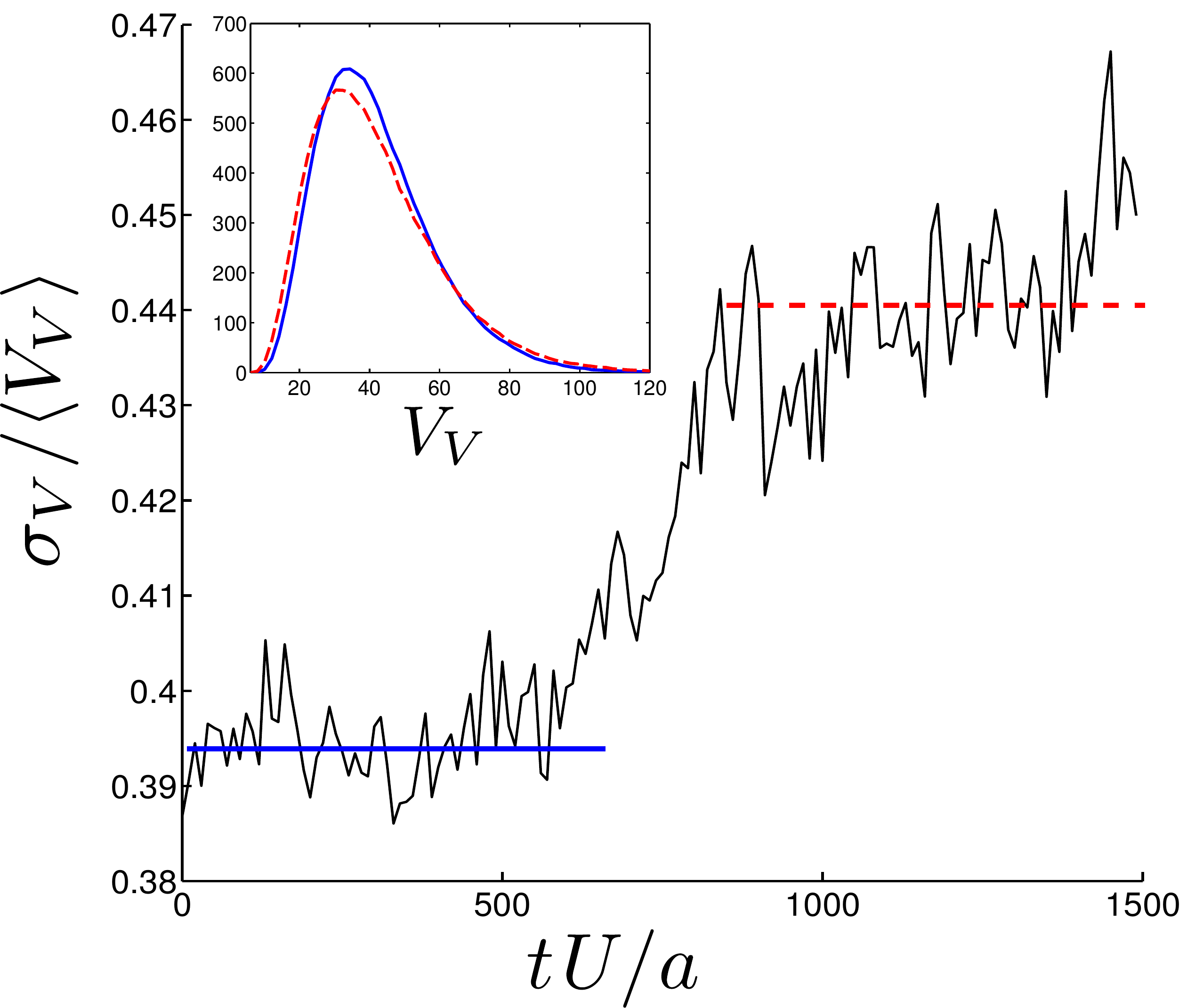}
\caption{Distribution of Voronoi cell volumes.
(Main figure): Time dependence of the standard deviation, $\sigma_V$, of the Voronoi volume distribution normalized by the mean value $\langle V_V \rangle$. 
(Inset): Time average of the Voronoi volume distribution before (\symbol{\solid}{}{10}{0}{blue}{black}) and after (\symbol{\dashed}{}{10}{0}{red}{black}) the the polar order transition. 
} 
\label{fig:STD_voro}
\end{figure}

Along with the orientational distribution, we also examine the spatial distribution of squirmers by computing the Voronoi tesselation for the set of points corresponding to the squirmers' centers. We have used the C++ library Voro++ \cite{Rycroft2009} to perform the computation and determine the volume, $V_V$, of each Voronoi cell.  Fig. \ref{fig:STD_voro} shows the evolution of the standard deviation, $\sigma_V$, of the $V_V$ distribution for the case where $N_p = 11,158$ and $L=78a$.  We see sudden increase in $\sigma_V$ as the suspension transitions to polar order, corresponding to a widening of the distribution.
The inset in Fig. \ref{fig:STD_voro} shows the time-averaged distribution before and after the transition.  Before polar ordering occurs, the mode of the distribution is $\bar{V_V}\approx 34a^3$, which is slightly less than the average value $\langle V_V \rangle = L^3/N_p = 42a^3$ that one might expect.  During the polar steady-state, we find that the mode decreases to $\bar{V_V}\approx 30a^3$, which is indicative of a slight clustering of the particles.  
\label{sec:clustering}

\begin{figure}
\centering
\subfloat[]{ \label{fig:rdf_r_theta} \includegraphics[height=7.5cm]{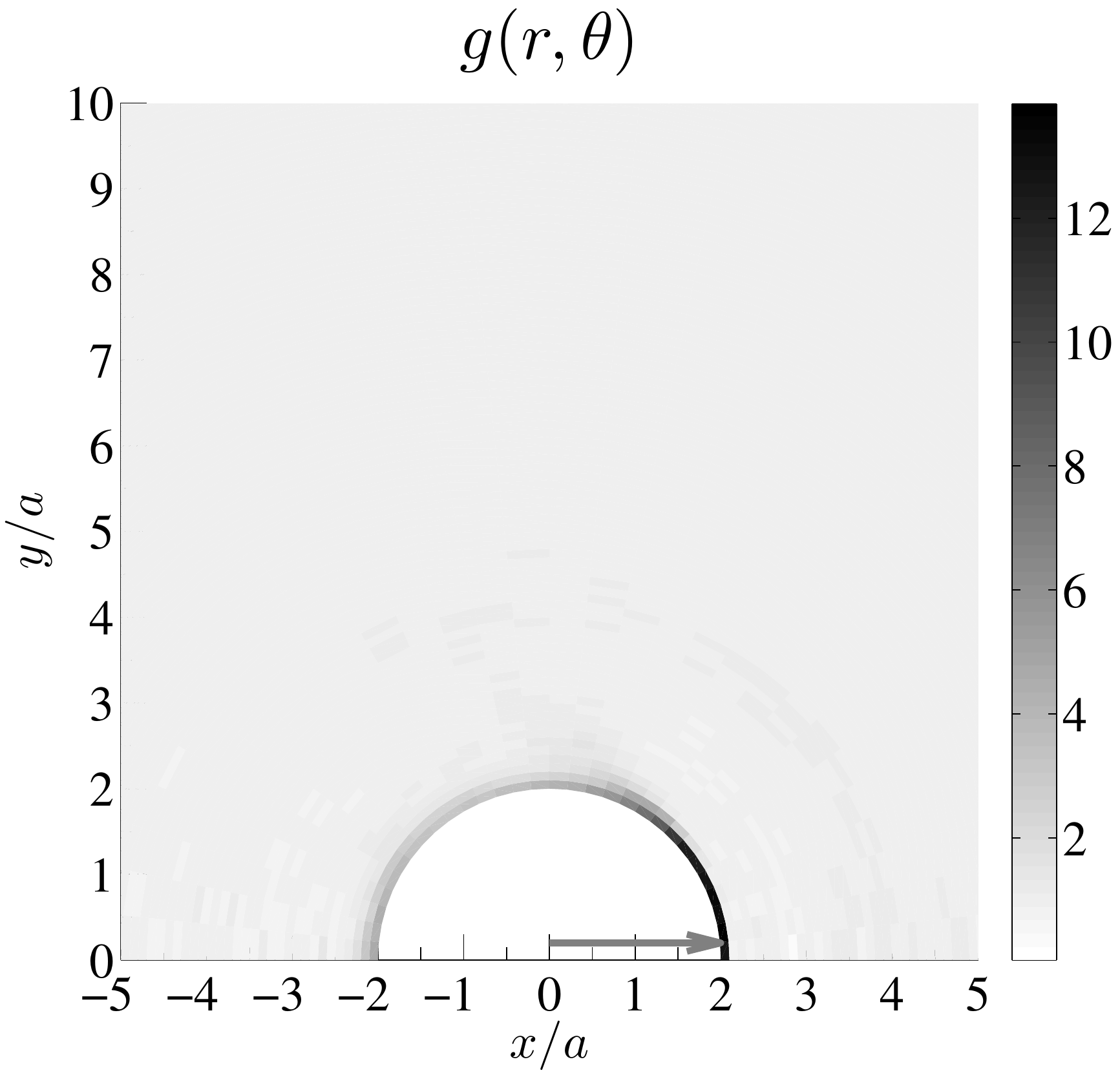}}
\subfloat[]{ \label{fig:rdf_r} \includegraphics[height=7cm]{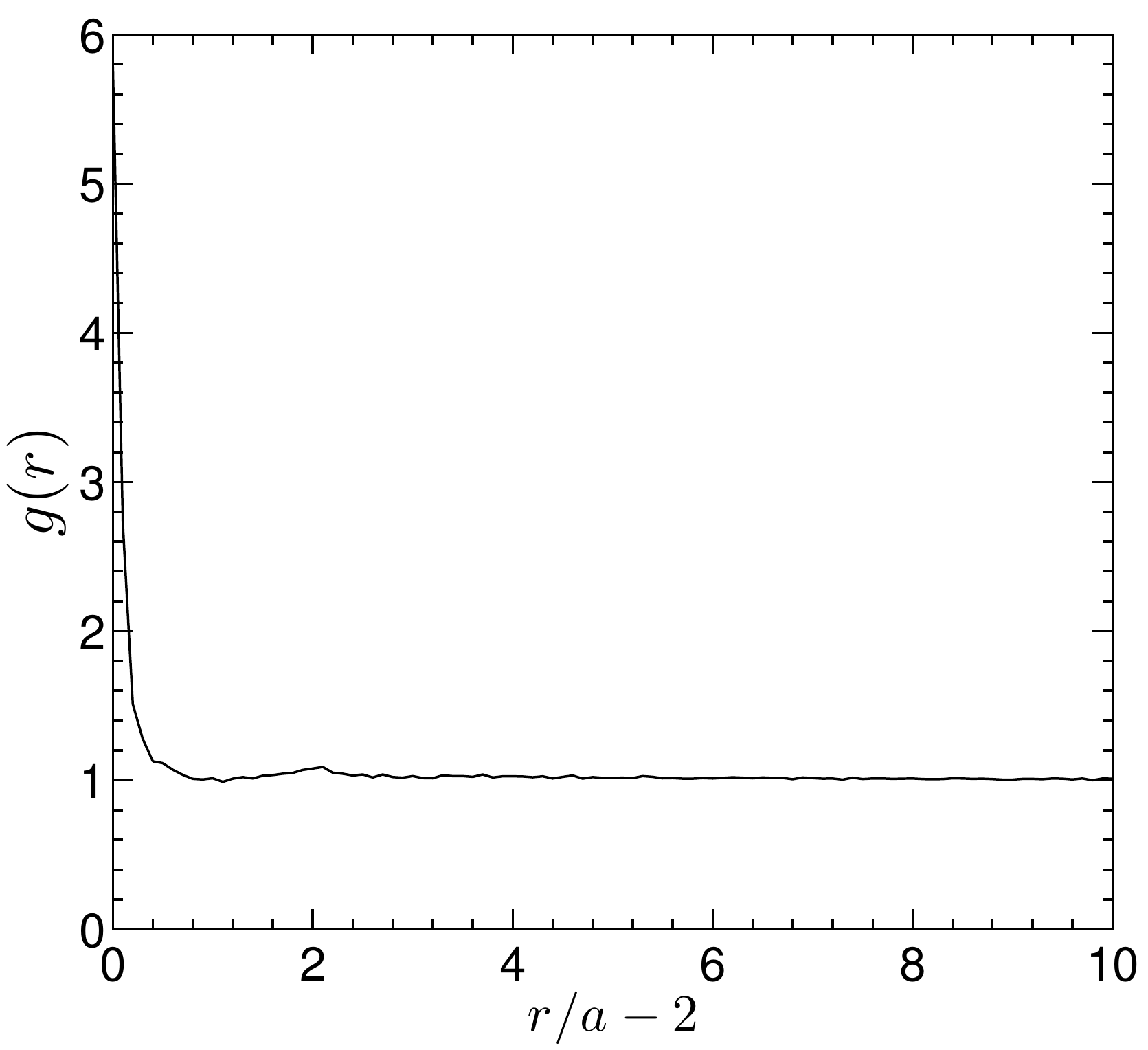}}

\caption{Steady state pair distribution function 
a) $g(r,\theta)$. 
b) $g(r)$.
} 
\label{fig:RDF}
\end{figure}

To further examine the spatial distribution, we compute the steady-state pair distribution function $g(r, \theta)$ which gives the probability of finding a squirmer at a distance $r = |\mathbf{r}|$ and with elevation angle $\theta = \cos^{-1}(\mathbf{r}\cdot \mathbf{p}/r)$ from another squirmer that has swimming direction $\mathbf{p}$.  Fig. \ref{fig:rdf_r_theta} shows $g(r, \theta)$ for the case where $N_p = 11,158$ and $L=78a$.   We see clearly that $g(r, \theta)$ depends not only on $r$, but on $\theta$ as well.  At steady-state, for a given squirmer there is a significantly higher probability of finding another squirmer in front of it ($\theta=0$) rather than behind it.  If we integrate $g(r, \theta)$ over $\theta$, we obtain the radial distribution function shown in Fig. \ref{fig:rdf_r}.  We find that $g(r)$ exhibits a peak at two radii from the swimmer surface. This peak suggests the existence of particle clusters whose sizes are greater than two individuals.

\subsubsection{Orientational correlations}
\begin{figure}
\centering
\subfloat[]{ \label{fig:polarcor_r_theta} \includegraphics[height=7.5cm]{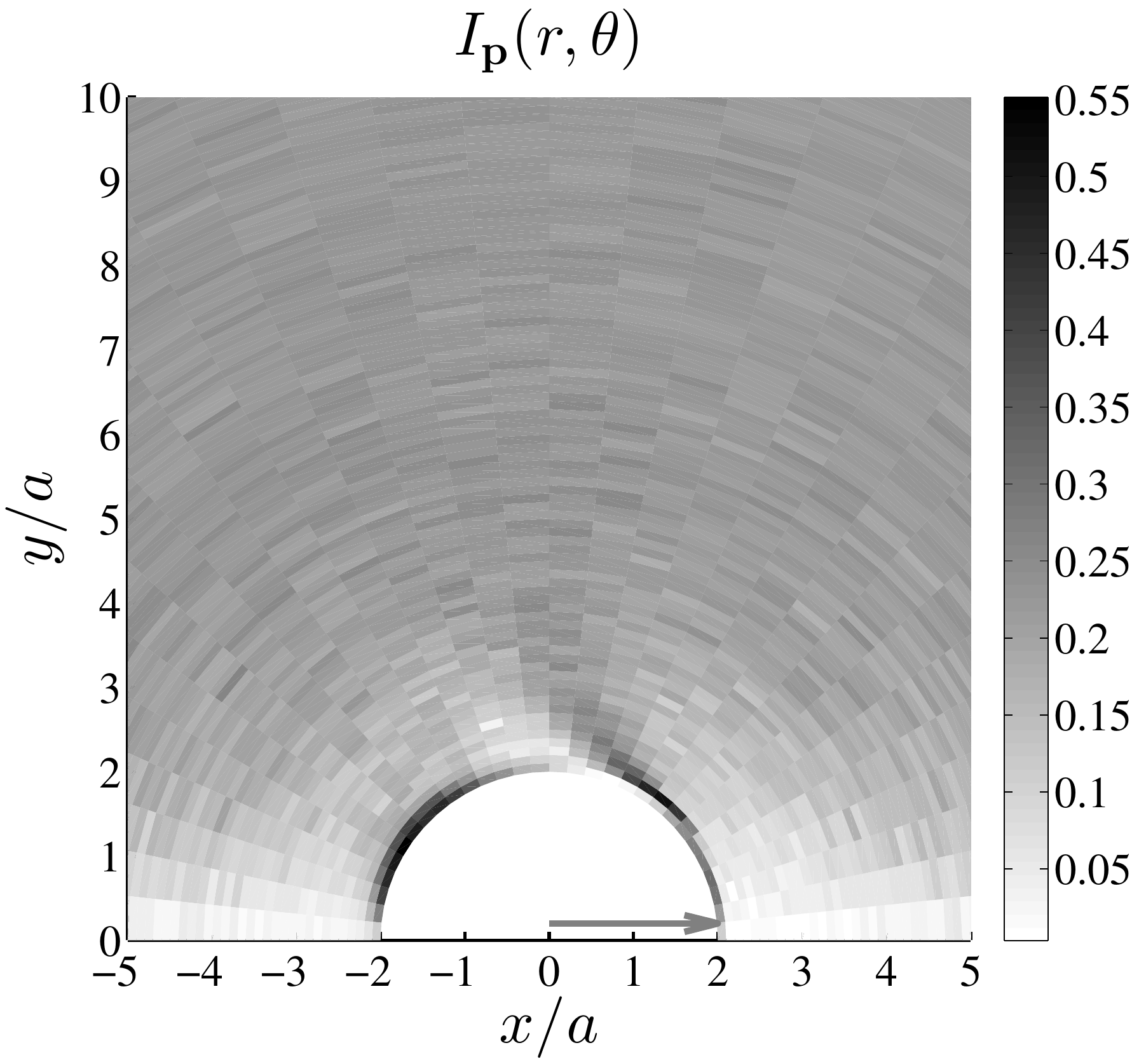}}
\subfloat[]{ \label{fig:polarcor_r} \includegraphics[height=7cm]{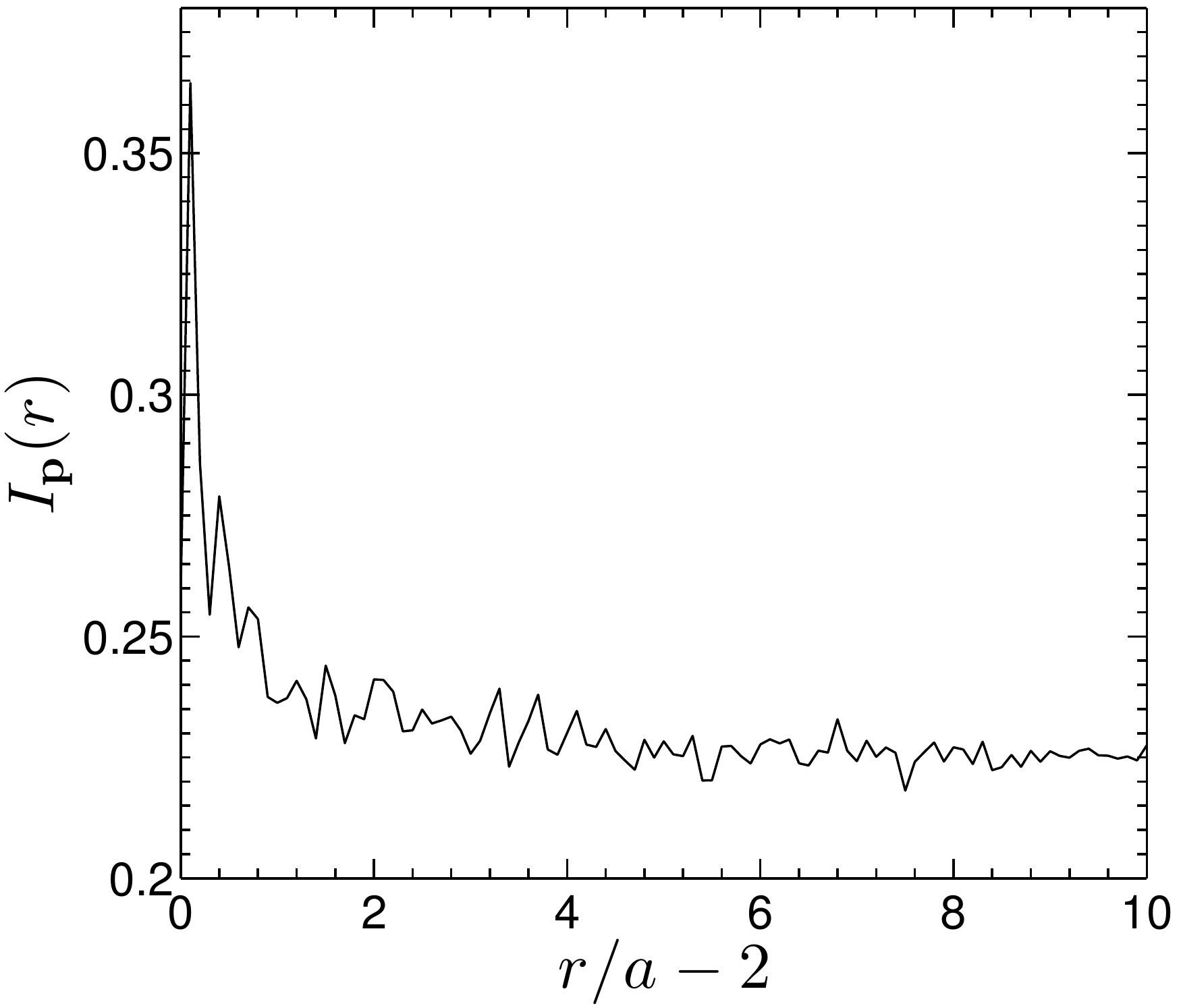}}

\caption{Correlations in squirmer orientation at steady state. 
a) $I_{\mathbf{p}}(r,\theta)$. 
b) $I_{\mathbf{p}}(r)$
} 
\label{fig:POLARCOR}
\end{figure}

The hydrodynamic and steric interactions between the squirmers also lead to correlations in their orientations.  We compute the steady-state orientational correlation function \cite{Ishikawa_2008}
\begin{equation}
 I_{\mathbf{p}}\left(\mathbf{r}\right) = \langle \mathbf{p(\mathbf{x})\cdot p(\mathbf{x+r})} \rangle,
\end{equation}
where the brackets $\langle \cdot \rangle$ denote the ensemble average that we compute by averaging over both time and squirmer pairs.  Fig. \ref{fig:polarcor_r_theta} shows the correlations in the frame of a squirmer located at the origin and with $\mathbf{p} = \mathbf{\hat z}$ and the definitions of $r$ and  $\theta$ are identical to those in Section \ref{sec:clustering}.  We find that the highest correlations occur near contact ($r \approx 2$) at the angles $\theta \approx \pi/3$ and $\theta \approx 3\pi/4$.  Despite the overall polar order, we can identify two distinct regions around $\theta = 0$ and $\theta = \pi$ where the orientations are uncorrelated. We do, however, find a positive value for the correlations, even far away from the origin.  This is most evident in the $\theta$-averaged correlation, $I_{\mathbf{p}}\left(r\right)$, shown in Fig. \ref{fig:polarcor_r} which approaches a finite value $I_{\mathbf{p}} = 0.22$ as $r$ increases. This is consistent with the observed long-range polar order of the suspension.  We also find that $I_{\mathbf{p}}\left(\mathbf{r}\right) > 0$ for all $\mathbf{r}$.  This again may be a result of the strong polar ordering of the suspension.  


\section{Extensions to ellipsoidal swimmers and time-dependent swimming gaits}
\label{sec:Spheroid_Time_dep}

In this section, we demonstrate the extension of our approach to both spheroidal swimmer shapes and time-dependent swimming gaits.  This illustrates the versatility of FCM while preserving good computational scalability and an accurate treatment of hydrodynamic interactions.  

\subsection{FCM for ellipsoidal particles}
\label{subsec:Spheroid_FCM}
To extend FCM to passive ellipsoidal particles, one simply modifies the Gaussian envelopes in Eq. (\ref{eq:FCM2}) \cite{Liu2009}.  
For example, taking the orthonormal vectors $\mathbf{\hat{e}}_1$, $\mathbf{\hat{e}}_2$, and $\mathbf{\hat{e}}_3$ to be aligned with the ellipsoid semi-axes having lengths $a_1$, $a_2$, and $a_3$ respectively, the Gaussian envelope corresponding to $\Delta_{n}(\mathbf{x})$ is
\begin{eqnarray}
\Delta^{ell}_n(\mathbf{x})&=&(2\pi)^{-3/2}(\sigma_{\Delta;1}\sigma_{\Delta;2}\sigma_{\Delta;3})^{-1}\textrm{exp}\left[-\frac{1}{2}(\mathbf{\mathbf{x} - \mathbf{Y}_n})^{T}\mathcal{Q}^{T}\bm{\Sigma}_{\Delta} \mathcal{Q} (\mathbf{\mathbf{x} - \mathbf{Y}_n})\right] \nonumber\\
\label{eq:Gauss_ell}
\end{eqnarray} 
where $\sigma_{\Delta;i} = a_i/\sqrt{\pi}$ for $i = 1, 2, 3$, $\mathcal{Q} = (\mathbf{\hat{e}}_1$  $\mathbf{\hat{e}}_2$  $\mathbf{\hat{e}}_3)$, and 
\begin{equation}
\bm{\Sigma}_{\Delta} = \left(
\begin{array}{c c c}
\sigma_{\Delta;1}^{-2} & 0 & 0 \\
0 & \sigma_{\Delta;2}^{-2} & 0 \\
0 & 0 & \sigma_{\Delta;3}^{-2}
\end{array}
\right).
\end{equation}
A similar expression is used for $\Theta^{ell}_n(\mathbf{x})$ with $\sigma_{\Theta;i} = a_i/\left(6\sqrt{\pi}\right)^{1/3}$.  
Beyond this, the underlying algorithm of projecting the particle forces onto the fluid and volume averaging the resulting fluid flow remains unchanged.  
The constraint that $\mathbf{E}_n = \mathbf{0}$ for each $n$ is still used to find the stresslets.  Thus, using FCM to compute the motion and hydrodynamic interactions of ellipsoids does not require any additional steps in the algorithm described in Section \ref{algorithm}.  As explained in Section \ref{sec:steric}, the modelling of steric interactions differs from that for spherical particles and our simple force barrier could not be used.  An extensive validation of FCM for ellipsoidal particles is presented in \cite{Liu2009} where many of the classical results for ellipsoidal particles in Stokes flow, e.g. Jeffery's orbits, are shown to be recovered exactly with FCM. We note that for very slender particles, FCM would not be appropriate and computations based on slender-body theories \cite{Tornberg2006} should be performed.\\

To extend FCM to active ellipsoidal particles requires adding the stresslet and possible potential dipole terms to the multipole expansion.  As for spherical particles (see Section \ref{sq_interactions} Eq. \eqref{eq:potdipint} and \eqref{eq:forcedipint}), these additional multipoles will lead to artificial, self-induced velocities and local rates-of-strain.  These effects must be subtracted away using the formula derived in Appendix A.  It is worth reiterating that these are the only steps that need to be added to the FCM algorithm for passive particles to simulate active ones.\\

\subsection{Spheroidal swimmer simulations}

\begin{figure}
\centering
\subfloat[]{ \label{fig:snapshot_ell} \includegraphics[height=6.8cm]{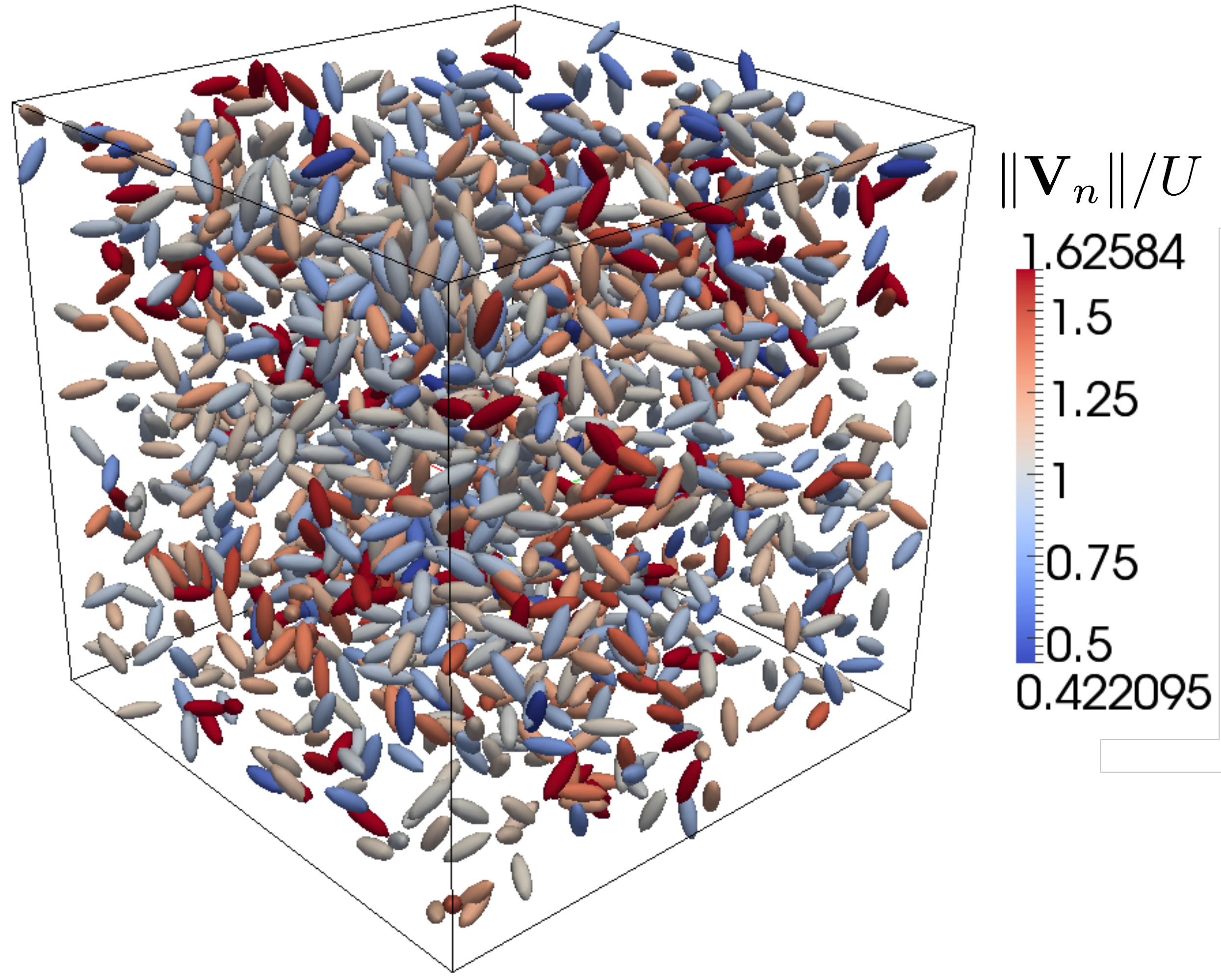}}
\subfloat[]{ \label{fig:P_t_ell} \includegraphics[height=6.8cm]{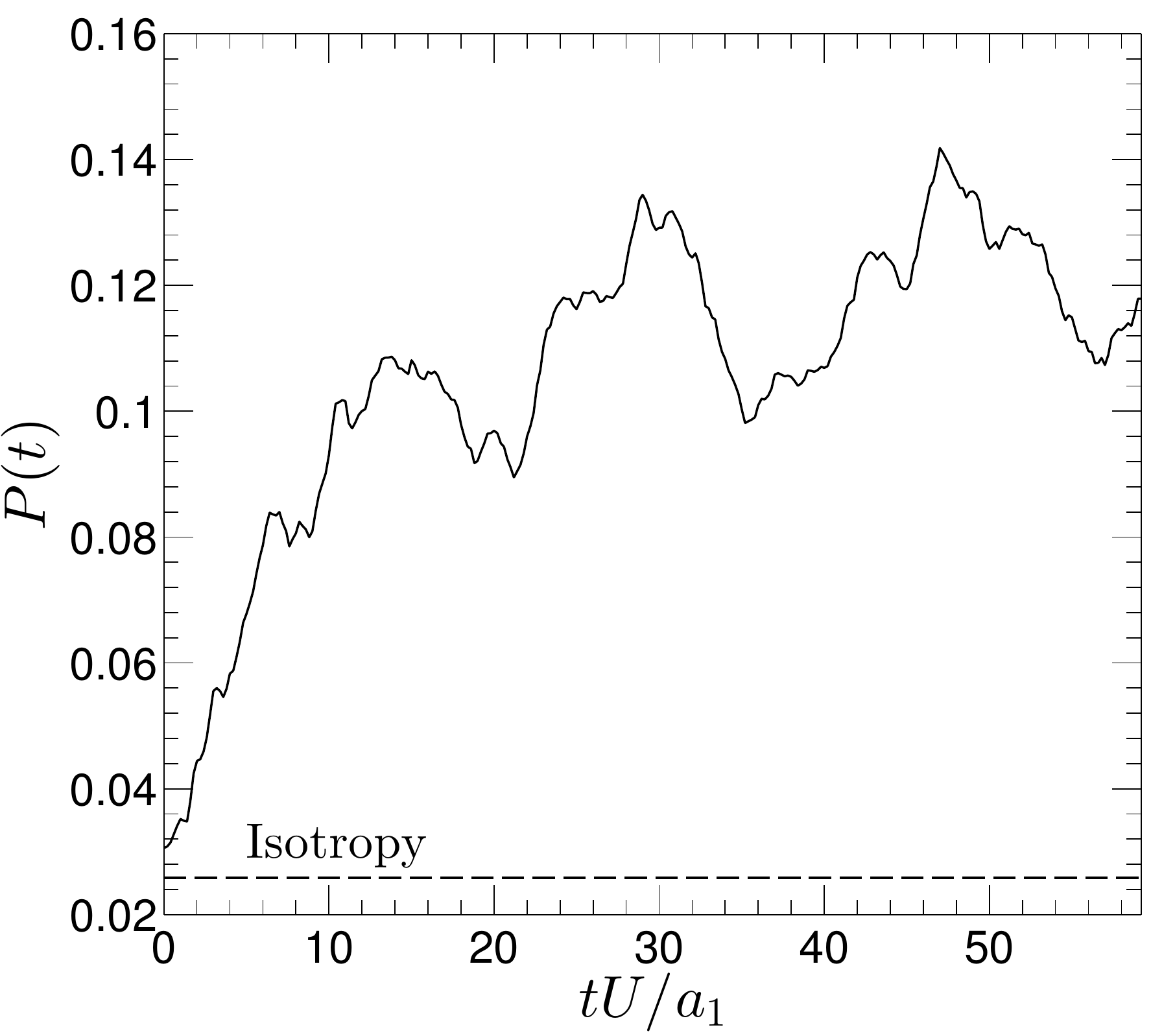}}
\caption{ Simulation of a dilute suspensions, $\phi_v = 0.05$, of prolate spheroidal pushers, $B_2 = -1.5$, $B_1 = 0$, with aspect ratio $\frac{a_1}{a_2}=\frac{a_1}{a_3}=3$.
a) Snapshot at $t = 30U/a_1$. 
Colors indicate the velocity magnitude normalized by the individual swimming speed.
b) Time evolution of polar order $P(t)$. 
\symbol{\dashed}{}{10}{0}{black}{black} : isotropic value of the polar order parameter $1/\sqrt{N_p} = 0.026$.
} 
\end{figure}

Previous particle-based simulations \cite{saintillan_Shelley_2012,Lushi2013} and continuum theory results \cite{Saintillan2008, Baskaran2009, Ezhilan2013} predict an unstable isotropic state for suspensions of prolate spheroidal pushers.  Using FCM, we simulate a dilute suspension, $\phi_v = 0.05$, of $N_p = 1500$ spheroidal pushers with aspect ratio $\frac{a_1}{a_2}=\frac{a_1}{a_3}=3$.  For these simulations, the stresslet parameter is $B_2 = -1.5$ and the degenerate quadrupole is set to zero, $B_1 = 0$.  Steric interactions are included by using a soft repulsive  potential with a $\sim r^{-5}$ decay and surface-to-surface distance approximated by the Berne-Pechukas range parameter \cite{Allen2006}.
Fig. \ref{fig:snapshot_ell} shows a snapshot of the suspension where clusters of swimmers have velocities nearly $1.6$ times larger than the isolated swimming speed.
As in \cite{saintillan_Shelley_2012, Lushi2013}, we see that the isotropic state is not stable and the polar order parameter increases with time (Fig. \ref{fig:P_t_ell}).  We do not see as large an increase in $P(t)$ as in \cite{saintillan_Shelley_2012, Lushi2013} due to the relatively low aspect ratio of our swimmers.  The tumbling effect due to Jeffery orbits and the steric torques that tend to align adjacent swimmers are much lower than they are for rod-like swimmers.

\subsection{Suspension dynamics of time-dependent swimmers}

Here, we show how to incorporate time-dependence into our model.  Using the procedure outlined in \cite{Ghose2014}, we determine the time-dependent multipole coefficients $B_1(t)$ and $B_2(t)$ from the experimental data provided in \cite{Guasto2010}.  We then perform suspension simulations that show time-dependence at the level of individual swimmers can affect the overall suspension properties.

\subsubsection{A single time-dependent swimmer}
\label{sec:Time_dep}
Recent experiments \cite{Guasto2010} quantified the periodic swimming gait and resulting flow field of the algae cell \emph{Chlamydomonas Rheinardtii}.  They extracted the swimming speed, the induced velocity field, and the power dissipation, showing also that all can be represented as periodic functions of time.  A recent theoretical investigation \cite{Ghose2014} showed that these quantities could be reproduced using a multipole-based model.  In their study, they considered three time-dependent multipoles: a stresslet, a degenerate quadrupole (or potential dipole) and a ``septlet." The stresslet decays as $r^{-2}$ whereas the degenerate quadrupole and the ``septlet" decay like $r^{-3}$.  Here, we utilize only the stresslet and degenerate quadrupole terms and find that they are sufficient to reproduce the features measured by \cite{Guasto2010}.

The measured swimming speed from \cite{Guasto2010}, $U(t)$, can be represented using the truncated Fourier series in time (see \cite{Ghose2014}):
\begin{equation}
 U(t)=a_{0}+a_{1}\cos\left(\omega t\right)+a_{2}\cos\left(2\omega t\right)+b_{1}\sin\left(\omega t\right)+b_{2}\sin\left(2\omega t\right)
\end{equation}
where $\omega$ is the frequency of the swimming gait.  The mean swimming speed over one beat period, $T=2\pi/\omega$, is given by $a_0 = 49.54a.s^{-1}$, where $a=2.5\mu m$ is the radius of the microorganism.  The values for the remaining coefficients are provided in the supplementary information of \cite{Ghose2014}.  From Eq. \eqref{eq:UB1A1}, we can immediately determine the time-dependent degenerate quadrupole strength
\begin{equation}
 B_1(t) = \dfrac{3}{2}U(t)
\label{eq:B1_U}
\end{equation}
in order to preserve the instantaneous force-free condition.  Unlike this term, there is more than one way to calibrate the stresslet strength, $B_2(t)$. For example, one could determine $B_2(t)$ using the power dissipation measurements from \cite{Guasto2010} and Eq. \eqref{eq:UB1A1} from \cite{Blake1972}
\begin{equation}
 \Pi_d(t) = \dfrac{2}{3} \pi \eta a \left( 8B_1(t)^2 + 4B_2(t)^2 \right).
 \label{eq:P_blake}
\end{equation}
for the power dissipated by a squirmer.  Using this approach, we found that our resulting flow field did not match the experimental results from \cite{Guasto2010}.  We instead determine $B_2(t)$ directly from the experimental flow field by fitting to the location of the moving stagnation point.  This is the same approach adopted by \cite{Ghose2014}.  We utilize three Fourier modes to describe the time evolution of $B_2$
\begin{equation}
\begin{array}{crl}
B_{2}(t) & = & c_{0}+c_{1}\cos\left(\omega t+\varphi_{c1}\right)+c_{2}\cos\left(2\omega t+\varphi_{c2}\right)+c_{3}\cos\left(3\omega t+\varphi_{c3}\right)\\
 &  & +s_{1}\sin\left(\omega t+\varphi_{s1}\right)+s_{2}\sin\left(2\omega t+\varphi_{s2}\right)+s_{3}\sin\left(3\omega t+\varphi_{s3}\right),
\end{array}
\label{eq:B2_emp}
\end{equation}
where the amplitudes $c_0,c_1,s_1,...$ and phases $\varphi_{c1}, \varphi_{s1},...$ are fitted manually. Appendix C gives the resulting values of these parameters.

\begin{figure}
\centering
\subfloat[]{ \includegraphics[height=8cm]{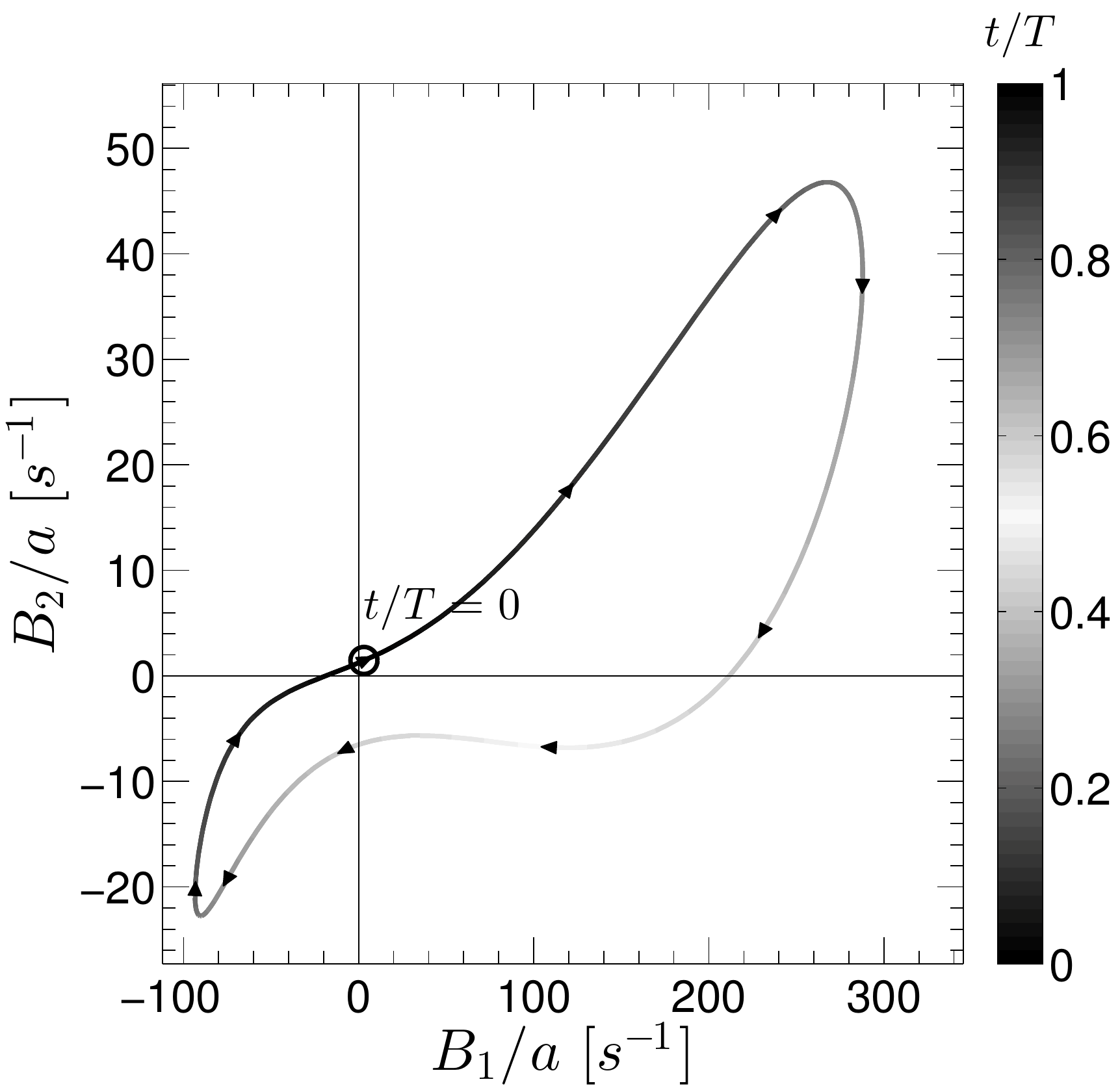} \label{fig:B1vsB2} }
\subfloat[]{ \includegraphics[height=8cm]{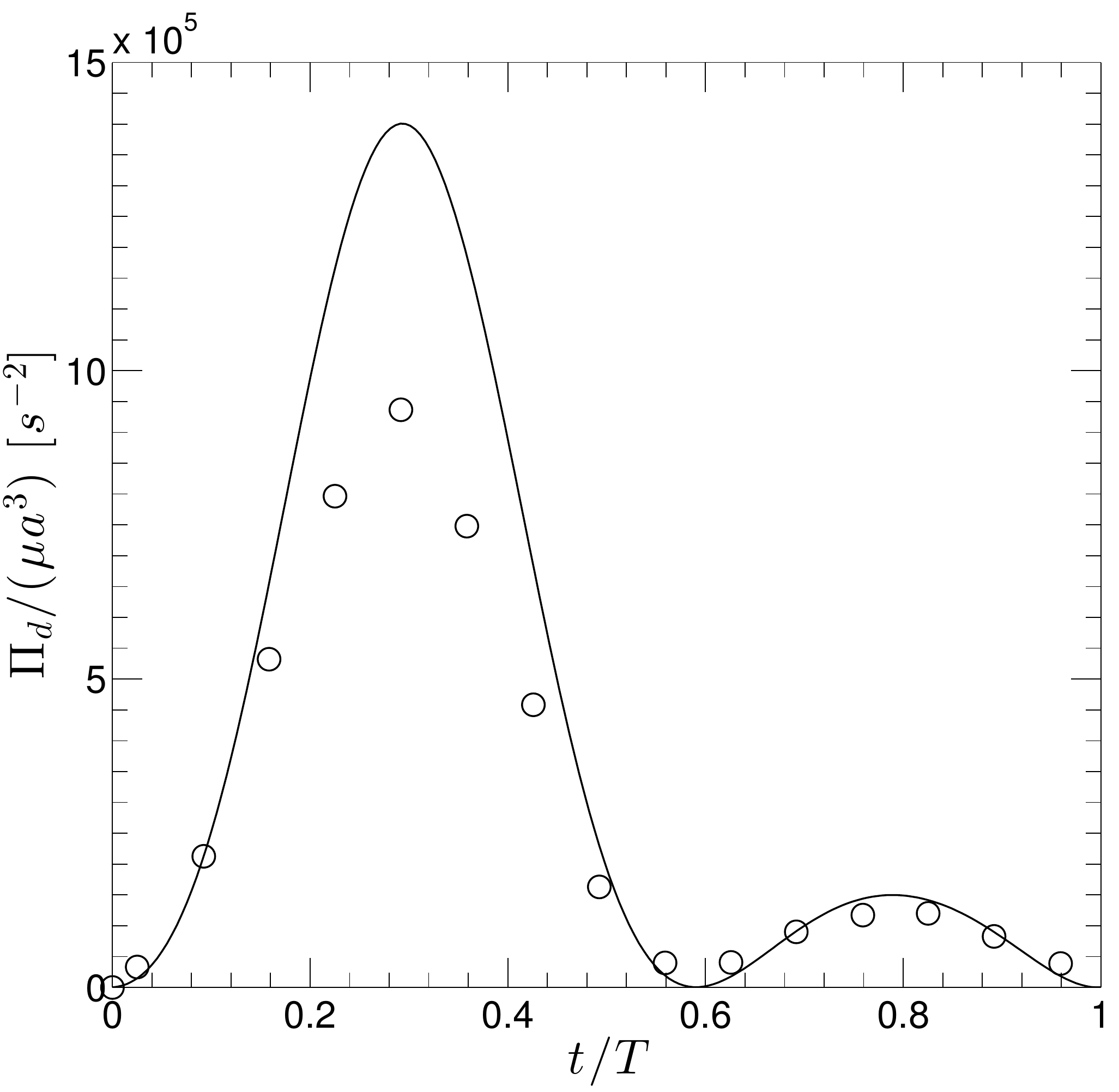} \label{fig:P_guasto}}
\caption{a) ($B_1(t)$, $B_2(t)$) phase diagram for one beat cycle. 
b) Power dissipation $\Pi_d(t)$ over one beat cycle. 
\symbol{\solid}{}{10}{0}{black}{black} : FCM,
\symbol{}{\bigcircle}{-5}{0}{black}{black} : results from \cite{Guasto2010}.
}
\end{figure}

The phase diagram in Fig. \ref{fig:B1vsB2} shows the value of $B_1$ versus $B_2$ and it is similar to that found by \cite{Ghose2014}.  We extract the average value of $\beta(t) = B_2(t)/B_1(t)$ over one beat cycle
\begin{equation}
\bar{\beta} = \dfrac{1}{T}\int\limits_{0}^{T} B_2(t)/B_1(t) dt = 0.1
\end{equation}
which corresponds to a puller squirmer with a relatively small stresslet magnitude.  Fig. \ref{fig:P_guasto} shows the resulting power dissipation as determined from Eq. \eqref{eq:P_blake}.  It reaches a peak value at $t/T \approx 0.3$, which coincides with the time at which the swimming speed reaches its maximum value.  We note that unlike \cite{Guasto2010} our swimmers generate axisymmetric flow fields and we are considering a 3D periodic domain.  Despite this, we achieve a qualitatively similar power dissipation profile with slightly greater values during the first half of the beat cycle.

\begin{figure}
\centering
\includegraphics[height=10cm]{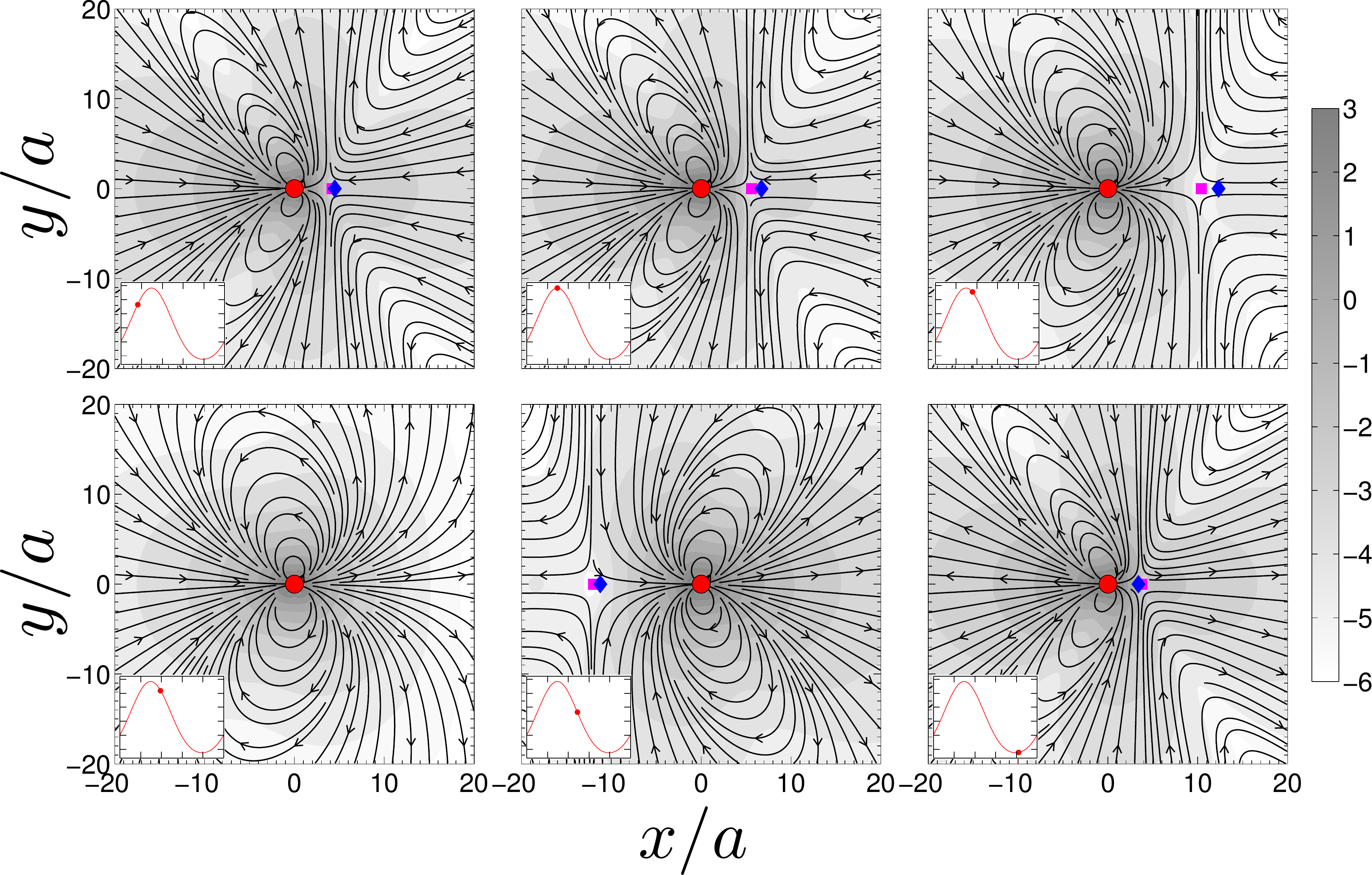}
\caption{Snapshots of the time-dependent flow field around a model \emph{Chlamydomonas}.
Background grey levels represent the natural logarithm of the norm of the velocity field in radii.$s^{-1}$.
\symbol{}{\blackssquare}{-5}{0}{black}{magenta}: position of the stagnation point given by the FCM model.
\symbol{}{\blacklosange}{-5}{0}{black}{blue}: position of the stagnation point measured by \cite{Guasto2010}.
(Insets) Swimming speed along the beat cycle.
}
\label{fig:Unsteady_usw} 
\end{figure}

Fig. \ref{fig:Unsteady_usw} shows the flow field around our  model of \emph{C. Rheinardtii} at six different times during its beat cycle. These time points are chosen to correspond to those in Fig. 3 of \cite{Guasto2010}.  We achieve very similar streamlines and, by construction, the position of the stagnation point matches the experimental data very well. We also note that our flow field is similar to that given by the multipole model found in \cite{Ghose2014} even though we do not include the rapidly decaying ``septlet" term in our model. These results  illustrate that our properly tuned, time-dependent squirmer model can yield flow fields very similar to those of real organisms.

\subsubsection{Interactions between two model C. Rheinardtii}

\begin{figure}
\centering
\includegraphics[height=11cm]{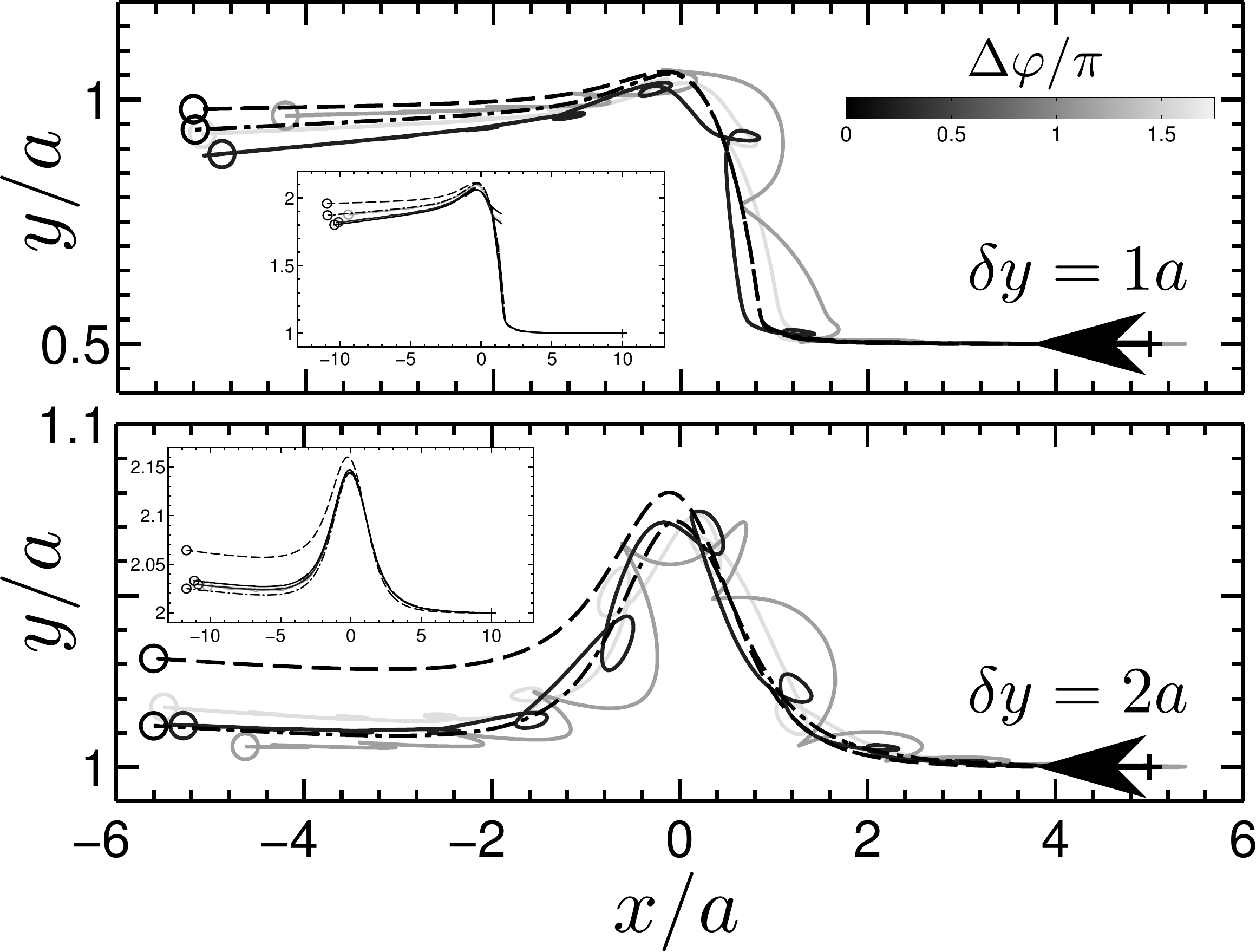}
\caption{Relative trajectories of two model \emph{C. Rheinardtii} swimming in opposite directions with stroke phase shift $\Delta \varphi$ and initial transverse distance $\delta y =1a$ and $2a$. The grey levels lighten as $\Delta \varphi$ increases from $\Delta \varphi = 0$ $\rightarrow$ $\Delta \varphi = 7\pi/4$.
\symbol{\dashdot}{}{7}{0}{black}{black}: trajectory of two steady squirmers with $\beta = 0$.
\symbol{\dashed}{}{7}{0}{black}{black}: trajectory of two steady squirmers with $\beta = 0.1$.
(Top) $\delta y = 1a$. 
(Bottom) $\delta y = 2a$
} 
\label{fig:Traj_unsteady}
\end{figure}

We first consider pairwise interactions between two model \emph{C. Rheinardtii} for two initial configurations, $\delta y = 1a$ and $2a$ (cf. Section \ref{Collision_squirmers}).  We introduce a phase shift, $\Delta \varphi$, between their swimming cycles.  When $\Delta \varphi = 0$, the swimmers are synchronized, whereas for $\Delta \varphi = \pi$ they are completely out of phase.  Fig. \ref{fig:Traj_unsteady} shows the effect of this phase shift on the trajectories.  For each case, we show only the trajectory of the swimmer labelled ``2" in Fig. \ref{fig:Sketch_Sq_coll}.  We also provide the trajectories of steady swimmers with $\beta = 0$ and $\beta = 0.1$ for comparison.  Recall that $\beta = 0.1$ corresponds to the average value of $B_2(t)/B_1(t)$ in our model.  We see that the trajectories, including the final positions and orientations, do depend on $\Delta \varphi$.  We find, however, that these variations are small compared to the swimmer separation distance (see the inset of Fig. \ref{fig:Traj_unsteady} for the relative trajectories in the frame of one swimmer).  The trajectories are also close to those for steady squirmers with $\beta = 0, 0.1$.

\subsubsection{Collective dynamics of time-dependent swimmers}
\begin{figure}
\centering
\includegraphics[width=15cm]{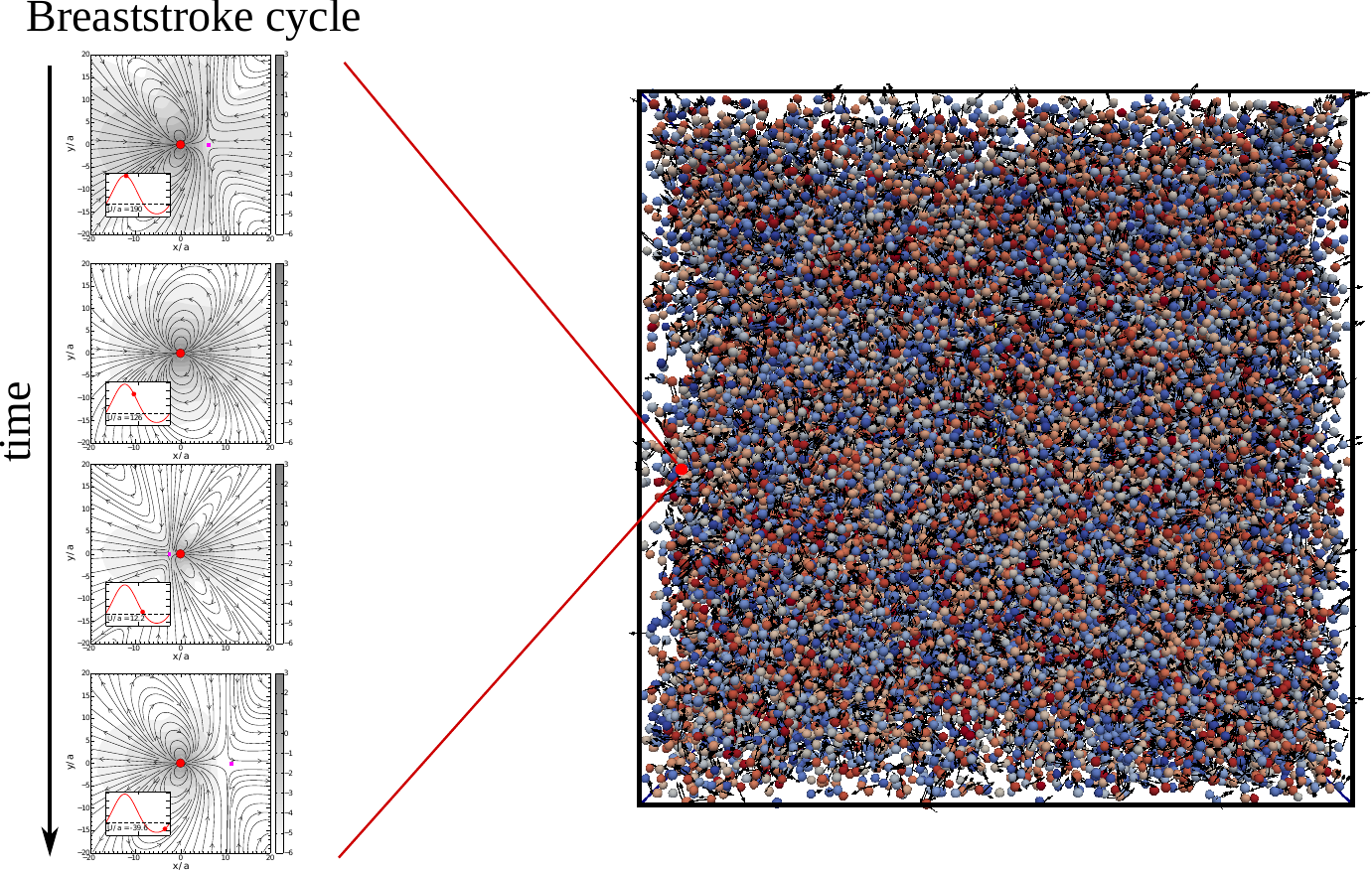}
\caption{Sketch of the simulations with time-dependent swimmers. Each swimmer $n$ generates time-dependent flow disturbances with the same amplitude and frequency, but the phase $\Delta \varphi_n$ may be different for each swimmer.
} 
\label{fig:sketch_time_dep}
\end{figure}

In this section, we present results from suspension simulations using our unsteady model for \emph{C. Rheinardtii}. To the best of our knowledge, similar results have not appeared in the literature.  We aim to show in this initial study that time-dependence of individual swimmers can affect their overall distribution.  Specifically, we show that the distribution of beat phases affects the steady-state polar order of the suspension. 
 Fig. \ref{fig:sketch_time_dep} shows a sketch of the simulations with time-dependent swimmers.

We consider a suspension of $N_p = 1395$ time-dependent swimmers distributed uniformly in a triply periodic domain.  The volume fraction is $\phi_v = 0.1$.  We take the initial swimming directions to be distributed uniformly over the unit-sphere.  For swimmer $n$, its gait is characterized by $B_1(t + \Delta \varphi_n)$ and $B_2(t + \Delta \varphi_n)$, where $\Delta \varphi_n$ is the value of its beat phase.  We consider two different distributions of the beat phases.  The phases are either uniformly distributed with $\Delta \varphi_n \in [0;2\pi]$, or $\Delta \varphi_n = 0$ for all $n$ such that all swimmers are synchronized.  We run the simulations for $3700$ dimensionless time units, corresponding to $4000$ beat cycles.  As mentioned in Section \ref{sec:Time_dep}, the mean velocity of \emph{C. Rheinardtii} is $U=49.54a.s^{-1}$.  Since we take the time-step $\Delta t = 0.0025a/U$, the total time for our simulations correspond to $1.5\times10^6 \Delta t$.

\begin{figure}
\centering
\subfloat[]{ \label{fig:P_t_comparison_with_beta_0_beta_0_12} \includegraphics[height=7cm]{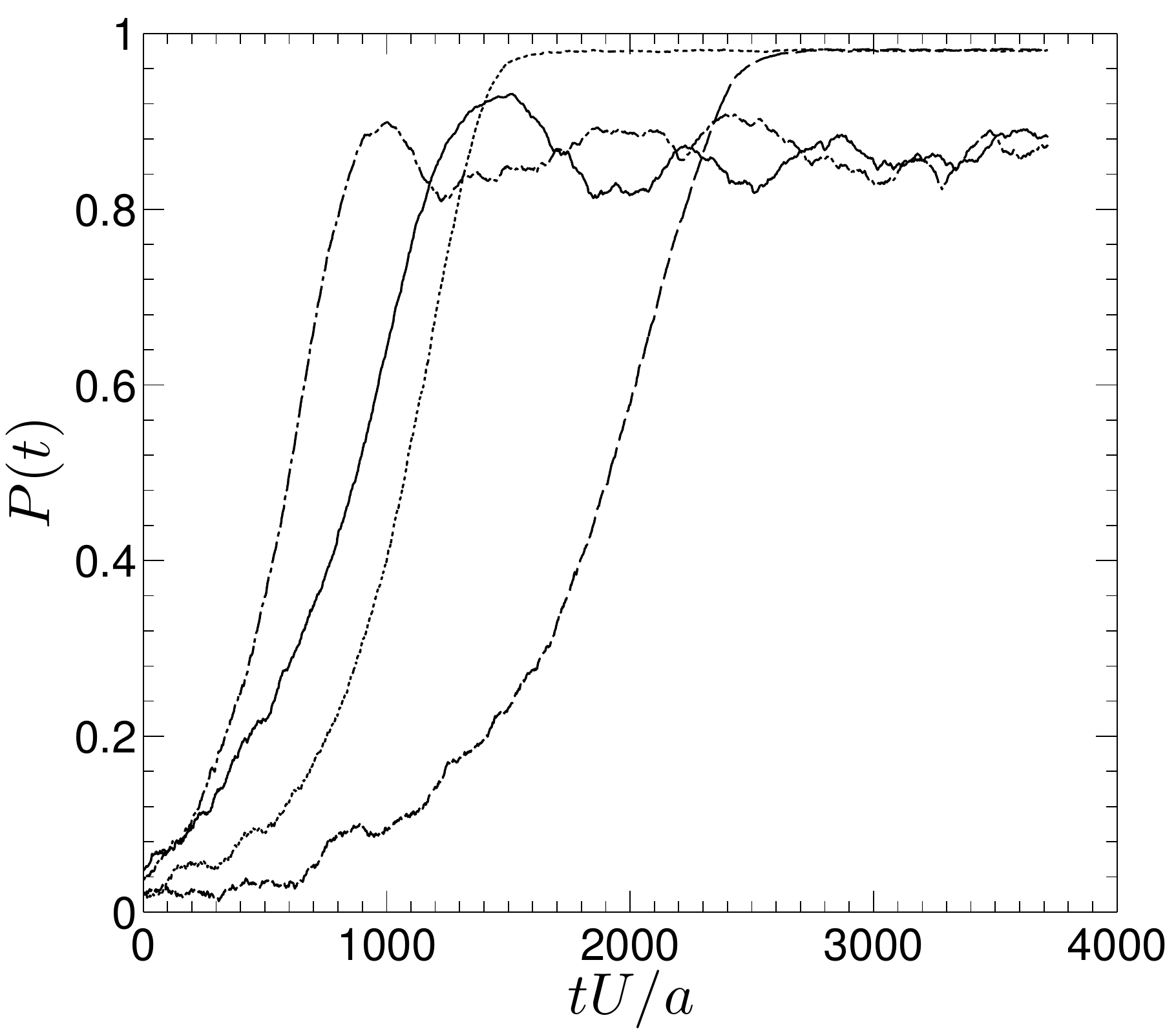}}
\subfloat[]{ \label{fig:Theta_pmean_unsteady} \includegraphics[height=7cm]{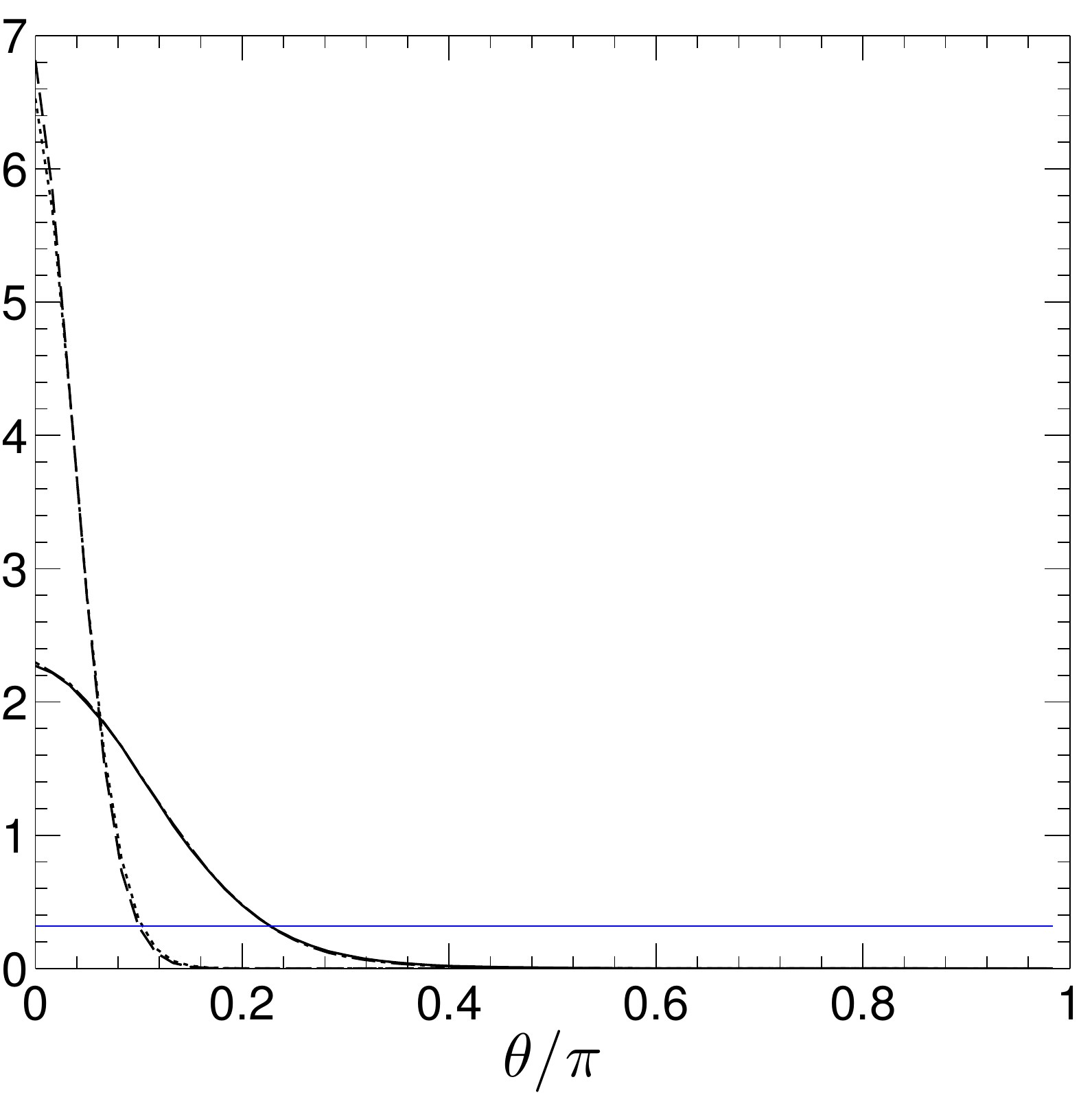}}
\caption{Polar order parameter and orientational distribution for suspensions of steady and time-dependent swimmers.
\symbol{\solid}{}{10}{0}{black}{black} : time-dependent swimmers with random phase;
\tdot \tdot \tdot \tdot \tdot \tdot : synchronized time-dependent swimmers;
\symbol{\dashed}{}{10}{0}{black}{black} : steady swimmers with $\beta = 0$;
\symbol{\dashdot}{}{10}{0}{black}{black} : steady swimmers with $\beta = 0.1$.
a) Time evolution of polar order.
b) Steady state orientation distribution around the mean director $\left.\Psi(\theta)\right|_{\left\langle \mathbf{p}\right\rangle }$.
\symbol{\solid}{}{10}{0}{blue}{black} : uniform distribution $\Psi_0(\theta) = 1/\pi$; 
} 
\label{fig:Orient_unsteady}
\end{figure}

Fig. \ref{fig:P_t_comparison_with_beta_0_beta_0_12} shows the evolution of the polar order parameter for these two distributions of beat phase.  Also shown is the polar order parameter for suspensions of steady swimmers with swimming speeds equal to the mean swimming speed for the time-dependent case and either $\beta = 0$ or $\beta = 0.1$.  We find that when there is a uniform distribution of the beat phase, we achieve results that match the steady case with $\beta = 0.1$.  On the other hand, if the swimmers are synchronized, we find that the polar order parameter matches that for the steady case with $\beta = 0$.  Fig. \ref{fig:Theta_pmean_unsteady} shows the steady-state orientational distributions $\left.\Psi(\theta)\right|_{\left\langle \mathbf{p}\right\rangle }$ about the mean director.  Again, we see that the distribution for case of a uniform distribution of beat phases matches that for the steady case with $\beta = 0.1$, while the synchronized suspension has the same distribution as the $\beta = 0$ case.  These new results show that the distribution of beat phases can strongly affect the orientational order of the suspension.  

\section{Conclusions}
In this study, we presented an extension of FCM to compute the hydrodynamic interactions between a large number ($10^4 - 10^5$) of active self-propelled particles in semi-dilute suspensions.  Our approach builds from the spherical squirmer model developed by Blake \cite{Blake1971} by including in the usual FCM force distribution the regularized singularities that correspond to the surface squirming modes.  We have shown that our model readily allows for different swimming gaits, as well as spheroidal swimmer shapes, and can also account for effects such as steric repulsion. 
We demonstrated the accuracy of our model by comparing velocity fields, pairwise interactions, and trajectories with the analytical or numerical results given by the full squirmer model.  
In our squirmer suspension simulations, we recovered the polar order found in other studies, but also quantified the effects of domain size on the final steady-state.  Specifically, we showed that for puller suspensions with $\beta = 1$, the final value of the polar order parameter decreases and the time needed to reach steady state increases as the system gets larger.  We did, however, find that these quantities appeared to converge to an asymptotic value and that polar order should be present even for an infinitely large domain.  The scale of our simulations allowed us to compile robust statistics and examine in detail the orientational distribution and local micro-structure of the suspension.  By extending the model to time-dependent swimming gaits, we also illustrated for the first time, to best of our knowledge, that time-dependence at the level of individual swimmers can affect the final steady-state distribution of the suspension.

We see that the squirmer model within the FCM framework provides an effective computational approach to simulate active suspensions, allowing for particle numbers that begin to approach continuum levels.  As a result, we feel that our computational model can both be used to complement experimental research, as well as inform the development of continuum models.  For example, we have shown that the time-dependency of swimming gaits can be readily included in our model by tuning it to available experimental data \cite{Guasto2010}.  By tuning our model to similar experimental data, but for a wider zoology of microorganisms \cite{Kiorboe2014}, we could assess the possible differences in collective dynamics exhibited by different species, or even look into how one species might interact with another.  In addition, our model could be an effective approach to address questions regarding the mixing of background scalar fields \cite{Lambert2013} or passive tracers dynamics \cite{Leptos2009, Lin2011} in active, time-dependent suspensions.  The effects of tracer Brownian motion \cite{Keaveny2014} can be included with squirmer model and the competition between tracer diffusion and advection due to the flows produced by the swimmers can be assessed.

\section*{Acknowledgements}
The authors would like to thank Prof. Martin Maxey for many insightful comments and acknowledge his instrumental role in bringing the authors together to carry out this work.  We would also like to acknowledge Prof. Raymond Goldstein, Dr. Ottavio Croze, Navish Wadhwa and Prof. Pierre Degond for interesting discussions. We are also grateful to Annaig Pedrono and Dr. Pascal Fede for providing helpful support on numerical issues. This work is developed within the MOTIMO ANR Project. Simulations were performed on the Calmip supercomputing mesocenter. We thank INPT for funding the international collaboration between IMFT and Imperial College.  We would also like to acknowledge travel support from EPSRC Mathematics Platform grant EP/I019111/1.

\renewcommand{\theequation}{A.\arabic{equation}}    
\setcounter{equation}{0}  

\section*{Appendix A: Self-induced effects for spheroidal swimmers}
In this Appendix, we show how to compute the artificial self-induced effects due to the squirming modes (see Section \ref{sq_interactions}) for the case of spheroidal swimmers described in Section \ref{subsec:Spheroid_FCM}.\\

In the following calculations, we consider an ellipsoidal squirmer located at the origin $\mathbf{Y}= \mathbf{0}$ with swimming direction $\mathbf{p} = \hat{\mathbf{e}}_1$.

\subsection*{Self-induced velocity $\mathbf{W}$ due to the degenerate quadrupole $\mathbf{H}$}

Using the FCM envelopes for ellipsoidal particles, the force-distribution corresponding to the degenerate quadrupole generated by the squirmer will be given by
\begin{equation}
\mathbf{f}_H = \mathbf{H}\nabla^{2}\Theta^{ell}(\mathbf{x}).
\end{equation}
where we also have $\boldsymbol{\sigma}_{\Theta}=\left(\sigma_{\Theta;1},\sigma_{\Theta;2},\sigma_{\Theta;3}\right)$, the width of the Gaussian envelope $\Theta^{ell}(\mathbf{x})$ for each of the semi-axes.  The resulting fluid flow can be obtained in Fourier space as
\begin{equation}
\widehat{u}_{i}=\displaystyle-\frac{1}{\mu}\left(\delta_{ij}-\frac{k_{i}k_{j}}{k^{2}}\right)\widehat{\Theta}^{ell}(\mathbf{k})H_{j}.
\end{equation}
The self-induced velocity, $\mathbf{W}$, will be given by volume averaging this fluid velocity against the monopole Gaussian envelope $\Delta^{ell}(\mathbf{x})$ for ellipsoidal particles.  Accordingly, the expression for $\mathbf{W}$ is
\begin{equation}
\begin{array}{cll}
W_{i} & = &  \displaystyle\frac{1}{8\pi^{3}}\int\limits_{\mathbb{R}^3}\widehat{u}_{i}\widehat{\Delta}^{ell}\left(\mathbf{k}\right)d^{3}\mathbf{k},\\
\\
 & =- &\displaystyle \frac{1}{8\pi^{3}}\int\limits_{\mathbb{R}^3}\frac{1}{\mu}\left(\delta_{ij}-\frac{k_{i}k_{j}}{k^{2}}\right)\widehat{\Theta}^{ell}(\mathbf{k})\widehat{\Delta}^{ell}\left(\mathbf{k}\right)d^{3}\mathbf{k}H_{j},
\end{array}\mbox{}
\end{equation}
where we write the width of the Gaussian envelope $\Delta^{ell}(\mathbf{x})$ for each semi-axes as $\boldsymbol{\sigma}_{\Delta}=\left(\sigma_{\Delta;1},\sigma_{\Delta;2},\sigma_{\Delta;3}\right)$.  This expression can be viewed as $W_{i} = M_{ij}^{HU} H_j$ where  
\begin{equation}
M_{ij}^{HU}=\displaystyle-\frac{1}{8\pi^{3}}\int\limits_{\mathbb{R}^3}\frac{1}{\mu}\left(\delta_{ij}-\frac{k_{i}k_{j}}{k^{2}}\right)\widehat{\Theta}^{ell}(\mathbf{k})\widehat{\Delta}^{ell}(\mathbf{k})d^{3}\mathbf{k}.
\end{equation}
is the FCM self-mobility matrix that relates the particle velocity to the degenerate quadrupole coefficient.  For a spheroidal particle ($\sigma_{\Delta;2}=\sigma_{\Delta;3}$ and $\sigma_{\Theta;2}=\sigma_{\Theta;3}$) and $\mathbf{H}$ in the direction $\hat{\mathbf{e}}_{1}$, we need only to consider $M_{11}^{HU}$ to obtain the self-induced effects.  This mobility matrix entry can be written as 
\begin{equation}
\begin{array}{ccl}
M_{11}^{HU} & = & \displaystyle-\frac{1}{8\pi^{3}}\int\limits_{\mathbb{R}^3}\frac{1}{\mu}\left(1-\frac{k_{1}^{2}}{k^{2}}\right)\widehat{\Theta}^{ell}(\mathbf{k})\widehat{\Delta}^{ell}(\mathbf{k})d^{3}\mathbf{k},\\
\\
 & = & \displaystyle-\frac{1}{8\pi^{3}}\int\limits_{\mathbb{R}^3}\frac{1}{\mu}\left(1-\frac{k_{1}^{2}}{k^{2}}\right)\mbox{exp}\left[-\frac{1+\chi^{2}}{2}\sigma_{\Delta;2}^{2}\left(\gamma^{2}k_{1}^{2}+k_{2}^{2}+k_{3}^{2}\right)\right]d^{3}\mathbf{k},
\end{array}
\label{eq:M11_QU_k}
\end{equation}
where $\gamma=\dfrac{a_{1}}{a_{2}}=\dfrac{\sigma_{\Theta;1}}{\sigma_{\Theta;2}}=\dfrac{\sigma_{\Delta;1}}{\sigma_{\Delta;2}}$ is the spheroid aspect ratio and  $\chi=\dfrac{\sigma_{\Theta;1}}{\sigma_{\Delta;1}}=\dfrac{\sigma_{\Theta;2}}{\sigma_{\Delta;2}}=\dfrac{\sigma_{\Theta;3}}{\sigma_{\Delta;3}}=\left(\dfrac{\pi}{6}\right)^{1/3}$. 

If we introduce cylindrical coordinates $\left(k_{1}=z,\,\, k_{2}=r\cos\theta,\,\, k_{3}=r\sin\theta\right)$, Eq. \eqref{eq:M11_QU_k} becomes
\begin{equation}
\begin{array}{ccl}
M_{11}^{HU} & = &\displaystyle -\frac{1}{8\pi^{3}}\intop\limits _{0}^{2\pi}\intop\limits _{-\infty}^{+\infty}\intop\limits _{0}^{+\infty}\frac{1}{\mu}\left(1-\frac{z^{2}}{r^{2}+z^{2}}\right)r\mbox{\,\ exp}\left[-\frac{1+\chi^{2}}{2}\sigma_{\Delta;2}^{2}\left(\gamma^{2}z^{2}+r^{2}\right)\right]drdzd\theta,\\
\\
 & = & \displaystyle -\frac{1}{4\pi^{2}}\intop\limits _{-\infty}^{+\infty}\intop\limits _{0}^{+\infty}\frac{1}{\mu}\left(1-\frac{z^{2}}{r^{2}+z^{2}}\right)r\mbox{\,\ exp}\left[-\frac{1+\chi^{2}}{2}\sigma_{\Delta;2}^{2}\left(\gamma^{2}z^{2}+r^{2}\right)\right]drdz.
\end{array}\label{eq:Eq_cylindrical}
\end{equation}
Let $t=\dfrac{z}{\sqrt{r^{2}+z^{2}}}$, which gives $z=\dfrac{rt}{\sqrt{1-t^{2}}}$
and $\dfrac{dz}{dt}=\dfrac{r}{\left(1-t^{2}\right)^{3/2}}$ and Eq. (\ref{eq:Eq_cylindrical}) becomes
\begin{equation}
\begin{array}{ccl}
M_{11}^{HU} & = & \displaystyle-\frac{1}{4\pi^{2}\mu}\intop\limits _{-1}^{1}\intop\limits _{0}^{+\infty}\left(1-t^{2}\right)\frac{r^{2}}{\left(1-t^{2}\right)^{3/2}}\mbox{\,\ exp}\left[-\frac{1+\chi^{2}}{2}\sigma_{\Delta;2}^{2}\left(\gamma^{2}\frac{r^{2}t^{2}}{1-t^{2}}+r^{2}\right)\right]drdt,\\
\\
 & = &\displaystyle -\frac{1}{4\pi^{2}\mu}\intop\limits _{-1}^{1}\left(1-t^{2}\right)^{-1/2}\intop\limits _{0}^{+\infty}r^{2}\mbox{\,\ exp}\left[-r^{2}\frac{1+\chi^{2}}{2}\sigma_{\Delta;2}^{2}\left(\gamma^{2}\frac{t^{2}}{1-t^{2}}+1\right)\right]drdt.
\end{array}
\end{equation}
After integration by parts with respect to $r$, we have  
\begin{equation}
M_{11}^{HU}=\displaystyle-\frac{1}{4\pi^{2}\mu}\intop_{-1}^{1}\left(1-t^{2}\right)\left[\frac{\sqrt{\pi}}{2}\frac{1}{2\sigma_{\Delta;2}^{3}\left(\frac{1+\chi}{2}\right)^{3/2}\left(\gamma^{2}\frac{t^{2}}{1-t^{2}}+1\right)^{3/2}}\right]dt,
\end{equation}
and eventually
\begin{equation}
M_{11}^{HU}=\displaystyle-\frac{1}{16\pi^{3/2}\mu\sigma_{\Delta;2}^{3}\left(\frac{1+\chi}{2}\right)^{3/2}}\intop_{-1}^{1}\frac{\left(1-t^{2}\right)^{2}}{\left(1+\left(\gamma^{2}-1\right)t^{2}\right)^{3/2}}dt.\label{eq:MQ_inetgrale_finale}
\end{equation}

From the integrals in the Appendix in \cite{Liu2009}, this maybe be expressed compactly as 
\begin{equation}
M_{11}^{HU}=-2C\left(I_{0}-I_{1}\right).
\end{equation}
where $C=\left(32\pi^{3/2}\mu\sigma_{\Delta;2}^{3}\left(\frac{1+\chi}{2}\right)^{3/2}\right)^{-1}$ and $I_{0}$ and $I_{1}$ are coefficients whose expressions are provided in the Appendix of \cite{Liu2009}.

The self-induced velocity $\mathbf{W}$ is then simply
\begin{equation}
\mathbf{W} = \mathbf{M}_{11}^{HU}\mathbf{H},
\end{equation}
and its artificial effect can be subtracted away from the total volume average velocity as done in Eq. \eqref{eq:part_vel_self_ind}.

\subsection*{Self-induced rate of strain $\mathbf{K}$  due to the squirming dipole $\mathbf{G}$}

In the frame of the swimmer, the non-zero entries of the self-induced rate of strain $\mathbf{K'}$ are given by
\begin{equation}
\begin{array}{cll}
\mathbf{K'}_{11} & = & M^{FCM}_{EG;1111}G_{11}+M^{FCM}_{EG;1212}G_{22}+M^{FCM}_{EG;1313}G_{33},\\
\mathbf{K'}_{22} & = & M^{FCM}_{EG;2222}G_{22}+M^{FCM}_{EG;2323}G_{33}+M^{FCM}_{EG;2121}G_{11},\\
\mathbf{K'}_{33} & = & M^{FCM}_{EG;3333}G_{33}+M^{FCM}_{EG;3232}G_{22}+M^{FCM}_{EG;3131}G_{11}.
\end{array}
\end{equation}
$\mathbf{M}^{FCM}_{EG}$ is the symmetric fourth-order FCM mobility tensor relating the particle rate-of-strain $\mathbf{E}$ to the swimming dipole coefficient $\mathbf{G}$ \cite{Liu2009}. Its components read
\begin{equation}
\begin{array}{lll}
M^{FCM}_{EG;1111} & = & -2C\left(I_1 - I_2\right),   \\
M^{FCM}_{EG;1212}= M^{FCM}_{EG;1313} & = & C\left(I_1 - I_2\right), \\
M^{FCM}_{EG;2222} =  M^{FCM}_{EG;3333} & = & -C\left(\dfrac{1}{4}I_0 + \dfrac{1}{2}I_1 - \dfrac{3}{4}I_2\right), \\
M^{FCM}_{EG;2323} & = &  -C\left(-\dfrac{1}{4}I_0 + \dfrac{1}{2}I_1 - \dfrac{1}{4}I_2\right), 
\end{array}
\end{equation}
where $C$ is given above and $I_2$ is a coefficient whose expression is detailed in the Appendix of \cite{Liu2009}. To obtain the self-induced rate of strain in the lab frame, $\mathbf{K}$, one just needs to apply a rotation operator onto $\mathbf{K'}$:
\begin{equation}
\boldsymbol{\mathbf{K}}=\mathcal{Q}\boldsymbol{\mathbf{K'}}\mathcal{Q}^{T},
\end{equation}
where $\mathcal{Q}=\left(\begin{array}{ccc}
\hat{\mathbf{e}}_{1} & \hat{\mathbf{e}}_{2} & \hat{\mathbf{e}}_{3}\end{array}\right)$ is the rotation matrix of the swimmer.

\renewcommand{\thetable}{B.\arabic{table}}    
\setcounter{table}{0}  

\section*{Appendix B: Repulsive force parameters}
The table below lists the parameter values used in our squirmer trajectory calculations.
\begin{table}[H]
\begin{centering}
\begin{tabular}{cccc}
\toprule
 & $R_{\mbox{ref}}/a$ & $F_{\mbox{ref}}/6\pi\eta aU$ & Exponent $\gamma$\tabularnewline
\midrule
\midrule  
 \symbol{\dashed}{}{10}{0}{blue}{black} & $2.2$ & $4$ & $2$\tabularnewline
\midrule 
 {\color{blue} \tdot  \tdot  \tdot \tdot \tdot \tdot \tdot}  & $2.4$ & $6$ & $1$\tabularnewline
\midrule 
\symbol{\solid}{}{10}{0}{blue}{black} & $3$ & $3$ & $5$\tabularnewline
\midrule 
 \symbol{\dashdot}{}{10}{0}{blue}{black} & $2.04$ & $5$ & $5$\tabularnewline
\bottomrule
\end{tabular}
\par\end{centering}

\caption{Parameters for the contact forces in Section \ref{Collision_squirmers} Fig. \ref{fig:Barriers_dz_1}}
\end{table}

\renewcommand{\thetable}{C.\arabic{table}}    
\setcounter{table}{0}  

\section*{Appendix C: Fitted values for \emph{C. Rheinardtii}}

This appendix provides the values used to tune the time-dependent squirmer model presented in Section \ref{sec:Time_dep}.

\begin{table}[H]
\begin{centering}
\begin{tabular}{ccccccc}
\toprule
 $c_0$ & $c_1$ & $c_2$ & $c_3$ & $s_1$ & $s_2$ & $s_3$ \tabularnewline
\midrule
\midrule  
 $4.5347$ & $64.053$ & $-84.7192$ & $-10.4545$ & $91.4529$ & $-92.6420$ & $-5.9849$ \tabularnewline
\bottomrule
\end{tabular}
\par\end{centering}

\caption{Magnitudes of the Fourier modes used to describe $B_2(t)$ in radii.$s^{-1}$.}
\end{table}

\begin{table}[H]
\begin{centering}
\begin{tabular}{cccccc}
\toprule
 $\varphi_{c_1}$ & $\varphi_{c_2}$ & $\varphi_{c_3}$ & $\varphi_{s_1}$ & $\varphi_{s_2}$ & $\varphi_{s_3}$ \tabularnewline
\midrule
\midrule  
 $1.7373$ & $3.5761$ & $-0.9154$ & $0.1666$ & $2.0054$ & -1.7125 \tabularnewline
\bottomrule
\end{tabular}
\par\end{centering}

\caption{Phases of the Fourier modes used to describe $B_2(t)$ in $rad$.}
\end{table}

\bibliographystyle{elsart-num}
\bibliography{FCMsquirm}
\end{document}